\documentclass[11pt, a4paper]{article}
\usepackage{styleBuding}
\usepackage{bm}
\usepackage{color}
\usepackage[usenames,dvipsnames,svgnames,table]{xcolor}
\usepackage{amsmath,amssymb}
\usepackage{graphicx}
\usepackage{slashed}
\usepackage{soul}
\usepackage{multirow}
\usepackage{longtable}
\usepackage{subfigure}
\usepackage{indentfirst}
\usepackage{diagbox}
\usepackage{siunitx} 
\usepackage{url} 
\usepackage[utf8]{inputenc}
\usepackage{cancel}

\pdfoutput=1
\usepackage{float}
\allowdisplaybreaks[4]

\begin{document}

\title{Interpreting Light Scalar Excesses and Heavy Scalar Cascades in the $\mu$-Term Extended NMSSM}

\author{Jingwei Lian$^{1,2}$}

\affiliation[1]{Henan Institute of Science and Technology, Xinxiang 453003, P.R. China}
\affiliation[2]{Department of Physics, Henan Normal University, Xinxiang 453007, P.R. China}

\emailAdd{lianjw@hist.edu.cn}

\date{}

\abstract{The hints for a scalar resonance near $95~{\rm GeV}$ in the LEP $b\bar b$ channel and in the LHC diphoton searches remain among the most persistent small deviations in the Higgs sector. At the same time, searches for a heavier resonance decaying into the observed Higgs boson and an additional scalar have become more restrictive, especially after the recent CMS analysis of the $b\bar b b\bar b$ final state. We study these observations in the $\mu$-term extended Next-to-Minimal Supersymmetric Standard Model ($\mu$NMSSM), where a singlet-like CP-even scalar can lie near $95~{\rm GeV}$ and the heavier doublet states can appear around the mass range probed by the CMS searches. After imposing Higgs data, extra Higgs limits, flavor constraints, electroweak precision observables and direct SUSY search bounds, we find viable regions that can accommodate the $95~{\rm GeV}$ diphoton and $b\bar b$ rates within the $2\sigma$ ranges. The viable points fall into two characteristic patterns. One gives a larger diphoton signal through a suppressed $h_s b\bar b$ coupling, while the other gives a larger LEP rate through stronger doublet mixing. The heavy CP-even scalar can decay through $H\to h h_s$ and $H\to h_s h_s$ with rates close to the reach of present and future searches. The CP-odd sector provides an additional channel, $A_2\to h A_1\to 4b$, which can populate the mass region around $(600,400)~{\rm GeV}$ where the latest CMS $4b$ analysis finds its largest local deviation, while remaining below its current limits. For the positive-$\mu$ subset, the inflation-inspired framework admits a gravitino as the lightest supersymmetric particle (LSP) dark matter candidate, with the lightest neutralino acting as the next-to-LSP and is typically long-lived at collider scales.}

\maketitle

\section{Introduction}\label{intro}
The discovery of a scalar boson with a mass of about 125 GeV at the Large Hadron Collider (LHC) has established the mechanism of electroweak symmetry breaking as a central component of the Standard Model (SM) of particle physics. 
Within the current experimental precision, the observed Higgs boson is remarkably consistent with the SM prediction.
Nevertheless, the scalar sector remains one of the most sensitive portals to physics beyond the Standard Model (BSM). The SM contains only one elementary scalar doublet, whereas many theoretically motivated extensions predict a richer Higgs spectrum containing additional CP-even, CP-odd and charged scalar states. Searching for such states is therefore a direct way of testing the structure of electroweak symmetry breaking and of probing the possible ultraviolet completion of the SM.

During the past several years, a number of mild but intriguing excesses have appeared in searches for additional scalar resonances. On the light scalar side, the LEP experiments first reported a mild excess in the process $e^+e^- \to Z^*\to Z +b\bar b$ for a scalar mass around $95$--$100~{\rm GeV}$, with signal strength~\cite{LEPWorkingGroupforHiggsbosonsearches:2003ing,Azatov:2012bz,Cao:2016uwt}
\begin{equation}
\mu_{b \bar b}^\text{exp} = 0.117 \pm 0.057.  \label{LEPrate}
\end{equation} 
More recently, searches in the diphoton channel at the LHC have shown an excess near $95~{\rm GeV}$, with the combined
ATLAS and CMS result favoring a signal strength with a local significance of $3.1\,\sigma$~\cite{CMS:2015ocq,CMS:2024yhz,Arcangeletti,Biekotter:2023oen}
\begin{equation}
\mu^{\rm exp}_{\gamma\gamma} \equiv  \mu^{\rm ATLAS+CMS}_{\gamma\gamma} = \frac{\sigma( gg \to \phi \to \gamma\gamma)}{\sigma_{\rm SM}(gg \to H_{\rm SM} \to \gamma\gamma)} = 0.24^{+0.09}_{-0.08},  \label{diphoton-rate}
\end{equation}
Here $\phi$ denotes a hypothetical non-SM scalar with mass $m_\phi = 95.4~{\rm GeV}$ responsible for the diphoton excess, and $\sigma_{\rm SM}$ refers to the cross section expected for an SM-like Higgs boson $H_{\rm SM}$ of the same mass. 
Additional phenomenological support for a state near $95~{\rm GeV}$ comes from a mild surplus in the $\tau^+\tau^-$ channel observed by CMS~\cite{CMS:2022goy} and a potential excess in $WW$ final states~\cite{Coloretti:2023wng}. 
Taken together, these observations motivate a closer look at a light scalar resonance around $95~{\rm GeV}$. 

A second class of anomalies involves resonant production of a heavier scalar state $X$ decaying into the observed $125~{\rm GeV}$ Higgs boson $H_{\rm 125}$ and an additional scalar $Y$. 
The CMS search in the $X \to H_{\rm 125}(\gamma\gamma) + Y(b\bar{b})$ process previously reported an excess with a local (global) significance of 3.8 (2.8) standard deviations for $M_X \simeq 650~{\rm GeV}$ and $M_Y$ in the mass range $90$--$100~{\rm GeV}$, with a best-fit cross section $\sigma_{\gamma\gamma b\bar b}=0.35^{+0.17}_{-0.13}~{\rm fb}$~\cite{CMS:2023boe}. 
This excess was not observed in later reports by CMS~\cite{CMS:2025qit} or ATLAS~\cite{ATLAS:2024auw,ATLAS:2025nda}. 
Those analyses showed good agreement with the SM background and set a $95\%$ C.L. upper limit on the signal cross section of about $0.2~{\rm fb}$. 
Nevertheless, the recent CMS search in the $\gamma\gamma\tau\tau$ final state reported a mild excess with a local significance of $2.3\,\sigma$ for the resonant topology $X\to Y(\gamma\gamma)H_{\rm 125}(\tau\bar\tau)$ at $(M_X,M_Y)=(650,95)~{\rm GeV}$~\cite{CMS:2025tqi}. 
This adds another experimental hint in the same mass region and motivates a closer look at correlated light and heavy scalar states.
Complementary final states further restrict this interpretation. 
A CMS search in the $\tau\bar{\tau}$ plus $b\bar{b}$ channel~\cite{CMS:2021yci} gives a stringent $95\%$ C.L. upper limit of approximately $3~{\rm fb}$ on $\sigma(X \to H_{\rm 125}(\tau\bar{\tau}) + \phi(b\bar{b}))$. 
This corresponds to an upper limit of about $0.1~{\rm fb}$ on $\sigma_{\gamma\gamma b\bar{b}}$~\cite{Ellwanger:2023zjc}, suggesting a tension between the observed excess and current bounds. 
A similar constraint on $\sigma_{\gamma\gamma b\bar{b}}$ follows from the most recent CMS analysis of the $b\bar{b}b\bar{b}$ final state~\cite{CMS:2026mwf}. 
That analysis did not observe a significant excess at the mass point associated with the earlier $\gamma\gamma b\bar b$ anomaly and gives an upper limit of about $20$--$25~{\rm fb}$ on $\sigma_{b\bar{b} b\bar{b}}$ for $M_X \simeq 650~{\rm GeV}$ and $M_Y \simeq 95~{\rm GeV}$. 
Instead, the largest deviation from the background expectation appears around $M_X \simeq 600~{\rm GeV}$ and $M_Y \simeq 400~{\rm GeV}$, where the upper limit on $\sigma_{b\bar{b} b\bar{b}}$ can reach about $300~{\rm fb}$. 
This new feature indicates that the scalar sector may be more intricate than a simple two-resonance picture involving only $95~{\rm GeV}$ and $650~{\rm GeV}$ states. It is therefore timely to examine whether a well-motivated BSM Higgs sector can address the persistent light scalar hints and the emerging heavy scalar structures while remaining compatible with the null results in other channels.

These observations have been investigated in a wide variety of BSM scenarios~\cite{Fan:2013gjf, Cao:2016uwt, Biekotter:2017xmf, Beskidt:2017dil, Fox:2017uwr,Haisch:2017gql,
Heinemeyer:2018wzl, Heinemeyer:2018jcd, Wang:2018vxp, Domingo:2018uim,Vega:2018ddp,
Cao:2019ofo,Biekotter:2019gtq, Choi:2019yrv,
Kundu:2019nqo,Biekotter:2019mib,Biekotter:2019kde,Sachdeva:2019hvk,
Biekotter:2020cjs,Abdelalim:2020xfk,Hollik:2020plc, Aguilar-Saavedra:2020wrj,
Biekotter:2021qbc, Biekotter:2021ovi,Heinemeyer:2021msz,
Biekotter:2022jyr,Li:2022etb, Iguro:2022dok, Iguro:2022fel, Benbrik:2022dja, Benbrik:2022azi, 
Li:2023hsr,Biekotter:2023oen,Biekotter:2023jld, Aguilar-Saavedra:2023vpd, Banik:2023ecr,Dutta:2023cig, Borah:2023hqw,Arcadi:2023smv,Ahriche:2023wkj, Chen:2023bqr,Dev:2023kzu,Ellwanger:2023zjc,Cao:2023gkc,Ahriche:2023hho,Belyaev:2023xnv, Azevedo:2023zkg, Ashanujjaman:2023etj,
Wang:2024bkg, Ellwanger:2024txc, Lian:2024smg, Cao:2024axg, Liu:2024cbr,Ellwanger:2024vvs, YaserAyazi:2024hpj, Gao:2024ljl, Gao:2024qag, Mondal:2024obd,Khanna:2024bah,Baek:2024cco,Dong:2024ipo, Kalinowski:2024uxe,
Xu:2025vmy, Abbas:2025ser, Li:2025tkm, Du:2025eop, Benbrik:2025hol,Benbrik:2025wkz,Khanna:2025cwq, Chang:2025bjt, Hmissou:2025riw, Chen:2025vtg,Bhatnagar:2025jhh,Lian:2025zoi,Dutta:2026tbc,Wang:2026rde}.  
The light scalar has also attracted attention in studies of future Higgs factories and lepton colliders, where a $95~{\rm GeV}$ state could be tested through cleaner production and decay channels~\cite{Sharma:2024vhv, Dong:2025orv, Dong:2025exu, Kumar:2025xng, Robens:2025kaa, Robens:2026xfc}. 
Such measurements would provide important constraints on the mixing that links the light scalar to the heavier states. 
It is therefore well motivated to go beyond an isolated fit of the light excess and to ask whether an extended Higgs sector can accommodate it together with the heavier resonances and cascade decay channels suggested by present LHC searches.
Several BSM studies have explored the possibility that the $95~{\rm GeV}$ and $650~{\rm GeV}$ hints arise from a common extended Higgs sector~\cite{Ellwanger:2023zjc, Banik:2023ecr, Azevedo:2023zkg, Benbrik:2025hol,Benbrik:2025wkz,Khanna:2025cwq, Lian:2025zoi, Arcadi:2025grl}. 
Supersymmetric (SUSY) extensions of the SM~\cite{Fayet:1974pd, Fayet:1976et, Fayet:1976cr, Fayet:1977yc, Farrar:1978xj,Martin:1997ns} provide a particularly appealing framework for this purpose, since they naturally contain enlarged Higgs sectors and viable dark matter (DM) candidates~\cite{ATLAS:2024lda,Constantin:2025mex,Jeanty:2026etw}. 
Among them, the Next-to-Minimal Supersymmetric Standard Model (NMSSM)~\cite{Fayet:1974pd, Ellwanger:2009dp, Maniatis:2009re} is especially suitable, because the additional singlet superfield gives rise to singlet-dominated scalar and pseudoscalar states while also modifying the neutralino sector. 
The $95~{\rm GeV}$ scalar interpretation has already been studied in the $\mathbb{Z}_3$ invariant NMSSM~\cite{Lian:2024smg,Ellwanger:2023zjc} and in the general NMSSM (GNMSSM)~\cite{Cao:2023gkc,Cao:2024axg}.
In the GNMSSM~\cite{Ellwanger:2009dp} the singlet bilinear and singlet tadpole terms provide additional independent mass parameters, which make the scalar and dark matter sectors more flexible. 
Here we take a more restricted route and keep only an explicit $\mu$ term of the Minimal Supersymmetric Standard Model type as the source of the departure from the $\mathbb{Z}_3$ invariant NMSSM. 
We refer to this setup as the $\mu$NMSSM, which can be regarded as a limit of the GNMSSM without singlet bilinear and singlet tadpole terms.

The explicit $\mu$ term is nevertheless important for the present analysis. In the $\mathbb{Z}_3$ invariant NMSSM one has $\mu_{\rm tot}=\mu_{\rm eff}=\lambda v_s/\sqrt{2}$, whereas in the $\mu$NMSSM $\mu_{\rm tot}=\mu_{\rm eff}+\mu$. 
The Higgsino-like chargino masses are then controlled by $\mu_{\rm tot}$ without fixing the singlet vacuum expectation value at the same time.
This separation is useful for the diphoton signal because a moderately light chargino can give a sizeable loop induced contribution to the $h_s\gamma\gamma$ while keeping the singlet-like scalar near 95 GeV. 
Moreover, the neutralino dark matter interpretation changes for the same reason. 
In the $\mathbb{Z}_3$ invariant NMSSM the singlino mass satisfies $m_N=2\kappa\mu_{\rm eff}/\lambda$, so a singlino dominated neutralino dark matter candidate is obtained only for $2|\kappa|<\lambda$. 
The observed relic abundance is achieved mainly through coannihilation with Higgsinos, while direct detection bounds favor a small $\lambda$~\cite{Cao:2018rix, Zhou:2021pit}. 
These conditions involve the same parameters that determine the singlet scalar masses and mixings. 
In the $\mu$NMSSM, $\mu_{\rm eff}$ and $\mu_{\rm tot}$ can play separate roles, relaxing this correlation and allowing the scalar excesses to be discussed without enforcing the same tight relation on the neutralino spectrum.

The additional $\mu$ term also leaves visible traces in the scalar spectrum.
In the CP-even sector, the singlet doublet mixing entries are proportional to $\lambda\mu_{\rm tot}v$, while the singlet dominated diagonal entry is parametrized by $m_B^2$ and the heavy doublet scale is mainly set by $m_A$. 
In the CP-odd sector, the explicit $\mu$ term enters the singlet dominated diagonal entry, and the doublet singlet mixing is controlled by $A_\lambda-m_N$. 
The light scalar mass near $95~{\rm GeV}$, the heavy doublet scale relevant for the CMS searches, and the required scalar mixing can therefore be adjusted with fewer correlations than in the $\mathbb{Z}_3$ invariant NMSSM, but with fewer independent singlet sector parameters than in the GNMSSM. 
The numerical analysis below tests whether this limited extension is sufficient for the light scalar rates and the correlated heavy scalar cascades.

The $\mu$NMSSM also has a useful interpretation in inflation-inspired supersymmetric constructions~\cite{Hollik:2018yek,Hollik:2020plc}. 
In that setting the $\mu$ term is not just a generic $\mathbb{Z}_3$ breaking parameter. 
It is induced when the NMSSM Higgs sector is embedded into supergravity and the Higgs doublets couple nonminimally to gravity. 
After the low energy theory is obtained, the induced bilinear term is related to the gravitino mass. 
This gives the positive $\mu$ region a special meaning, although we do not impose the inflationary construction as a boundary condition on the full scan. 
We allow $\mu$ to take either sign when studying the scalar resonances, and then examine the $\mu>0$ subset as the part that can be connected more directly to the inflation-inspired picture. 
In this subset the gravitino can be the lightest supersymmetric particle (LSP), while the lightest neutralino becomes the next-to-lightest supersymmetric particle (NLSP). 
This is different from Ref.~\cite{Lian:2025zoi}, where the neutralino was required to provide the observed DM abundance.

The purpose of this work is to test whether the $\mu$NMSSM can account for the present hints for both light and heavy scalar states. 
We focus on a CP-even scalar near $95~{\rm GeV}$ dominated by the singlet component, which contributes to the $b\bar b$ and $\gamma\gamma$ excesses, together with heavier scalar states that can enter searches for a SM-like Higgs boson plus an additional scalar. 
We ask in particular whether the model can remain compatible with the earlier CMS structure near $(650,95)~{\rm GeV}$ and with the newer feature near $(600,400)~{\rm GeV}$ in the CMS $4b$ analysis, without violating the current Higgs and collider limits.

The rest of the paper is organized in the following way. In Sec.~\ref{Section-Model}, we summarize the $\mu$NMSSM ingredients needed for the analysis. In Sec.~\ref{Section-explanation}, we describe the scan, discuss the numerical results, and study the implications for the light and heavy resonance searches, including the $\mu>0$ subset motivated by the inflation-inspired construction. We conclude in Sec.~\ref{conclusion}.

\section{Theoretical preliminaries}\label{Section-Model}
\subsection{The Superpotential of the $\mu$NMSSM}
In this section we briefly introduce the $\mu$NMSSM and summarize the features of its Higgs sector that are relevant for interpreting the light and heavy scalar resonance structures. Compared with the $\mathbb{Z}_3$ invariant NMSSM, the model contains an explicit MSSM-like bilinear term in the superpotential~\cite{Ellwanger:2009dp,Hollik:2018yek,Hollik:2020plc},
\begin{equation}
 W_{\mu{\rm NMSSM}} = W_{\rm Yukawa} + \left(\mu + \lambda \hat S \right) \hat H_u \cdot \hat H_d + \frac{\kappa}{3} \hat S^3 ,
 \label{eq:superpotential}
\end{equation}
where $\hat H_u$ and $\hat H_d$ are the two Higgs doublet superfields, $\hat S$ is a gauge singlet chiral superfield, and $W_{\rm Yukawa}$ contains the usual quark and lepton Yukawa interactions. The explicit parameter $\mu$ breaks the $\mathbb{Z}_3$ symmetry of the scale invariant NMSSM, but no terms proportional to $\hat S^2$ or $\hat S$ are introduced. In this sense the model can be regarded as a restricted version of the GNMSSM. It is more economical than the full GNMSSM while retaining the singlet extended Higgs dynamics needed for light and heavy scalar phenomenology. The origin of the additional $\mu$ term depends on the underlying theoretical framework.
In GNMSSM constructions, the dimensionful bilinear and tadpole parameters may arise from the breaking of an underlying discrete $\mathbb{Z}_4^R$ symmetry after supersymmetry breaking~\cite{Ross:2011xv,Lee:2010gv,Lee:2011dya,Ross:2012nr}. These terms are essential for resolving the tadpole problem~\cite{Ellwanger:1983mg, Ellwanger:2009dp} and the cosmological domain wall problem that plagues the $\mathbb{Z}_3$-NMSSM~\cite{Abel:1996cr, Kolda:1998rm, Panagiotakopoulos:1998yw}. 

An alternative realization arises in the inflation-inspired supersymmetric construction, where the additional $\mu$ term emerges from a superconformal embedding of the Higgs sector into supergravity. In this framework, the Higgs doublets couple nonminimally to Einstein gravity through the supergravity frame operator $X(\Phi) R$ of chiral superfields $X(\Phi)$ and the curvature multiplet $R$ with $X(\Phi)=\chi \hat H_u \cdot \hat H_d$, where $\chi$ denotes the dimensionless coupling governing the Higgs-gravity interaction~\cite{Einhorn:2009bh}. After a K\"ahler transformation, the nonminimal coupling induces the $\mu$ term in Eq.~(\ref{eq:superpotential}) with $\mu = \frac{3}{2}m_{3/2}\chi$, where $m_{3/2}$ is the gravitino mass~\cite{Ferrara:2010in}. 
The nonminimal coupling $\chi$ is not arbitrary. To reproduce the observed amplitude of primordial density perturbations, it is typically required to be large and approximately related to the NMSSM coupling $\lambda$ as $\chi \simeq 10^5 \lambda$~\cite{Lee:2010hj,Ferrara:2010in}. Thus, for phenomenologically relevant values of $\lambda$, an electroweak scale or TeV scale $\mu$ parameter naturally points to a gravitino mass well below the electroweak scale. This motivates scenarios in which the gravitino is the LSP DM candidate, while the neutralino behaves as the NLSP at collider scales. From a phenomenological perspective, however, the low energy theory can be treated more generally as a $\mu$NMSSM without imposing the strict assumptions of the inflationary scenario. In the present work, we therefore allow $\mu$ to take either sign in order to explore the Higgs phenomenology of the model in a broad parameter region, while the inflation-inspired construction provides an additional motivation for the $\mu > 0$ regime.

\subsection{Higgs Sector}
The soft SUSY-breaking Lagrangian for the Higgs fields in the $\mu$NMSSM takes the form
\begin{equation}
\begin{aligned}
 -\mathcal{L}_{soft} = &\Bigg[\lambda A_{\lambda} S H_u \cdot H_d + \frac{1}{3} \kappa A_{\kappa} S^3+ B_{\mu} \mu H_u\cdot H_d + h.c.\Bigg]   \\
& + m^2_{H_u}|H_u|^2 + m^2_{H_d}|H_d|^2 + m^2_{S}|S|^2 , \label{softL}
\end{aligned}
\end{equation}
Here $H_u$, $H_d$, and $S$ denote the scalar components of the Higgs superfields, and $m^2_{H_u}$, $m^2_{H_d}$, and $m^2_{S}$ are their soft-breaking masses, respectively. 
The corresponding tree-level Higgs potential can be written as a function of the scalar fields $H_u$, $H_d$, and $S$ as~\cite{Hollik:2018yek}
\begin{equation}
\begin{aligned}
V_{\rm Higgs}=&
\left(m_{H_u}^2+\left|\mu+\lambda S\right|^2\right)\left|H_u\right|^2
+\left(m_{H_d}^2+\left|\mu+\lambda S\right|^2\right)\left|H_d\right|^2
+m_S^2\left|S\right|^2  \\
&+\left|\lambda H_u\cdot H_d+\kappa S^2\right|^2
+\frac{g_1^2+g_2^2}{8}\left(\left|H_u\right|^2-\left|H_d\right|^2\right)^2
+\frac{g_2^2}{2}\left|H_u^\dagger H_d\right|^2  \\
&+\left[
\lambda A_\lambda S H_u\cdot H_d
+\frac{1}{3}\kappa A_\kappa S^3
+B_\mu\mu\,H_u\cdot H_d
+h.c.
\right] .
\label{eq:higgspotential}
\end{aligned}
\end{equation}
We denote the electroweak vacuum expectation values (vevs) as $\left\langle H_u^0 \right\rangle = v_u/\sqrt{2}$, $\left\langle H_d^0 \right\rangle = v_d/\sqrt{2}$, and $\left\langle S \right\rangle = v_s/\sqrt{2}$ with $v = \sqrt{v_u^2+v_d^2}\simeq 246~\mathrm{GeV}$ and $\tan{\beta} \equiv v_u/v_d$. By minimizing the scalar potential, the soft-breaking masses $m^2_{H_u}$, $m^2_{H_d}$, and $m^2_{S}$ can be traded for the vevs. The Higgs-sector inputs may then be chosen as $\tan{\beta}$, $v_s$, the Yukawa couplings $\lambda$ and $\kappa$, the soft-breaking trilinear coefficients $A_\lambda$ and $A_\kappa$, the bilinear mass parameter $\mu$ and the soft-breaking parameter $B_{\mu}$. 

To describe the Higgs states, it is convenient to use the rotated field combinations $H_{\rm SM} \equiv \sqrt{2} {\rm Re}(\sin\beta H_u^0+\cos\beta H_d^0)$, $H_{\rm NSM} \equiv \sqrt{2} {\rm Re}(\cos\beta H_u^0-\sin\beta H_d^0)$, and $A_{\rm NSM} \equiv \sqrt{2} {\rm Im}(\cos\beta H_u^0-\sin\beta H_d^0)$. Here $H_{\rm SM}$ behaves like the SM Higgs field, while $H_{\rm NSM}$ and $A_{\rm NSM}$ describe the additional doublet fields~\cite{Cao:2012fz}. The singlet field remains unrotated and is written as $\sqrt{2} S \equiv H_{\rm S}+i A_S$.

In the basis $\left(H_{\rm NSM}, H_{\rm SM}, {\rm Re}[S]\right)$, the elements of the symmetric squared mass matrix for the CP-even Higgs bosons can be written as~\cite{Ellwanger:2009dp, Miller:2003ay}
\begin{equation}
\begin{aligned}
  {\cal M}^2_{S, 11}&= m^2_A + \frac{1}{2} (2 m_Z^2- \lambda^2v^2)\sin^22\beta,  \\
  {\cal M}^2_{S, 12}&=-\frac{1}{4}(2 m_Z^2-\lambda^2v^2)\sin4\beta, \\
  {\cal M}^2_{S, 13}&= \sqrt{2} \lambda \mu_{tot} v (\delta -1) \cot{2\beta},  \\
  {\cal M}^2_{S, 22}&=m_Z^2\cos^22\beta+ \frac{1}{2} \lambda^2v^2\sin^22\beta,  \\
  {\cal M}^2_{S, 23}&= \sqrt{2} \lambda \mu_{tot} v \delta, \\
  {\cal M}^2_{S, 33}&= m_B^2,
 \label{CPevenHiggsMass}
\end{aligned} 
\end{equation}
while those for the CP-odd Higgs bosons in the basis $\left(A_{\rm NSM}, A_S\right)$ are
\begin{equation}
\begin{aligned}
{\cal M}^2_{P,11} &= m_A^2, \qquad
{\cal M}^2_{P,12} = \frac{\lambda v}{\sqrt{2}}(A_\lambda - m_N),\\
{\cal M}^2_{P,22} &= \frac{(A_\lambda + 2 m_N)\sin 2\beta - 2\mu}{4\mu_{eff}}\lambda^2 v^2 - \frac{3}{2}m_NA_\kappa.
\end{aligned}
\label{CPoddHiggsMass}
\end{equation}
We introduce a $\delta$ factor in ${\cal M}^2_{S, 13}$ and ${\cal M}^2_{S, 23}$, defined as
\begin{equation}
 \delta \equiv 1- \frac{(A_\lambda + m_N)\sin2\beta}{2\mu_{tot}}, \label{delta}
 \end{equation}
where $\mu_{tot} \equiv \mu_{eff}+\mu = \lambda v_s/\sqrt{2} + \mu$ is the Higgsino mass and $m_N \equiv \sqrt{2} \kappa v_s$ is the singlino mass. This definition simplifies the mass matrix and allows the mixings of $H_{\rm S}$ with $H_{\rm SM}$ and $H_{\rm NSM}$ to be controlled directly. The factor $\delta$ is expected to be tiny in the alignment without decoupling limit~\cite{Carena:2015moc, Biekotter:2021qbc}. 
The parameter $m_A$ denotes the doublet-like pseudoscalar mass scale, while $m_B$ characterizes the singlet-like CP-even scalar mass scale. The soft-breaking parameters $A_\lambda$, $A_\kappa$ and $B_{\mu}\mu$ can then be expressed in terms of the physical input parameters,
\begin{eqnarray} 
A_\lambda &=& \frac{2(1-\delta)\mu_{tot}}{\sin{2\beta}}-m_N,\nonumber\\
A_\kappa &=& \frac{2 m^2_B}{m_N} + \frac{\lambda^3v^2}{4\kappa\mu^2_{eff}} \left[ 2\mu -A_\lambda\sin 2\beta \right] - 2m_N, \\
B_{\mu}\mu &=& \frac{1}{2}\left[ m^2_A \sin{2\beta} - \mu_{eff}(2A_\lambda + m_N) \right]. \nonumber 
\label{Simplify-1}
\end{eqnarray} 
Diagonalizing the squared mass matrix with mixing angles denoted by $V^j_{h_i}$ yields three CP-even Higgs mass eigenstates,
  \begin{eqnarray}
    h_i & = & V_{h_i}^{\rm NSM} H_{\rm NSM}+V_{h_i}^{\rm SM} H_{\rm SM}+V_{h_i}^{\rm S} H_{\rm S},
   \label{Vij}
  \end{eqnarray}   
with $h_i=\{h_{\rm s},h,H\}$ ordered by ascending mass. The second CP-even mass eigenstate $h$ corresponds to the observed $125~{\rm GeV}$ SM-like Higgs boson. This state is predominantly composed of the $H_{\rm SM}$ component, and current Higgs data restrict the combined $H_{\rm NSM}$ and $H_{\rm S}$ admixture to roughly the ten percent level~\cite{ATLAS:2022vkf,CMS:2022dwd}. We therefore require $\sqrt{\left(V_h^{\rm NSM}\right)^2+\left(V_h^{\rm S}\right)^2}\lesssim0.1$ and $|V_h^{\rm SM}|\sim1$. The imaginary components $A_S$ and $A_{\rm NSM}$ mix into two CP-odd Higgs mass eigenstates $a_i=\{A_H,A_s\}$, while the charged components give rise to a pair of charged Higgs bosons $H^\pm$. The doublet-like CP-odd state $A_H$ and the charged Higgs bosons $H^\pm$ are nearly degenerate with the heavy CP-even boson $H$.

The present study assumes that the lightest CP-even Higgs boson $h_s$ is dominated by the singlet component and is responsible for both the $\gamma\gamma$ and $b\bar{b}$ excesses near $95.4~{\rm GeV}$. The signal strengths normalized to their SM predictions are~\cite{Cao:2023gkc}
\begin{eqnarray}
	\mu_{\gamma\gamma}|_{m_{h_s} = 95.4~{\rm GeV}} &=&
  \frac{\sigma_{\rm SUSY}(\mathrm{pp} \to h_s)}
       {\sigma_{\rm SM}(\mathrm{pp} \to h_s )} \times
       \frac{{\rm Br}_{\rm SUSY}(h_s \to \gamma \gamma)}
       {{\rm Br}_{\rm SM}(h_s \to \gamma \gamma)},  \label{muCMS}  \\
  	\mu_{b\bar{b}}|_{m_{h_s} = 95.4~{\rm GeV}} &=&
  \frac{\sigma_{\rm SUSY}(e^+e^-\to Z h_s)}
       {\sigma_{\rm SM}(e^+e^-\to Z h_s)} \times
       \frac{{\rm Br}_{\rm SUSY}(h_s\to b\bar{b})}
       {{\rm Br}_{\rm SM}(h_s \to b\bar{b})},  \label{muLEP}
  \end{eqnarray}
where the mass of $h_s$ is fixed at $95.4~{\rm GeV}$. The production rate $\sigma(\mathrm{pp} \to h_s)$ and the decay branching ratio ${\rm Br}(h_s \to \gamma \gamma)$ labeled with the subscript `SUSY' refer to the predictions from the model, whereas those with the subscript `SM' assume SM couplings for $h_s$. 

The $\gamma\gamma b\bar{b}$ signal near $650~{\rm GeV}$ can be attributed to the heavy CP-even Higgs boson $H$ dominated by the doublet component, which decays into a SM-like Higgs boson $h$ and a Higgs boson $h_s$ dominated by the singlet component. The cross section of the resonance in the $\gamma\gamma b\bar{b}$ channel is
\begin{equation}
 \sigma_{\gamma\gamma b\bar{b}} = \sigma(\mathrm{gg} \to H) \times {\rm Br}_{\rm SUSY}(H \to h h_s) \times {\rm Br}_{\rm SUSY}(h \to \gamma \gamma) \times {\rm Br}_{\rm SUSY}(h_s \to b \bar{b}),  \label{XSbbrr}
\end{equation}
where the masses of $H$, $h$ and $h_s$ are fixed at approximately $650~{\rm GeV}$, $125~{\rm GeV}$ and $95.4~{\rm GeV}$, respectively.  

\subsection{Neutralino Sector}

The fermionic partners of the neutral gauge and Higgs fields form five neutralino states. 
In the gauge eigenstate basis $(\tilde B,\tilde W^0,\tilde H_d^0,\tilde H_u^0,\tilde S)^T$,
the neutralino mass matrix can be written as
\begin{equation}
\mathcal{M}_{\tilde\chi^0}=
\begin{pmatrix}
M_1 & 0 & -\frac{g_1v_d}{2} & \frac{g_1v_u}{2} & 0 \\
 & M_2 & \frac{g_2v_d}{2} & -\frac{g_2v_u}{2} & 0 \\
 &  & 0 & -\mu_{\rm tot} & -\frac{\lambda v_u}{\sqrt{2}} \\
 &  &  & 0 & -\frac{\lambda v_d}{\sqrt{2}} \\
 &  &  &  & m_N
\end{pmatrix}.
\label{eq:neutralino-matrix}
\end{equation}
Here $M_1$ and $M_2$ are the bino and wino soft masses, while the Higgsino and singlino mass parameters are the same combinations that enter the Higgs-sector discussion. Thus the explicit $\mu$ term shifts the Higgsino mass entry directly, whereas the singlino mass is controlled by $\kappa v_s$. The symmetric Majorana mass matrix in Eq.~(\ref{eq:neutralino-matrix}) is diagonalized by a unitary matrix $N$,
\begin{equation}
  N^*\,\mathcal{M}_{\tilde\chi^0}\,N^\dagger
  ={\rm diag}\left(m_{\tilde\chi^0_1},m_{\tilde\chi^0_2},
  m_{\tilde\chi^0_3},m_{\tilde\chi^0_4},m_{\tilde\chi^0_5}\right),
\end{equation}
and the mass eigenstates are ordered by increasing mass as
\begin{equation}
  \tilde\chi_i^0
  =N_{i1}\tilde B+N_{i2}\tilde W^0+N_{i3}\tilde H_d^0
  +N_{i4}\tilde H_u^0+N_{i5}\tilde S .
\label{eq:neutralino-mixing}
\end{equation}
The entries $N_{ij}$ shown in the benchmark tables follow this convention. 
Their signs depend on the phase convention chosen for the Majorana fields, while $\left|N_{i1}\right|^2$, $\left|N_{i2}\right|^2$, $\left|N_{i3}\right|^2+\left|N_{i4}\right|^2$, and $\left|N_{i5}\right|^2$ measure the bino, wino, Higgsino, and singlino fractions of $\tilde\chi_i^0$, respectively. 

For a generic sign of the explicit $\mu$ parameter, one may instead consider the case in which the lightest neutralino is the LSP and accounts for the observed dark matter abundance. 
This possibility was studied in Ref.~\cite{Lian:2025zoi}, where the scalar resonance interpretation was combined with neutralino DM constraints in the GNMSSM. 
The viable DM solutions were found to require a dominantly bino-like LSP. 
The reason is that neutralinos dominated by Higgsino or singlino components in the relevant scalar resonance region usually have sizable couplings to the $Z$ boson, the SM-like Higgs, or the light singlet-like scalars. 
Such couplings tend either to make the annihilation too efficient, leading to an underabundant relic density, or to enhance the spin-independent direct-detection rate. 
A bino-like LSP suppresses these couplings and can still obtain an acceptable relic abundance through resonant annihilation, coannihilation, or a controlled Higgsino/singlino admixture. 
Since this issue has already been analyzed in detail, we do not impose the neutralino relic-density, direct-detection, or indirect-detection constraints in the present work.
Instead, the neutralino mixing information is particularly relevant for the positive-$\mu$ points, where $\tilde\chi^0_1$ acts as the NLSP in the gravitino DM interpretation discussed below.

\section{Explanation of the excesses}\label{Section-explanation}
We now turn to the numerical analysis of the scalar resonance interpretation. 
We require the samples to reproduce the light scalar signals in the $b\bar b$ and $\gamma\gamma$ channels, to give the appropriate size of the heavy scalar cascade rates, and to satisfy Higgs data, flavor observables, electroweak precision tests, extra Higgs searches and direct SUSY searches.
The numerical calculation proceeds in several steps. 
We first implement the $\mu$NMSSM with \textsf{SARAH-4.15.4}~\cite{SARAH_Staub2008,SARAH3_Staub2012,SARAH4_Staub2013,SARAH_Staub2015}, which provides the analytical expressions needed for the mass matrices, vertices and renormalization group equations. 
The generated \textsf{SPheno-4.0.6} spectrum generator~\cite{Porod2003SPheno,Porod2011SPheno3} is then used to compute the SUSY spectrum, Higgs masses, mixing matrices, partial widths and branching fractions. 
Flavor observables are evaluated with \textsf{FlavorKit}~\cite{Porod:2014xia}. 
The resulting sample is analyzed with both posterior probability density (PDF) and profile likelihood (PL) methods~\cite{Fowlie:2016hew}. 
The posterior density indicates where the viable parameter volume accumulates, whereas the profile likelihood is less sensitive to volume effects and is useful for identifying the best fit regions in the projected planes.

\subsection{Strategy in scanning the parameter space}\label{scanStrategy}
The scan is organized around the parameters that most directly determine the Higgs spectrum and the dominant radiative correction to the SM-like Higgs mass. 
The superpotential couplings $\lambda$ and $\kappa$ control the singlet interactions and the size of the tree-level Higgs mass contribution. 
The quantities $\mu_{\rm tot}$ and $\mu_{\rm eff}$ determine the Higgsino mass scale and the singlet vev, while $\tan\beta$ fixes the relative size of the two doublet vevs. 
The parameters $m_A$ and $m_B$ are used to place the doublet-like and singlet-like mass scales in the regions relevant for the heavy and light scalar signals.

Instead of sampling the soft parameters $A_\lambda$, $A_\kappa$ and $B_\mu$ directly, we trade them for the more transparent quantities $\delta$, $m_B$ and $m_A$ defined in Eqs.~(\ref{delta}) and~(\ref{Simplify-1}).  
Directly scanning the soft terms would generate many points with Higgs masses far away from the target regions, while the variables used here make it easier to explore spectra with $m_{h_s}\simeq95~{\rm GeV}$ and $m_H$ around the heavy resonance window. 
The parameter $\delta$ is especially useful because it controls the mixing between the SM-like Higgs direction and the singlet direction. 
The trilinear parameters of the third generation squarks are taken as $A_t=A_b$. 
They are kept free because the stop sector gives an important loop correction to the SM-like Higgs mass.

The nine scanned parameters and their ranges are shown in Table~\ref{tab:scan}. 
The intervals were chosen after broader exploratory scans. 
Parameters that do not play a direct role in the Higgs resonance interpretation are fixed at values safely above present LHC limits. 
We also require perturbativity between the electroweak and GUT scales. 
In practice, we impose the usual NMSSM estimate $\lambda^2+\kappa^2\lesssim0.5$~\cite{Miller:2003ay}. 
This criterion can be used for the $\mu$NMSSM because the additional $\mu$ term does not change the scale dependent beta functions up to two-loop order~\cite{Ellwanger:2009dp, King:1995vk, Masip:1998jc}. 

\begin{table}[tbp]
\caption{Input ranges used in the scan. Flat priors are assigned to all scanned parameters. The parameters $\delta$, $m_A$ and $m_B$ are used instead of the original soft terms $A_\lambda$, $A_\kappa$ and $B_\mu$, as described in the text. We set $A_t=A_b$, fix the gluino mass to $3~{\rm TeV}$, and take the remaining SUSY mass parameters not central to the present analysis to be $2~{\rm TeV}$. All inputs are defined at $Q_{\rm input}=1~{\rm TeV}$.
\label{tab:scan}}
\centering
\vspace{0.3cm}
\resizebox{0.7\textwidth}{!}{
\begin{tabular}{c|c|c|c}
\hline\hline
Parameter  & Range & Parameter  & Range   \\
\hline
$\lambda$  & $0.3 \sim 0.7$ & $\mu_{tot}/{\rm GeV}$  & $100.0 \sim 1000 $ \\
$\kappa$  & $-0.5 \sim 0.5$ & $\mu_{eff}/{\rm GeV}$  & $-2000 \sim 2000$ \\
$\delta$  & $-0.05 \sim 0.05$ & $m_A/{\rm GeV}$  & $500.0 \sim 650.0$\\
$\tan \beta$  & $1.0 \sim 4.0$ & $m_B/{\rm GeV}$  & $90.0 \sim 120$ \\
			&		&  $A_t/{\rm GeV}$  & $1000 \sim 3000$ \\
\hline\hline
\end{tabular}}
\end{table}

The parameter space is explored with the MultiNest algorithm~\cite{MultiNest2009,Importance2019}. 
We use $n_{\rm live}=5000$ live points. 
In nested sampling, the live points determine how well successive likelihood contours are resolved, so this choice is useful for following the narrow regions selected by the light and heavy scalar mass requirements. 
The likelihood guiding the scan is written as
$\mathcal{L}=\mathcal{L}_{\rm anomaly}\mathcal{L}_{\rm Res}$. 
The first factor measures the agreement with the central values of the $95~{\rm GeV}$ diphoton and $b\bar b$ excesses and with the earlier $\gamma\gamma b\bar b$ excess near $650~{\rm GeV}$,
\begin{equation}
\begin{split}
\chi^2_{650 + 95} = \left( \frac{ \sigma_{\gamma\gamma b\bar{b}} - 0.35~{\rm fb}}{0.13~{\rm fb}}\right)^2 + \left( \frac{\mu_{\gamma\gamma} - 0.24}{0.08}\right)^2 + \left( \frac{\mu_{b\bar{b}} - 0.117}{0.057}\right)^2 .   
\end{split}
\label{chi2-excesses}
\end{equation}
with $\mathcal{L}_{\rm anomaly}=\exp(-\chi^2_{650+95}/2)$. 
The second factor enforces the basic physical and experimental requirements that are not included as Gaussian terms in Eq.~(\ref{chi2-excesses}). 
We take $\mathcal{L}_{\rm Res}=1$ when all restrictions are satisfied and $\mathcal{L}_{\rm Res}=\exp(-100)$ otherwise. 
The restrictions are listed below.
\begin{itemize}
\item \textbf{Masses of the three scalars:} 
To interpret the observed excesses, we impose the following mass requirements on the three CP-even Higgs bosons.
\begin{itemize}
\item Light scalar ($h_s$): $m_{h_s}$ is required to lie in the range $95.4 \pm 3~{\rm GeV}$.
\item Heavy scalar ($H$): $m_H$ is required to lie in the range $650 \pm 25~{\rm GeV}$, covering the mass region relevant to the earlier CMS excess~\cite{Ellwanger:2023zjc, CMS:2023boe}.
\item SM-like Higgs ($h$): $m_h$ is required to agree with the measured Higgs mass within a $\pm 3~{\rm GeV}$ window, allowing for both experimental and theoretical uncertainties.
\end{itemize}
\item \textbf{SM-like Higgs data fit:} 
The Higgs signal rates are tested with \textsf{HiggsTools} code that incorporates \textsf{HiggsSignals-2}~\cite{HS2013xfa,HSConstraining2013hwa,HS2014ewa,HS2020uwn}. 
The resulting $\chi^2$ value is included in the likelihood. 
When results are shown in two-dimensional planes, points satisfying $\Delta\chi^2_{125}\equiv \chi^2-\chi^2_{125,\mathrm{SM}}\lesssim6.18$ are treated as compatible with the Higgs data at about the $2\,\sigma$ level, with $\chi^2_{125,\mathrm{SM}}=153$~\cite{Muhlleitner:2020wwk}.
\item \textbf{Extra Higgs searches:} 
The predicted rates of the non-SM Higgs bosons $h_s$, $H$, $A_H$ and $H^\pm$ are required to satisfy the LEP, Tevatron and LHC limits implemented in \textsf{HiggsBounds-5.10.2} through \textsf{HiggsTools}~\cite{Bahl:2022igd,HB2008jh,HB2011sb,HBHS2012lvg,HB2013wla,HB2020pkv}.
\item \textbf{$B$-physics observables:} The branching ratios $B_s \to \mu^+ \mu^-$ and $B \to X_s \gamma$ are required to agree with current experimental measurements at the $2\sigma$ level~\cite{ParticleDataGroup:2024cfk}.
\item \textbf{Electroweak precision observables:}
For each viable parameter point, the oblique parameters $S$, $T$ and $U$ are evaluated using the \textsf{SPheno} package and confronted with the global electroweak fit~\cite{ParticleDataGroup:2024cfk},
\begin{equation}
S = 0.021 \pm 0.096, \qquad
T = 0.040 \pm 0.120, \qquad
U = 0.008 \pm 0.092 . \nonumber
\end{equation}
The correlations are taken to be $\rho_{ST}=0.91$, $\rho_{SU}=-0.62$, and $\rho_{TU}=-0.83$.
We construct the corresponding $\chi^2$ with three degrees of freedom and require agreement with the fit within $2\,\sigma$.
\end{itemize}

Vacuum stability is an essential theoretical consistency requirement because the scalar potential of supersymmetric models may contain charge- or color-breaking extrema, as well as Higgs-singlet directions in which a global minimum deeper than the desired electroweak vacuum is generated. 
If such a deeper minimum exists, the electroweak vacuum is not absolutely stable but metastable. 
The point can nevertheless remain acceptable if the quantum tunneling time from the electroweak vacuum to the deeper global minimum is longer than the age of the Universe, in which case it is referred to as long-lived; otherwise, it is short-lived and excluded by the vacuum-stability requirement. 
Although this requirement is not used to filter the full scan, we apply it as a posterior check of the representative benchmark points using the \textsf{Vevacious} framework~\cite{VPP2014,Camargo-Molina:2013qva}.
For each benchmark point, the tunneling analysis is performed with both the tree-level scalar potential and the one-loop effective potential. 
The tree-level input contains the Higgs potential in Eq.~(\ref{eq:higgspotential}), together with the remaining scalar directions needed by \textsf{Vevacious} to search for charge- and color-breaking extrema. 
The one-loop potential is written as
\begin{equation}
  V_{\rm 1-loop}=V_{\rm tree-level}+V_{\rm CW}+V_{\rm CT},
\end{equation}
where $V_{\rm CW}$ denotes the Coleman-Weinberg correction and $V_{\rm CT}$ is the counter-term potential introduced to maintain the desired electroweak minimum conditions~\cite{Camargo-Molina:2013qva}. 
The entries $V_{\rm tree-level}$ and $V_{\rm 1-loop}$ shown in Tables~\ref{tab:benchmark-p1-p2}--\ref{tab:benchmark-p7-p8} report the vacuum classification obtained from these two potentials. 
We display both labels because the loop-corrected tunneling calculation can be sensitive to gauge dependence, field-dependent mass hierarchies, and large logarithms in regions where the perturbative expansion is less stable~\cite{Hollik:2018wrr,Andreassen:2016cvx}. 
The tree-level and one-loop results should therefore be read as complementary diagnostics of the vacuum structure.

As a final collider check, we test the viable spectra against direct LHC searches for SUSY particles. 
In the present setup these constraints are expected to be weak since the electroweakino sector is dominantly Higgsino-like and lies in a relatively heavy mass range, while the colored superpartners, sleptons, and the remaining gaugino states are taken to be heavy and decoupled. 
We quantify this expectation with \texttt{SModelS-3.1.1}, using the simplified-model database and its recent extension~\cite{Khosa:2020zar,Altakach:2024jwk}. 
For each spectrum, \texttt{SModelS} decomposes the production and cascade decays of the $R$-parity-odd particles into simplified-model elements, computes the corresponding $\sigma\times{\rm BR}$ predictions, and matches them to the ATLAS and CMS upper-limit or efficiency-map results collected in the database. 
The exclusion test is summarized by the largest ratio $R_{\rm SModelS}$ between the predicted signal rate and the experimental $95\%$ C.L. limit; points with $R_{\rm SModelS}>1$ are excluded. 
The samples considered here satisfy $R_{\rm SModelS}<1$, confirming that current electroweakino searches do not significantly affect the scalar-resonance interpretation.

\subsection{Numerical Results}
\begin{figure}[tbp]
\centering
\includegraphics[width=0.98\textwidth]{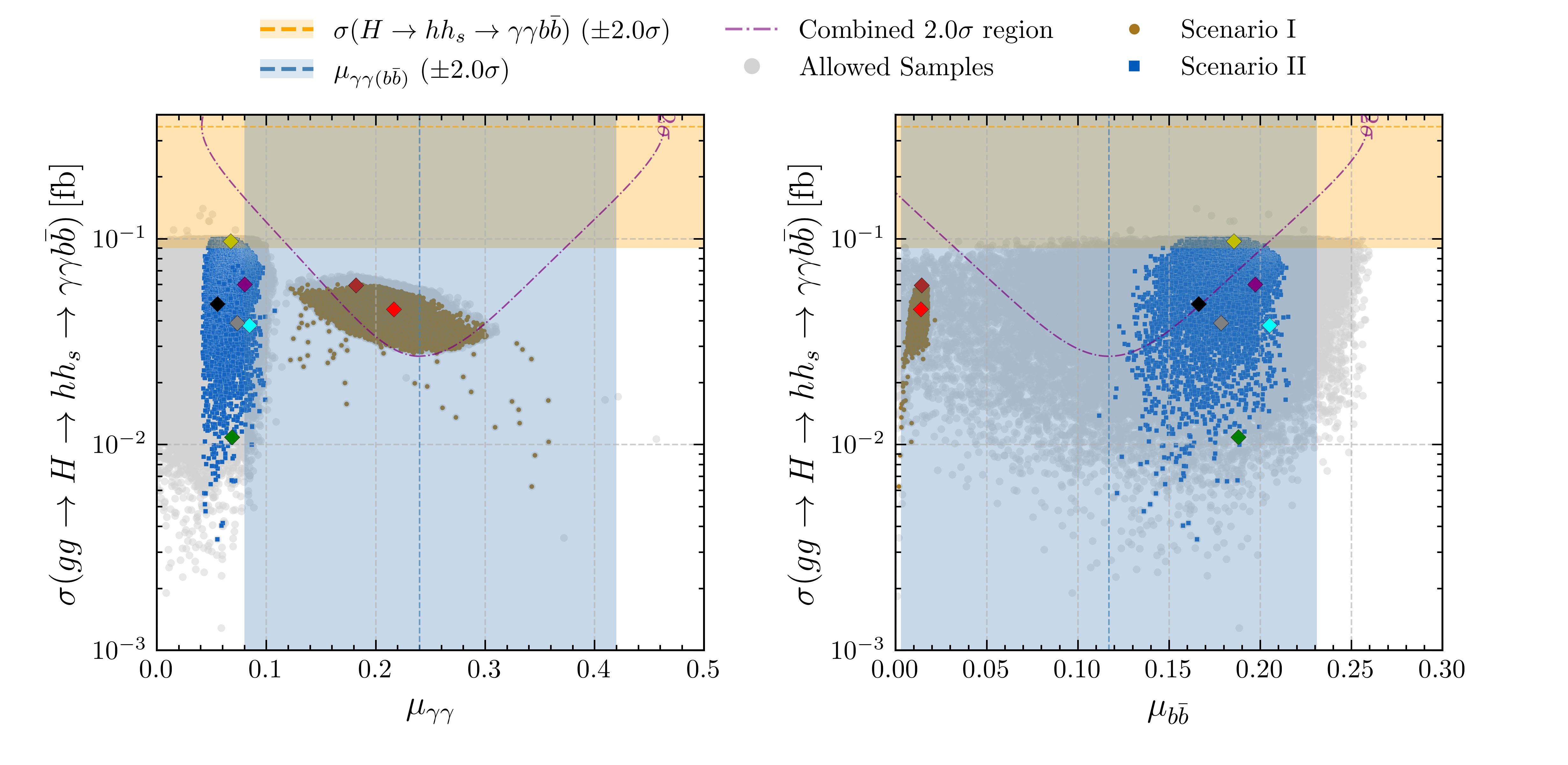}
\vspace{-0.6cm}
\caption{Distribution of viable samples in the planes of $\mu_{\gamma\gamma}$ and $\mu_{b\bar b}$ versus the heavy scalar cascade rate $\sigma(\mathrm{gg}\to H\to h h_s\to \gamma\gamma b\bar b)$. 
The grey points denote all allowed samples, while the brown circles and blue squares denote \textbf{Scenario I} and \textbf{Scenario II}, respectively. 
The colored bands show the corresponding $2\sigma$ intervals for the light-scalar and heavy-scalar excesses, and the dot-dashed curve indicates the combined $2\sigma$ region. 
The diamond markers denote representative benchmark points, with P1 shown in red, P2 in brown, P3 in purple, P4 in black, P5 in cyan, P6 in green, P7 in grey, and P8 in yellow. 
Their detailed information is given in Tables~\ref{tab:benchmark-p1-p2}--\ref{tab:benchmark-p7-p8}.}
\label{fig:mu-sigma}
\end{figure}

\begin{figure}[t]
\centering
\makebox[\textwidth][c]{%
\includegraphics[width=1.15\textwidth]{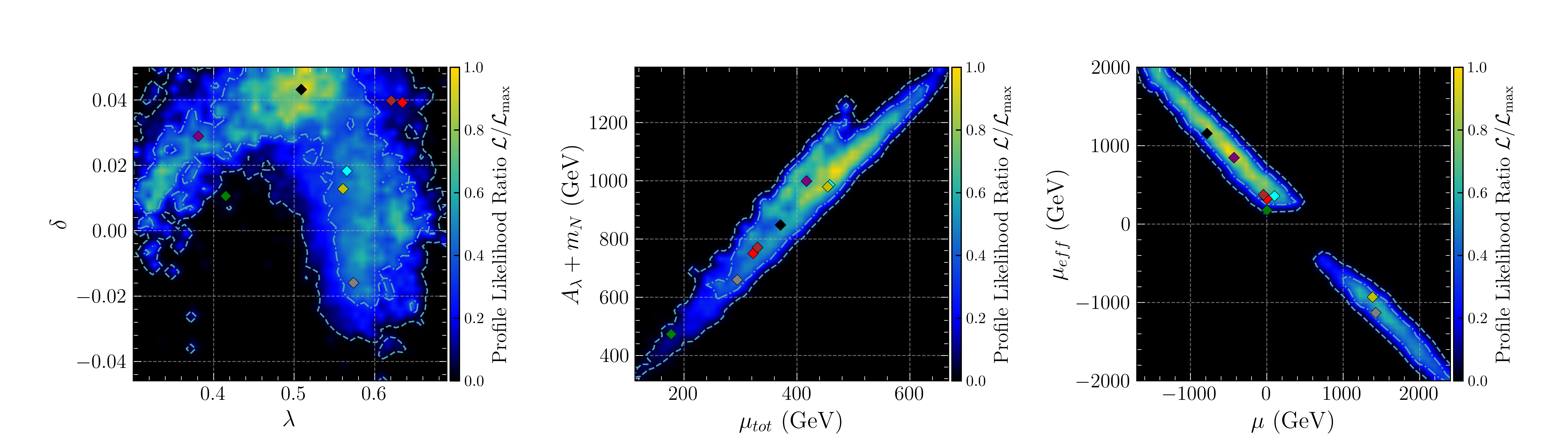}}
\vspace{-0.8cm}
\caption{Profile likelihood distributions in representative input-parameter planes. 
The left panel shows the correlation between $\lambda$ and the singlet-doublet mixing parameter $\delta$; the middle panel shows the relation between $\mu_{\rm tot}$ and $A_\lambda+m_N$; the right panel shows the relation between the explicit $\mu$ parameter and $\mu_{\rm eff}$. 
The contour lines indicate the $1\sigma$ and $2\sigma$ confidence intervals of the profile likelihood. 
The diamond markers denote the same benchmark points as in Fig.~\ref{fig:mu-sigma}.}
\label{fig:parameter-planes}
\end{figure}

\begin{figure}[tbp]
\centering
\makebox[\textwidth][c]{%
\includegraphics[width=1.15\textwidth]{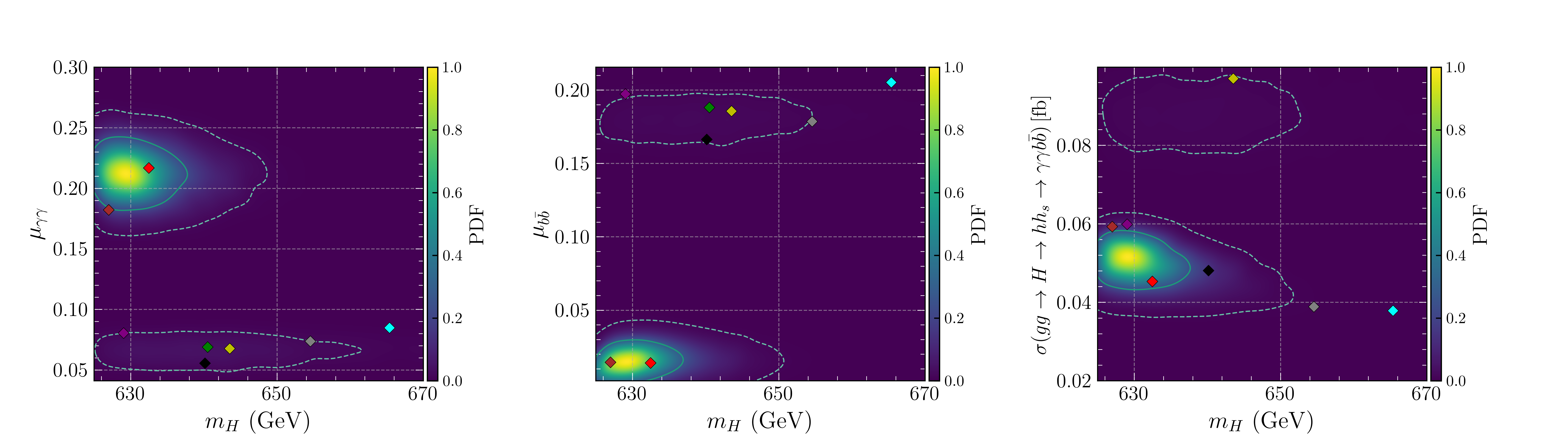}}
\vspace{-0.8cm}
\caption{Posterior probability density (PDF) distributions for the light- and heavy-scalar signal observables as functions of the heavy CP-even mass $m_H$. 
The panels show $\mu_{\gamma\gamma}$, $\mu_{b\bar b}$, and $\sigma(\mathrm{gg}\to H\to h h_s\to\gamma\gamma b\bar b)$, respectively. 
The contour lines denote the $1\sigma$ and $2\sigma$ credible regions of the posterior PDF. 
The diamond markers denote representative benchmark points as in Fig.~\ref{fig:mu-sigma}.}
\label{fig:heavy-signal}
\end{figure}

\begin{figure}[tbp]
\centering
\includegraphics[width=0.98\textwidth]{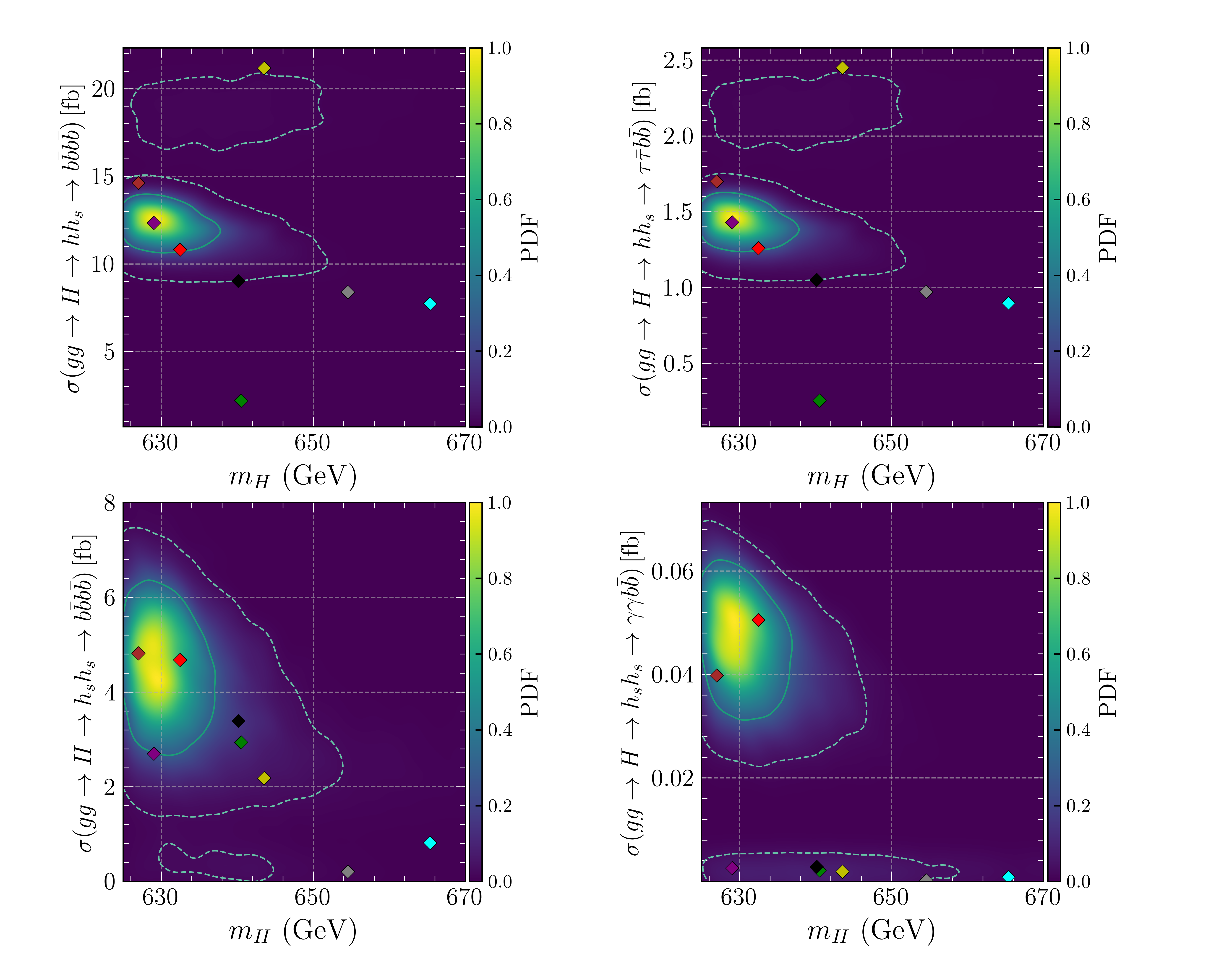}
\vspace{-0.5cm}
\caption{Posterior probability density distributions for selected heavy CP-even cascade rates as functions of $m_H$. 
The upper panels show the $H\to h h_s$ modes with $h h_s\to b\bar b b\bar b$ and $\tau\bar\tau b\bar b$, respectively. 
The lower panels show the $H\to h_s h_s$ modes with $h_s h_s\to b\bar b b\bar b$ and $\gamma\gamma b\bar b$, respectively. 
The diamond markers denote representative benchmark points as in Fig.~\ref{fig:mu-sigma}.}
\label{fig:H-cascade-channels}
\end{figure}

\begin{figure}[tbp]
\centering
\makebox[\textwidth][c]{%
\includegraphics[width=1.15\textwidth]{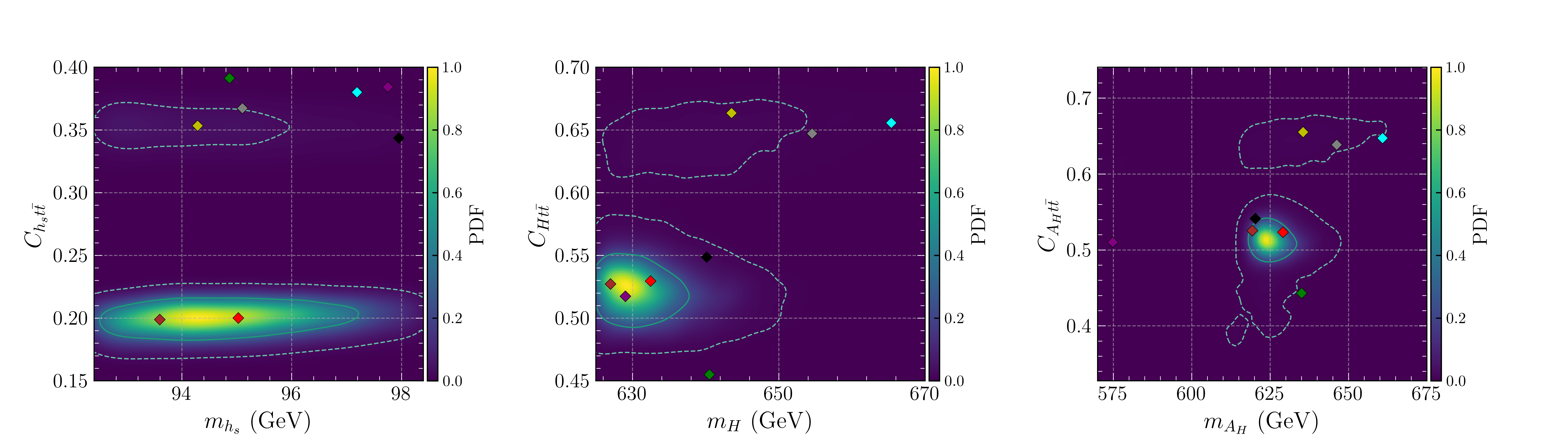}}
\vspace{-0.8cm}
\caption{Posterior probability density distributions of selected reduced couplings. 
The left panel shows the relation between the light scalar mass and its reduced top-quark coupling $C_{h_s t\bar t}$. 
The middle panel shows the heavy CP-even reduced top-quark coupling $C_{H t\bar t}$ as a function of $m_H$. 
The right panel shows the reduced top-quark coupling of the heavy CP-odd state $A_H$ as a function of $m_{A_H}$. 
The diamond markers denote representative benchmark points as in Fig.~\ref{fig:mu-sigma}.}
\label{fig:couplings}
\end{figure}

\begin{figure}[tbp]
\centering
\includegraphics[width=0.98\textwidth]{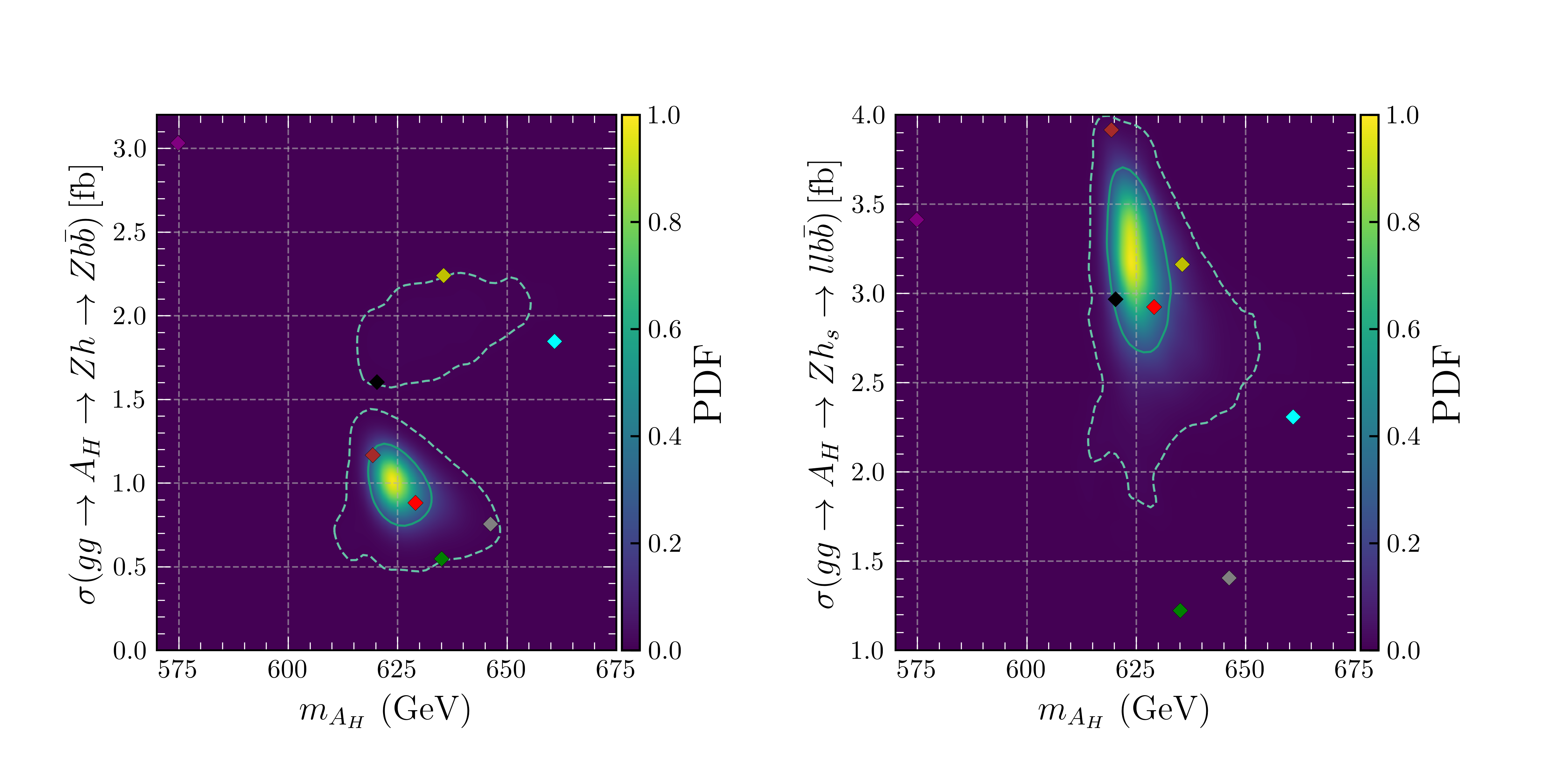}
\vspace{-0.6cm}
\caption{Posterior probability density distributions for heavy pseudoscalar cascade rates as functions of $m_{A_H}$. 
The left panel shows $\sigma(\mathrm{gg}\to A_H\to Zh\to Zb\bar b)$, while the right panel shows $\sigma(\mathrm{gg}\to A_H\to Zh_s\to \ell\ell b\bar b)$. 
The diamond markers denote representative benchmark points as in Fig.~\ref{fig:mu-sigma}.}
\label{fig:AH-Zh}
\end{figure}

\begin{figure}[tbp]
\centering
\includegraphics[width=0.7\textwidth]{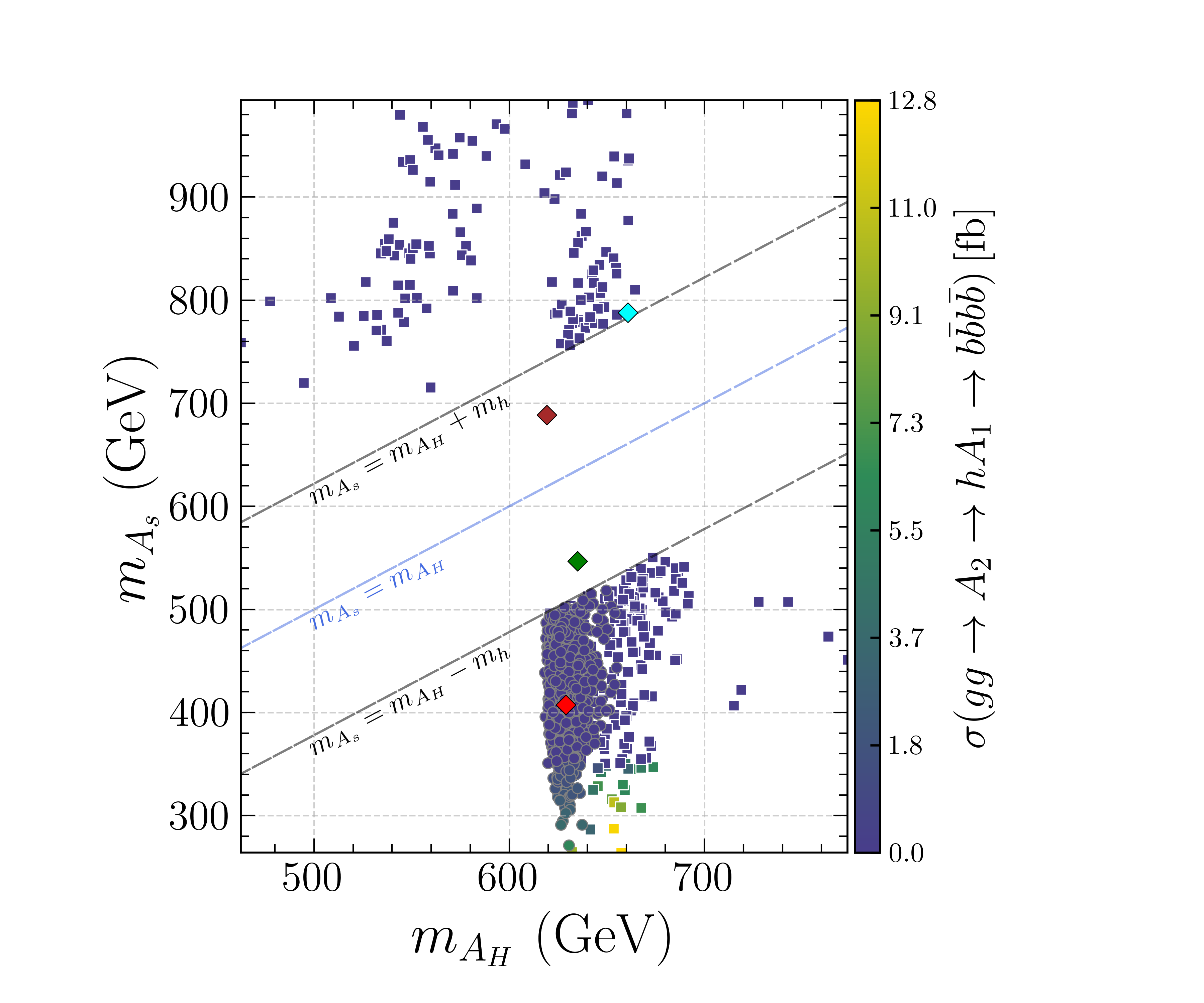}
\vspace{-0.6cm}
\caption{Distribution of viable samples in the $(m_{A_H},m_{A_s})$ plane for the subset with $m_{A_s}<1~{\rm TeV}$. 
The circular and square points denote \textbf{Scenario I} and \textbf{Scenario II} samples, respectively. 
The color indicates $\sigma(\mathrm{gg}\to A_2\to h A_1\to b\bar b b\bar b)$, where $A_2$ and $A_1$ are the heavier and lighter states among $A_H$ and $A_s$. 
The grey dashed lines show the thresholds $m_{A_s}=m_{A_H}\pm m_h$, the blue line shows $m_{A_s}=m_{A_H}$, and the diamond markers denote the benchmark points that lie in the displayed mass range.}
\label{fig:pseudoscalar-plane}
\end{figure}

After applying the theoretical and experimental requirements described above, we obtain $81,611$ viable samples that can accommodate a light scalar around $95~{\rm GeV}$ while keeping the heavy scalar sector close to the mass region relevant for the CMS searches for heavy scalars. 
Among them, more than $26,600$ samples can simultaneously explain the $95~{\rm GeV}$ diphoton and $b\bar b$ excesses within the $2\sigma$ ranges. 
The following scalar-resonance analysis mainly focuses on this subset, whose properties are projected onto seven planes in Figs.~\ref{fig:mu-sigma}--\ref{fig:pseudoscalar-plane}. 
The positive-$\mu$ subset is then used to illustrate the gravitino DM interpretation in Fig.~\ref{fig:gravitino-widths}. 
The diamond markers in the figures denote eight representative benchmark points selected to cover the characteristic regions of the viable parameter space. 
They are not isolated solutions, but rather typical examples of the two broad phenomenological patterns that emerge from the scan. 
Their detailed spectra, couplings, branching ratios and consistency tests are listed in Tables~\ref{tab:benchmark-p1-p2}--\ref{tab:benchmark-p7-p8}.

A salient feature of the result is that the viable samples split into two scenarios according to the predicted diphoton signal strength of the light scalar. 
Following the notation used in the figures, \textbf{Scenario I} corresponds to the region with
\begin{equation}
  \mu_{\gamma\gamma} > 0.11 ,
\end{equation}
whereas \textbf{Scenario II} corresponds to
\begin{equation}
  \mu_{\gamma\gamma} < 0.11 .
\end{equation}
This division is not imposed as a prior condition, but arises from the structure of the scalar mixing and the resulting pattern of reduced couplings of $h_s$. 
In \textbf{Scenario I}, the $95~{\rm GeV}$ scalar has a suppressed coupling to $b\bar b$, which reduces its total width and enhances the branching ratio into photons. 
Consequently, this scenario can naturally provide a sizeable LHC diphoton rate, while the LEP $b\bar b$ rate remains small. 
In \textbf{Scenario II}, by contrast, the light scalar has a larger doublet component and therefore a larger coupling to the $Z$ boson and to bottom quarks. 
This enhances the LEP rate, but at the same time increases the total width of $h_s$ and suppresses the diphoton branching fraction. 
The two scenarios therefore realize two different compromises between the LEP and LHC light-scalar hints.

Fig.~\ref{fig:mu-sigma} displays the most direct phenomenological consequence of this split. The grey points denote all allowed samples, while the brown circles and blue squares denote \textbf{Scenario I} and \textbf{Scenario II}, respectively.
The left panel shows that the samples with $\mu_{\gamma\gamma}>0.11$ form a broad island around $\mu_{\gamma\gamma}\simeq 0.12$--$0.30$, which overlaps well with the preferred diphoton region in Eq.~(\ref{diphoton-rate}). 
For these points, the cascade rate $\sigma(\mathrm{gg}\to H\to h h_s\to \gamma\gamma b\bar b)$ is typically at the level of a few times $10^{-2}~{\rm fb}$, and can reach approximately $6\times 10^{-2}~{\rm fb}$. 
Although this value is below the original best-fit value of the earlier CMS excess, it lies close to the region still allowed by the later null searches. 
This pattern reflects the tension already mentioned in Sec.~\ref{intro}. Current constraints disfavor a very large $\gamma\gamma b\bar b$ rate, but still allow a non-negligible rate that can be tested in future searches.

The right panel of Fig.~\ref{fig:mu-sigma} shows the complementary behavior in the LEP channel. 
\textbf{Scenario I} predicts a small $\mu_{b\bar b}$, usually around $\mathcal{O}(10^{-2})$, because the $h_s ZZ$ and $h_s b\bar b$ couplings are both suppressed. 
\textbf{Scenario II} instead forms a dense island around $\mu_{b\bar b}\simeq 0.15-0.20$, in good agreement with the LEP excess. 
In this scenario, the heavy scalar cascade rate can reach the lower edge of the original CMS $\sigma(\gamma\gamma b\bar{b})$ band, with maximal values around $0.1~{\rm fb}$. 
This improvement comes with a smaller diphoton rate, mostly below the central value of the ATLAS+CMS combination. 
Taken together, Fig.~\ref{fig:mu-sigma} shows two preferred ways of arranging the light-scalar signals. 
The first keeps the LEP rate small and fits the diphoton excess more directly, while the second improves the LEP interpretation at the cost of a smaller diphoton signal. 
The benchmark points follow this separation, with P1 and P2 on the diphoton-favored branch and P3--P8 on the LEP-favored branch.

The origin of the two scenarios can be seen more clearly from the parameter correlations shown in Fig.~\ref{fig:parameter-planes}. 
The left panel displays the profile likelihood in the $(\lambda,\delta)$ plane. 
The viable samples prefer a moderately large singlet-doublet coupling and a high-likelihood concentration around $\lambda\simeq 0.60-0.64$ for \textbf{Scenario I}. 
This is expected because a sizeable $\lambda$ is needed to generate the required mixing among $H_{\rm SM}$, $H_{\rm NSM}$ and $H_S$, and also helps to raise the SM-like Higgs mass without relying on an excessively heavy stop sector. 
At the same time, $\lambda$ cannot be pushed to arbitrarily large values because of perturbativity up to the GUT scale and because too large a mixing would spoil the observed Higgs couplings. 
The parameter $\delta$ remains small, mostly positive, and often close to the upper side of the scanned interval. 
This agrees with the alignment-motivated expectation discussed around Eq.~(\ref{delta}), where the $H_{\rm SM}$--$H_S$ mixing must be controlled to keep the $125~{\rm GeV}$ state SM-like.

The middle panel of Fig.~\ref{fig:parameter-planes} illustrates the approximate linear relation between $\mu_{\rm tot}$ and $A_\lambda+m_N$. 
This correlation is a direct consequence of Eq.~(\ref{delta}), since a small value of $\delta$ implies
\begin{equation}
  A_\lambda+m_N \simeq \frac{2\mu_{\rm tot}}{\sin 2\beta}.
\end{equation}
Because the successful samples favor low $\tan\beta$, typically between $1.5$ and $2.0$, the coefficient $2/\sin2\beta$ is of order a few. 
The high-likelihood band therefore follows a narrow diagonal trajectory. 
\textbf{Scenario I} occupies a relatively narrow window of low $\mu_{\rm tot}$, roughly $200$--$400~{\rm GeV}$, while \textbf{Scenario II} covers a broader interval from about $150~{\rm GeV}$ to $650~{\rm GeV}$. 
The low $\mu_{\rm tot}$ region is important because it corresponds to lighter Higgsino-like charginos, which can enhance the loop-induced $h_s\gamma\gamma$ coupling. 
The benchmark points make this point clear. 
P1 (red) and P2 (brown) have $\mu_{\rm tot}=323$ and $330~{\rm GeV}$, respectively, with $m_{\tilde\chi^\pm_1}=321$ and $329~{\rm GeV}$. 
Their chargino-loop contributions to $C_{h_s\gamma\gamma}$ are about $0.11$ and $0.10$ in magnitude, more than half of the total reduced coupling $C_{h_s\gamma\gamma}\simeq 0.20$, which helps produce $\mu_{\gamma\gamma}=0.217$ and $0.182$. 
A light chargino is not sufficient by itself. 
P6 (green) and P7 (grey) have even smaller chargino masses, $m_{\tilde\chi^\pm_1}=177$ and $293~{\rm GeV}$, but their diphoton rates remain modest, $\mu_{\gamma\gamma}=0.069$ and $0.074$. 
The reason is that they belong to \textbf{Scenario II}, where the light scalar has a much larger bottom coupling. 
For P6 and P7 one finds $C_{h_s b\bar b}=0.55$ and $0.51$, compared with $0.084$ and $0.092$ for P1 and P2. 
The corresponding branching ratios ${\rm Br}(h_s\to b\bar b)$ are about $0.86$ for P6 and P7, so the enhanced bottom width dominates over the chargino-loop enhancement and suppresses the diphoton branching fraction.

The right panel of Fig.~\ref{fig:parameter-planes} shows the relation between the explicit $\mu$ parameter and $\mu_{\rm eff}$. 
Since the physical Higgsino mass is governed by $\mu_{\rm tot}=\mu+\mu_{\rm eff}$, the viable samples lie close to diagonal bands where cancellations or additions between $\mu$ and $\mu_{\rm eff}$ produce an electroweak-scale $\mu_{\rm tot}$. 
The benchmark points make this structure more transparent. 
P1 (red) has a small positive explicit term, $\mu=2.6~{\rm GeV}$, together with $\mu_{\rm eff}=319.9~{\rm GeV}$, while P2 (brown) reaches a similar $\mu_{\rm tot}$ with a slightly negative $\mu=-42.7~{\rm GeV}$. 
The two negative-$\mu$ points in \textbf{Scenario II}, P3 (purple) and P4 (black), instead rely on cancellations between a negative explicit $\mu$ and a large positive $\mu_{\rm eff}$. 
The positive-$\mu$ points P5--P8 show that \textbf{Scenario II} is also compatible with the sign preferred by the inflation-inspired construction. 
P5 (cyan) and P6 (green) have positive $\mu_{\rm eff}$ and positive or very small $\mu$, whereas P7 (grey) and P8 (yellow) use the opposite cancellation pattern, with large positive $\mu$ compensated by negative $\mu_{\rm eff}$. 
Thus the collider fit mainly selects the combination $\mu_{\rm tot}$ rather than the sign of the explicit $\mu$ term, leaving the positive-$\mu$ benchmark points available for the gravitino interpretation discussed later.

Fig.~\ref{fig:heavy-signal} presents the posterior probability density in the planes involving the heavy CP-even scalar mass $m_H$. 
The viable samples cluster in the range $m_H\simeq 625$--$670~{\rm GeV}$, with the densest regions below $650~{\rm GeV}$. 
This mild preference for the lower side of the scanned heavy mass window can be understood from the rate balance. 
At smaller $m_H$, the gluon-fusion production cross section of the doublet-like heavy scalar is larger, and the phase space for $H\to h h_s$ is still ample. 
As $m_H$ increases, the production rate decreases and the total width can receive larger contributions from other channels, making it harder to maintain a sizeable $\gamma\gamma b\bar b$ final state.

The left panel of Fig.~\ref{fig:heavy-signal} again shows the two islands in $\mu_{\gamma\gamma}$. 
The high-density region of \textbf{Scenario I} is centered near $\mu_{\gamma\gamma}\simeq 0.20$ and $m_H\simeq 625$--$635~{\rm GeV}$. 
This region provides the best fit to the LHC diphoton excess. 
The lower island, associated with \textbf{Scenario II}, has $\mu_{\gamma\gamma}\simeq 0.05$--$0.08$ and extends to somewhat larger $m_H$. 
Although its diphoton rate is farther from the ATLAS+CMS central value, it is not excluded because the excess remains statistically limited and because the LEP channel is better reproduced.

The middle panel confirms the complementarity in the LEP signal strength. 
\textbf{Scenario I} predicts $\mu_{b\bar b}$ close to zero or at most a few times $10^{-2}$, whereas \textbf{Scenario II} gives $\mu_{b\bar b}\simeq 0.17-0.20$. 
This pattern is controlled mainly by the reduced couplings of $h_s$. 
In \textbf{Scenario I}, $h_s$ is more singlet-like and its coupling to $ZZ$ is small, so the LEP production process $e^+e^-\to Zh_s$ is suppressed. 
In \textbf{Scenario II}, the mixing with the SM-like direction is larger, giving a visible LEP rate. 
However, because the same mixing also increases the $h_s\to b\bar b$ partial width, the diphoton branching fraction is reduced. 
This explains why the two excesses are difficult to fit simultaneously at their central values in the $\mu$NMSSM under the present constraints.

The right panel of Fig.~\ref{fig:heavy-signal} shows the probability density for the $\gamma\gamma b\bar b$ cascade rate. 
The highest posterior density lies around the same lower side of $m_H$, with $\sigma_{\gamma\gamma b\bar b}$ between $4\times 10^{-2}~{\rm fb}$ and $6\times 10^{-2}~{\rm fb}$. 
The upper island, corresponding mostly to \textbf{Scenario II}, can reach nearly $0.1~{\rm fb}$, which is close to the lower edge of the original $2\sigma$ region of the CMS excess and comparable to the limits inferred from other final states. 
The benchmark points demonstrate that both scenarios may produce observable heavy-scalar cascades, but the more LEP-like \textbf{Scenario II} tends to predict a larger $\gamma\gamma b\bar b$ rate because of the enhanced $h_s\to b\bar b$ branching fraction.

Fig.~\ref{fig:H-cascade-channels} displays four cascade channels of the heavy CP-even state, but they correspond to two different decay topologies. 
The two upper panels are induced by $H\to h h_s$, while the two lower panels are induced by $H\to h_s h_s$. The upper-left panel shows the $H\to h h_s\to b\bar b b\bar b$ rate. 
This is directly related to the most recent CMS search for a heavy scalar resonance in the $X\to YH\to b\bar b b\bar b$ final state~\cite{CMS:2026mwf}. 
For the mass configuration relevant to the earlier $\gamma\gamma b\bar b$ excess, namely $M_X\simeq 650~{\rm GeV}$ and $M_Y\simeq 95~{\rm GeV}$, that analysis gives an upper limit of about $20$--$25~{\rm fb}$ on the $4b$ cross section, and no significant excess is observed at this point. 
The predicted rate in our samples is typically around $10$--$13~{\rm fb}$ in the highest-density region, while some benchmark points approach the current upper limit. 
Thus the $4b$ channel already provides one of the most relevant constraints on the $H\to h h_s$ interpretation. 

The upper-right panel shows the corresponding $H\to h h_s\to \tau\bar\tau b\bar b$ rate, where the SM-like Higgs decays into $\tau\bar\tau$ and the light scalar decays into $b\bar b$. 
This final state is constrained by the CMS search in~\cite{CMS:2021yci}. 
For $m_H\simeq 650~{\rm GeV}$ and $m_{h_s}\simeq 95~{\rm GeV}$, the model-independent $95\%$ C.L. upper limit is approximately $3~{\rm fb}$, as discussed in the Introduction. 
The high-probability region of our scan predicts $\sigma(\mathrm{gg}\to H\to h(\tau\bar\tau)h_s(b\bar b))$ around $1$--$1.5~{\rm fb}$, with benchmark points reaching roughly $2.5~{\rm fb}$. 
This channel is therefore still allowed, but the largest predicted rates are already close to the current CMS sensitivity. 
Since the same $H\to h h_s$ decay controls the $\gamma\gamma b\bar b$ final state, the $\tau\tau b\bar b$ bound also prevents the cascade rate from being increased arbitrarily.

The lower-left panel instead corresponds to $H\to h_s h_s\to b\bar b b\bar b$. 
The predicted rate is nevertheless sizable, typically a few fb and occasionally larger, because both light scalars decay dominantly into bottom quarks in much of the viable parameter space. 
A dedicated search for a heavy resonance decaying into two equal-mass light scalars near $95~{\rm GeV}$ in the four-bottom final state would therefore provide a direct test of this part of the model.

The lower-right panel shows $H\to h_s h_s\to \gamma\gamma b\bar b$. 
Although its rate is smaller than that of the four-bottom mode, it can reach the level of several times $10^{-2}~{\rm fb}$ in the high-diphoton region. 
This final state is especially characteristic of \textbf{Scenario I}, where the suppression of the $h_s b\bar b$ width enhances ${\rm Br}(h_s\to\gamma\gamma)$. 
Compared with the usual $X\to H_{\rm 125}(\gamma\gamma)Y(b\bar b)$ topology, this signal would contain a narrow diphoton peak near $95~{\rm GeV}$ together with a $b\bar b$ resonance at the same mass. 
The simultaneous reconstruction of $m_{\gamma\gamma}$, $m_{b\bar b}$ and $m_{\gamma\gamma b\bar b}$ makes the continuum $\gamma\gamma+{\rm jets}$ background relatively smooth and reducible, although the small ${\rm Br}(h_s\to\gamma\gamma)$ keeps the signal rate low. 
A dedicated $X\to h_s h_s\to\gamma\gamma b\bar b$ search would therefore provide a clean probe of the double-light-scalar cascade, complementing the existing constraints on $H\to h h_s$ and directly testing heavy resonances decaying into two light scalars around $95~{\rm GeV}$.

The left panel of Fig.~\ref{fig:couplings} shows the reduced top-quark coupling of the light scalar, $C_{h_s t\bar t}$, as a function of $m_{h_s}$. 
The high-density region is concentrated near $m_{h_s}\simeq 93$--$97~{\rm GeV}$ and $C_{h_s t\bar t}\simeq 0.18$--$0.21$. 
A coupling of this size is large enough to generate gluon-fusion production at the LHC, but small enough to preserve the singlet-dominated nature of the light scalar. 
The benchmarks with larger $C_{h_s t\bar t}$ correspond to points with enhanced doublet mixing. 
These points typically have a stronger LEP signal and therefore belong to, or lie close to, \textbf{Scenario II}. 
The same mixing also tends to increase the bottom coupling and suppress the diphoton branching fraction.

The middle panel displays the reduced top coupling of the heavy CP-even scalar $H$. 
The preferred region lies around $C_{H t\bar t}\simeq 0.50$--$0.55$ for $m_H\simeq 625$--$635~{\rm GeV}$. 
Since the gluon-fusion production rate scales approximately with the square of this coupling, this range is crucial for generating visible heavy-scalar cascades. 
At the same time, the coupling cannot be too large because the heavy scalar would then acquire excessive production rates in channels such as $H\to t\bar t$, $H\to ZZ$, or $H\to hh$, which are constrained by LHC searches. 
The viable points therefore occupy a relatively narrow strip in which $H$ remains mostly non-SM doublet-like but still has enough coupling to top quarks to be produced.

The right panel shows the corresponding reduced top coupling of the heavy CP-odd state $A_H$. 
For a heavy doublet-like pseudoscalar, this coupling is approximately
\begin{equation}
  C_{A_H t\bar t} \simeq |P^{\rm NSM}_{A_H}|\cot\beta ,
\end{equation}
up to the sign convention of the pseudoscalar mixing matrix. 
The benchmark points confirm this simple scaling. 
For all eight points one finds $C_{A_H t\bar t}/\cot\beta\simeq 0.99$--$1.00$, showing that $A_H$ is almost purely doublet-like and that the spread in the right panel is mainly driven by $\tan\beta$. 
The three benchmarks with $C_{A_H t\bar t}>0.6$, namely P5 (cyan), P7 (grey) and P8 (yellow), have $\tan\beta=1.54$, $1.57$ and $1.53$, giving $C_{A_H t\bar t}=0.647$, $0.639$ and $0.655$, respectively. 
The other benchmarks have $\tan\beta\simeq 1.85$--$2.25$, for which $\cot\beta\lesssim 0.54$, and their $A_H t\bar t$ couplings stay below $0.6$. 
P6 (green) illustrates this point most clearly. 
Although its heavy pseudoscalar is also nearly doublet-like, its larger $\tan\beta=2.25$ lowers the coupling to $C_{A_H t\bar t}=0.443$.

This low-$\tan\beta$ edge is close to the sensitivity of the CMS search for heavy scalar and pseudoscalar resonances decaying into $t\bar t$ at $13~{\rm TeV}$ with $35.9~{\rm fb}^{-1}$~\cite{CMS:2019pzc}.  
The largest benchmark values, $C_{H t\bar t}=0.663$ and $C_{A_H t\bar t}=0.655$ for P8, remain below the upper limits of roughly 0.735 on $C_{H t\bar t}$ and 0.675 on $C_{A_H t\bar t}$ for masses near $650~{\rm GeV}$~\cite{Ellwanger:2023zjc}, while P5 and P7 are only slightly smaller. 
Thus the present $t\bar t$ searches do not exclude the benchmark points, but improved analyses with larger data sets can directly test the low-$\tan\beta$ part of the preferred region. 
This comparison should be interpreted with some care because the heavy states also have cascade and electroweakino decay modes, and because the experimental $t\bar t$ analysis includes interference with the SM continuum background. 

Fig.~\ref{fig:AH-Zh} focuses on the $A_H\to Z h$ and $A_H\to Z h_s$ modes. 
The left panel shows that the rate for $\mathrm{gg}\to A_H\to Zh\to Zb\bar b$ is typically around $1~{\rm fb}$ in the high-density region, although some benchmark points can reach larger values. 
This channel is controlled by the mixing between the heavy doublet pseudoscalar and the CP-even states. 
The $Zh$ mode is sensitive to the departure from exact alignment, because a perfectly SM-like $h$ would suppress the relevant coupling. 
The rate can nevertheless reach the fb level, indicating that the small non-SM components of the $125~{\rm GeV}$ Higgs still allowed by HiggsSignals can have visible consequences in heavy pseudoscalar decays.

The right panel shows the $A_H\to Zh_s\to \ell\ell b\bar b $ rate. 
Although \textbf{Scenario I} has a suppressed $h_s b\bar b$ coupling, the total width of $h_s$ is also reduced, so ${\rm Br}(h_s\to b\bar b)$ remains at the level of $0.59-0.63$ for P1 and P2. 
More importantly, these two points have relatively large pseudoscalar cascade branching ratios, ${\rm Br}(A_H\to Zh_s)=9.8\%$ and $11.2\%$, compared with about $2.5-5.8\%$ for most of the \textbf{Scenario II} benchmarks. 
Together with sizeable gluon-fusion production, $\sigma(\mathrm{gg}\to A_H)=0.51~{\rm pb}$ for P1 and $0.56~{\rm pb}$ for P2, this gives large $\ell\ell b\bar b$ rates. 
Even though the \textbf{Scenario II} benchmarks have a large $h_s\to b\bar b$ branching fraction, their $Zh_s$ rates are lower because either the pseudoscalar production rate or ${\rm Br}(A_H\to Zh_s)$ is reduced. 
Thus the right panel does not separate the two scenarios simply by the size of the $h_s b\bar b$ coupling. 

The final projection, shown in Fig.~\ref{fig:pseudoscalar-plane}, addresses the possibility that the pseudoscalar sector contributes to heavy-resonance structures in multi-bottom final states. 
Only samples with a singlet-like pseudoscalar below $1~{\rm TeV}$ are shown. 
The circles and squares correspond to \textbf{Scenario I} and \textbf{Scenario II}, respectively. 
The color scale denotes the rate for the ordered cascade
\begin{equation}
  \mathrm{gg}\to A_2\to h A_1\to b\bar b b\bar b ,
\end{equation}
where $A_2$ is the heavier one of $A_H$ and $A_s$, and $A_1$ is the lighter one. 
Thus, below the lower grey dashed line one has $m_{A_H}>m_{A_s}+m_h$ and the relevant decay is $A_H\to hA_s$. 
Above the upper grey dashed line one has $m_{A_s}>m_{A_H}+m_h$ and the relevant decay is $A_s\to hA_H$. 
The region between the two grey dashed lines is kinematically closed for the two-body decay $A_2\to hA_1$.

The four visible benchmark points illustrate these possibilities. 
P1 (red) lies below the lower threshold line, with $m_{A_H}=629~{\rm GeV}$ and $m_{A_s}=407~{\rm GeV}$, so the channel $A_H\to hA_s$ is open. 
Its $4b$ rate is nevertheless small because the lighter pseudoscalar is above the $t\bar t$ threshold and decays mainly into top quarks, leaving ${\rm Br}(A_s\to b\bar b)\simeq 0.5\%$. 
P5 (cyan) lies above the upper threshold line, with $m_{A_s}=788~{\rm GeV}$ and $m_{A_H}=661~{\rm GeV}$, so the opposite ordering $A_s\to hA_H$ is allowed, but only with limited phase space. 
In addition, the daughter state $A_H$ decays dominantly into $t\bar t$, which keeps the $4b$ rate tiny. 
P2 (brown) and P6 (green) lie between the two grey lines. 
They therefore represent the two closed-threshold cases, one on each side of the blue diagonal.

The largest color values in the scan occur mainly below the lower grey line, where $A_H\to hA_s$ is open and the singlet-like pseudoscalar is light enough that its decay into $t\bar t$ is absent or suppressed. 
In this part of the plane the daughter pseudoscalar can still decay appreciably into $b\bar b$, allowing $\sigma(\mathrm{gg}\to A_2\to hA_1\to 4b)$ to reach the level of several fb and in some cases about $10~{\rm fb}$. 
By contrast, points above the upper grey line often have a heavy doublet-like daughter $A_H$, whose dominant decay is $t\bar t$, so the same ordered cascade gives a much smaller $4b$ rate.

Two implications follow from this result. 
First, the heavy-resonance phenomenology of the $\mu$NMSSM is not limited to the CP-even cascade $H\to h h_s$. 
The CP-odd sector can generate additional resonant structures involving the SM-like Higgs and a lighter pseudoscalar, with the relevant decay chain determined by the ordering of $A_H$ and $A_s$. 
Second, this topology is directly relevant to the most recent CMS search in the $b\bar b b\bar b$ final state~\cite{CMS:2026mwf}. 
That analysis finds its largest local deviation around $(M_X,M_Y)\simeq(600,400)~{\rm GeV}$, rather than at the earlier $(650,95)~{\rm GeV}$ point. 
The pseudoscalar cascade shown in Fig.~\ref{fig:pseudoscalar-plane} naturally populates a nearby mass configuration, especially when the heavy parent is close to the doublet scale and the lighter pseudoscalar lies at a few hundred GeV. 
The predicted $4b$ rates in our scan remain below the current CMS upper limits, while a dedicated fit to this channel could further test whether the CP-odd cascade can account for this local excess.

\subsection{Gravitino dark matter in the inflation-inspired scenario}
\label{subsec:gravitino-dm}

\begin{figure}[tbp]
\centering
\makebox[\textwidth][c]{%
\includegraphics[width=1.15\textwidth]{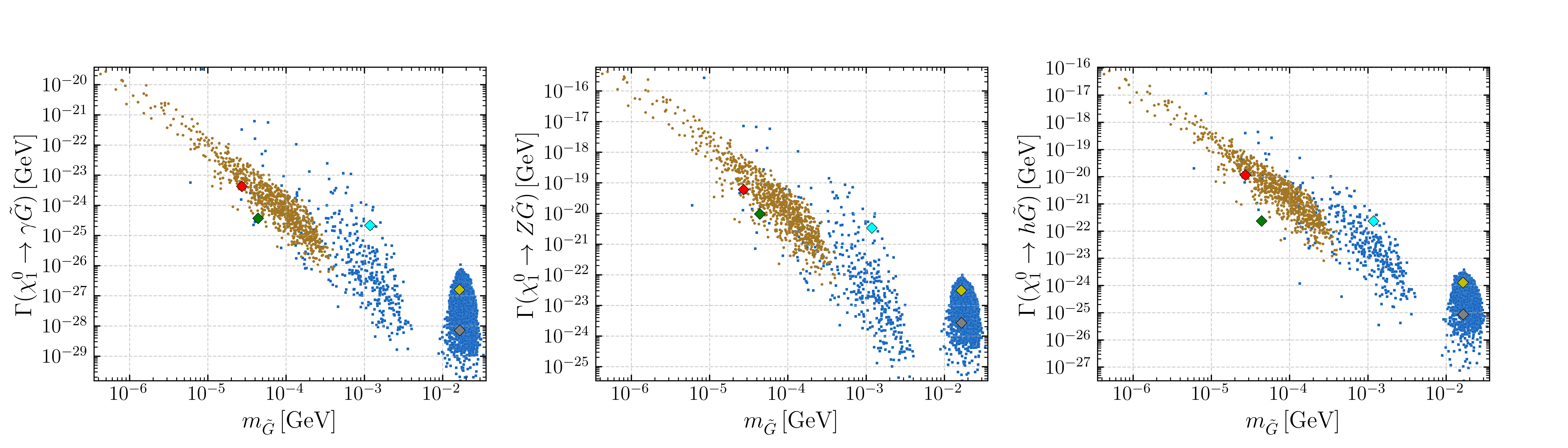}}
\vspace{-0.8cm}
\caption{Partial decay widths of the neutralino NLSP into the gravitino LSP for the positive-$\mu$ samples, shown as functions of the gravitino mass $m_{\tilde G}$. 
The three panels correspond to $\tilde\chi^0_1\to\gamma\tilde G$, $\tilde\chi^0_1\to Z\tilde G$, and $\tilde\chi^0_1\to h\tilde G$, respectively. 
The colored points follow the same scenario classification as in the previous figures, and the diamond markers denote representative benchmark points as in Fig.~\ref{fig:mu-sigma}.}
\label{fig:gravitino-widths}
\end{figure}

After the collider analysis, we now focus on the subset with a positive explicit $\mu$ term. 
This region is singled out by the inflation-inspired supergravity construction discussed in Sec.~\ref{Section-Model}. 
In that framework, the nonminimal Higgs-gravity coupling introduced to realize Higgs-sector inflation also induces the low energy $\mu$ term after the superconformal theory is matched to the $\mu$NMSSM. 
Combining the induced-$\mu$ relation with the inflationary normalization of the Higgs-gravity coupling gives a gravitino mass
\begin{equation}
  m_{\tilde G}=m_{3/2}\simeq \frac{2\mu}{3\times 10^5\lambda}.
\end{equation}
We therefore do not impose this relation on the full scan, but use it as a motivated dark matter interpretation of the positive-$\mu$ samples. 
For the viable points, it maps the collider-motivated Higgs parameter space to a light gravitino with a mass ranging from the keV scale to several tens of MeV. 
The lightest neutralino $\tilde\chi^0_1$ is then no longer the stable dark matter particle, but becomes the NLSP and eventually decays into the gravitino.

Fig.~\ref{fig:gravitino-widths} shows the partial decay widths of $\tilde\chi^0_1$ into $\gamma\tilde G$, $Z\tilde G$, and $h\tilde G$. 
In the neutralino convention of Eq.~(\ref{eq:neutralino-mixing}), and writing $S_{hd}$ and $S_{hu}$ for the $\sqrt{2}{\rm Re}H_d^0$ and $\sqrt{2}{\rm Re}H_u^0$ components of the SM-like Higgs boson $h$, the three widths used in Fig.~\ref{fig:gravitino-widths} are~\cite{Ambrosanio:1996jn,Feng:2003uy,Covi:2009bk,Hasenkamp:2009zz}
\begin{equation}
\begin{aligned}
\Gamma(\tilde\chi^0_1\to \gamma\tilde G)
&= \frac{|N_{11}c_W+N_{12}s_W|^2}{48\pi M_{\rm Pl}^2}
\frac{m_{\tilde\chi^0_1}^5}{m_{\tilde G}^2},\\
\Gamma(\tilde\chi^0_1\to Z\tilde G)
&= \frac{|-N_{13}s_W+N_{14}c_W|^2}{48\pi M_{\rm Pl}^2}
\frac{m_{\tilde\chi^0_1}^5}{m_{\tilde G}^2}
\left(1-\frac{m_Z^2}{m_{\tilde\chi^0_1}^2}\right)^4,\\
\Gamma(\tilde\chi^0_1\to h\tilde G)
&= \frac{|N_{13}S_{hd}+N_{14}S_{hu}|^2}{96\pi M_{\rm Pl}^2}
\frac{m_{\tilde\chi^0_1}^5}{m_{\tilde G}^2}
\left(1-\frac{m_h^2}{m_{\tilde\chi^0_1}^2}\right)^4 .
\end{aligned}
\label{eq:neutralino-gravitino-widths}
\end{equation}
Here $s_W=\sin\theta_W$, $c_W=\cos\theta_W$, and $M_{\rm Pl}$ is the reduced Planck mass. 
The common goldstino scaling, $\Gamma_{\tilde\chi^0_1}\propto m_{\tilde\chi^0_1}^5/(M_{\rm Pl}^2m_{\tilde G}^2)$, makes the widths decrease rapidly as $m_{\tilde G}$ increases. 
The photon mode is controlled by the small bino-wino admixture of the NLSP, while the $Z\tilde G$ and $h\tilde G$ modes are mainly controlled by its Higgsino components. 
This explains why the photon width is usually the smallest one in Fig.~\ref{fig:gravitino-widths}.

The scenario dependence enters indirectly through the positive-$\mu$ spectra. 
Among the benchmark points, P1 is the positive-$\mu$ representative of \textbf{Scenario I}. 
It has a mostly singlino NLSP, but its gravitino mass is only of order $10~{\rm keV}$, so the $m_{\tilde G}^{-2}$ enhancement keeps the $Z\tilde G$ and $h\tilde G$ widths relatively large. 
The positive-$\mu$ representatives of \textbf{Scenario II}, P5--P8, show a broader pattern. 
P5 and P6 have lighter gravitinos and sizeable Higgsino components, leading to larger $Z\tilde G$ widths, whereas P7 and P8 have $m_{\tilde G}$ around a few $10~{\rm MeV}$ and are therefore much longer-lived even though their NLSPs are Higgsino-like. 
Thus Fig.~\ref{fig:gravitino-widths} is mainly shaped by the explicit $\mu$ parameter, the neutralino mixing and the gravitino mass, rather than by the light-scalar decay pattern that separates \textbf{Scenario I} from \textbf{Scenario II}. 
Numerically, the largest widths are at most around $10^{-16}~{\rm GeV}$, while many samples have partial widths several orders of magnitude smaller. 
Using
\begin{equation}
  c\tau_{\tilde\chi^0_1}\simeq 1.97~{\rm m}\left(\frac{10^{-16}~{\rm GeV}}{\Gamma_{\rm tot}}\right),
\end{equation}
one finds that the neutralino NLSP is typically a long-lived particle at collider scales. 
Only the very light-gravitino edge of the parameter space can lead to decays inside the main LHC detectors, and even there displaced decays are more likely than prompt decays.

This lifetime pattern limits the impact of current collider searches. 
Prompt searches for gauge-mediated SUSY with photons, $Z$ bosons, Higgs bosons and missing transverse momentum are therefore not directly applicable to most of our points~\cite{ATLAS:2018nud}. 
On the other hand, for larger $m_{\tilde G}$ the neutralino often escapes the detector before decaying and behaves effectively as missing energy. 
The electroweakino sector in the present scan is dominantly Higgsino/singlino-like and relatively compressed, with the lightest chargino and the two lightest neutralinos typically at a few hundred GeV. 
As already checked with \textsf{SModelS}, existing electroweakino simplified-model searches do not exclude the viable samples. 
Dedicated long-lived searches for delayed photons, displaced $Z/h$ decays plus missing energy, and disappearing or soft tracks associated with the charged Higgsino can nevertheless probe parts of this scenario in the future. 
The high-luminosity LHC, timing layers, and proposed far detectors would be especially useful for the region with decay lengths from meters to kilometers, while a future lepton collider could test the Higgsino-like states directly if the available collision energy is sufficient.

The gravitino relic abundance is controlled primarily by the reheating temperature $T_R$ after inflation. 
For thermal production through scatterings in the plasma, a useful estimate is~\cite{Bolz:2000fu,Pradler:2006qh}
\begin{equation}
  \Omega_{\tilde G}^{\rm TP} h^2 \simeq 0.27
  \left(\frac{T_R}{10^{10}~{\rm GeV}}\right)
  \left(\frac{100~{\rm GeV}}{m_{\tilde G}}\right)
  \left(\frac{m_{\tilde g}}{1~{\rm TeV}}\right)^2 ,
\end{equation}
where $m_{\tilde g}$ denotes the gluino mass. 
Taking the gluino mass fixed in our scan, $m_{\tilde g}=3~{\rm TeV}$, and requiring $\Omega_{\tilde G}h^2\simeq 0.12$ gives the rough relation
\begin{equation}
  T_R \simeq 5\times 10^4~{\rm GeV}
  \left(\frac{m_{\tilde G}}{10~{\rm MeV}}\right)
  \left(\frac{3~{\rm TeV}}{m_{\tilde g}}\right)^2 .
\end{equation}
Therefore, the gravitino mass interval shown in Fig.~\ref{fig:gravitino-widths} corresponds approximately to reheating temperatures from a few GeV for a keV gravitino to $10^5~{\rm GeV}$ for a gravitino of several tens of MeV. 
This estimate should be regarded as indicative, since the high-temperature expression is modified if $T_R$ is comparable to or below the superpartner masses. 
The nonthermal contribution from late NLSP decays is suppressed by the mass ratio,
\begin{equation}
  \Omega_{\tilde G}^{\rm NTP}h^2
  = \frac{m_{\tilde G}}{m_{\tilde\chi^0_1}}\,
  \Omega_{\tilde\chi^0_1}^{\rm FO}h^2 ,
\end{equation}
and is usually subdominant for a gravitino from the keV to MeV scale unless the neutralino freeze-out abundance before decay is very large~\cite{Feng:2003uy}.

Several cosmological constraints must be kept in mind. 
For gravitino masses near the keV scale, free streaming and structure formation bounds on warm DM can become important, depending on the production mechanism and the reheating history. 
For the heavier part of the range, the gravitino behaves as cold DM, but the neutralino NLSP may decay close to the epoch of Big Bang nucleosynthesis. 
Hadronic and electromagnetic energy injection from the $Z\tilde G$ and $h\tilde G$ modes can then constrain the combination of the NLSP lifetime and its freeze-out abundance~\cite{Kawasaki:2004yh}. 
The widths in Fig.~\ref{fig:gravitino-widths} suggest that much of the low-$m_{\tilde G}$ region decays before BBN, while the upper-MeV region may require a dedicated BBN analysis. 
Such a study would need the full neutralino yield, branching fractions, thermal production rate, and possible entropy dilution, and is beyond the scope of the present work. 
We therefore regard Fig.~\ref{fig:gravitino-widths} as a first consistency check and leave a detailed treatment of gravitino DM and reheating cosmology for future work.

The eight benchmark points in Tables~\ref{tab:benchmark-p1-p2}--\ref{tab:benchmark-p7-p8} provide the detailed spectra and constraints behind the colored diamonds in Figs.~\ref{fig:mu-sigma}--\ref{fig:gravitino-widths}. 
Besides the signal rates already discussed above, these tables show that all eight points satisfy the electroweak precision test, with $S$, $T$ and $U$ in the ranges used in the scan, and give acceptable SM-like Higgs fits with $\chi^2_{\rm HiggsSignals}\simeq 152$--$159$. 
They also pass the simplified-model SUSY constraints, since $R_{\rm SModelS}<1$ in each case and even the largest values are only about $0.45$. 
The entry labelled Signal Region denotes the experimental signal region that gives the quoted maximal $R_{\rm SModelS}$ value. 
\newcommand{\BPCascadeNote}{A dash in the $A_2\to h A_1$ row means that the parent pseudoscalar exceeds $1~{\rm TeV}$, outside the range used for this calculation, or that the decay is kinematically closed. Dashes in the $m_{3/2}$ and neutralino partial-width rows means that the positive-$\mu$ supergravity relation used for the gravitino interpretation is not imposed.}
\newcommand{\BPTableSkip}{\vspace{0.12cm}}

\begin{table}[!p]
\centering
\caption{\label{tab:benchmark-p1-p2} Details of benchmark points P1 and P2 in \textbf{Scenario I}. Input dimensional parameters and cross sections are given in GeV and fb, respectively; branching ratios are in percent. \BPCascadeNote}
\BPTableSkip
\resizebox{1\textwidth}{!}{%
\begin{tabular}{lrlr|lrlr}
\hline\hline
\multicolumn{4}{c|}{\bf Benchmark Point P1 (red)} & \multicolumn{4}{c}{\bf Benchmark Point P2 (brown)} \\
\hline
\multicolumn{8}{l}{\textbf{Parameters}} \\
\hline
$\lambda$ & $0.635$ & $\kappa$ & $0.236$ & $\lambda$ & $0.621$ & $\kappa$ & $0.336$ \\
$\delta$ & $0.039$ & $\tan\beta$ & $1.893$ & $\delta$ & $0.040$ & $\tan\beta$ & $1.903$ \\
$\mu_{\rm tot}$ & $322.5~\mathrm{GeV}$ & $\mu$ & $2.6~\mathrm{GeV}$ & $\mu_{\rm tot}$ & $330.5~\mathrm{GeV}$ & $\mu$ & $-42.7~\mathrm{GeV}$ \\
$\mu_{\rm eff}$ & $319.9~\mathrm{GeV}$ & $B_\mu$ & $-44432.2~\mathrm{GeV}$ & $\mu_{\rm eff}$ & $373.2~\mathrm{GeV}$ & $B_\mu$ & $-59551.4~\mathrm{GeV}$ \\
$A_\lambda$ & $512.4~\mathrm{GeV}$ & $A_\kappa$ & $-437.4~\mathrm{GeV}$ & $A_\lambda$ & $366.3~\mathrm{GeV}$ & $A_\kappa$ & $-775.5~\mathrm{GeV}$ \\
$A_t$ & $1069.1~\mathrm{GeV}$ & $m_N$ & $237.9~\mathrm{GeV}$ & $A_t$ & $1060.9~\mathrm{GeV}$ & $m_N$ & $404.3~\mathrm{GeV}$ \\
$m_A$ & $617.7~\mathrm{GeV}$ & $m_B$ & $111.9~\mathrm{GeV}$ & $m_A$ & $608.7~\mathrm{GeV}$ & $m_B$ & $112.9~\mathrm{GeV}$ \\
\multicolumn{8}{l}{\textbf{Spectrum}} \\
\hline
$m_{\tilde\chi^0_1}$ & $241.3~\mathrm{GeV}$ & $m_{h_s}$ & $95.0~\mathrm{GeV}$ & $m_{\tilde\chi^0_1}$ & $315.3~\mathrm{GeV}$ & $m_{h_s}$ & $93.6~\mathrm{GeV}$ \\
$m_{\tilde\chi^0_2}$ & $330.6~\mathrm{GeV}$ & $m_h$ & $127.8~\mathrm{GeV}$ & $m_{\tilde\chi^0_2}$ & $-350.3~\mathrm{GeV}$ & $m_h$ & $126.7~\mathrm{GeV}$ \\
$m_{\tilde\chi^0_3}$ & $-346.5~\mathrm{GeV}$ & $m_H$ & $632.5~\mathrm{GeV}$ & $m_{\tilde\chi^0_3}$ & $422.8~\mathrm{GeV}$ & $m_H$ & $627.0~\mathrm{GeV}$ \\
$m_{\tilde\chi^0_4}$ & $1006.7~\mathrm{GeV}$ & $m_{A_s}$ & $407.3~\mathrm{GeV}$ & $m_{\tilde\chi^0_4}$ & $1006.7~\mathrm{GeV}$ & $m_{A_s}$ & $688.5~\mathrm{GeV}$ \\
$m_{\tilde\chi^0_5}$ & $1040.1~\mathrm{GeV}$ & $m_{A_H}$ & $629.1~\mathrm{GeV}$ & $m_{\tilde\chi^0_5}$ & $1040.2~\mathrm{GeV}$ & $m_{A_H}$ & $619.3~\mathrm{GeV}$ \\
$m_{\tilde\chi^\pm_1}$ & $320.5~\mathrm{GeV}$ & $m_{H^\pm}$ & $621.7~\mathrm{GeV}$ & $m_{\tilde\chi^\pm_1}$ & $328.9~\mathrm{GeV}$ & $m_{H^\pm}$ & $616.0~\mathrm{GeV}$ \\
$m_{\tilde\chi^\pm_2}$ & $1039.6~\mathrm{GeV}$ & \multicolumn{2}{l|}{} & $m_{\tilde\chi^\pm_2}$ & $1039.7~\mathrm{GeV}$ & \multicolumn{2}{l}{} \\
\multicolumn{8}{l}{\textbf{Signals}} \\
\hline
\multicolumn{2}{l}{$\mu_{\gamma\gamma}$} & \multicolumn{2}{l|}{$0.217$} & \multicolumn{2}{l}{$\mu_{\gamma\gamma}$} & \multicolumn{2}{l}{$0.182$} \\
\multicolumn{2}{l}{$\mu_{b\bar b}$} & \multicolumn{2}{l|}{$0.014$} & \multicolumn{2}{l}{$\mu_{b\bar b}$} & \multicolumn{2}{l}{$0.014$} \\
\multicolumn{2}{l}{$\sigma(gg\to H\to h h_s\to \gamma\gamma b\bar b)$} & \multicolumn{2}{l|}{$0.045~\mathrm{fb}$} & \multicolumn{2}{l}{$\sigma(gg\to H\to h h_s\to \gamma\gamma b\bar b)$} & \multicolumn{2}{l}{$0.059~\mathrm{fb}$} \\
\multicolumn{2}{l}{$\sigma(gg\to H\to h h_s\to b\bar b\tau\bar{\tau})$} & \multicolumn{2}{l|}{$1.258~\mathrm{fb}$} & \multicolumn{2}{l}{$\sigma(gg\to H\to h h_s\to b\bar b\tau\bar{\tau})$} & \multicolumn{2}{l}{$1.699~\mathrm{fb}$} \\
\multicolumn{2}{l}{$\sigma(gg\to H\to h h_s\to b\bar b b\bar b)$} & \multicolumn{2}{l|}{$10.805~\mathrm{fb}$} & \multicolumn{2}{l}{$\sigma(gg\to H\to h h_s\to b\bar b b\bar b)$} & \multicolumn{2}{l}{$14.612~\mathrm{fb}$} \\
\multicolumn{2}{l}{$\sigma(gg\to H\to h_s h_s\to b\bar b\tau\bar{\tau})$} & \multicolumn{2}{l|}{$0.518~\mathrm{fb}$} & \multicolumn{2}{l}{$\sigma(gg\to H\to h_s h_s\to b\bar b\tau\bar{\tau})$} & \multicolumn{2}{l}{$0.531~\mathrm{fb}$} \\
\multicolumn{2}{l}{$\sigma(gg\to H\to h_s h_s\to b\bar b b\bar b)$} & \multicolumn{2}{l|}{$4.675~\mathrm{fb}$} & \multicolumn{2}{l}{$\sigma(gg\to H\to h_s h_s\to b\bar b b\bar b)$} & \multicolumn{2}{l}{$4.813~\mathrm{fb}$} \\
\multicolumn{2}{l}{$\sigma(gg\to A_2\to h A_1\to b\bar b b\bar b)$} & \multicolumn{2}{l|}{$0.027~\mathrm{fb}$} & \multicolumn{2}{l}{$\sigma(gg\to A_2\to h A_1\to b\bar b b\bar b)$} & \multicolumn{2}{l}{$\text{\ttfamily -----}$} \\
\multicolumn{2}{l}{$\sigma(gg\to A_H\to Zh\to Zb\bar b)$} & \multicolumn{2}{l|}{$0.881~\mathrm{fb}$} & \multicolumn{2}{l}{$\sigma(gg\to A_H\to Zh\to Zb\bar b)$} & \multicolumn{2}{l}{$1.166~\mathrm{fb}$} \\
\multicolumn{2}{l}{$\sigma(gg\to A_H\to Zh_s\to \ell\ell b\bar b)$} & \multicolumn{2}{l|}{$2.923~\mathrm{fb}$} & \multicolumn{2}{l}{$\sigma(gg\to A_H\to Zh_s\to \ell\ell b\bar b)$} & \multicolumn{2}{l}{$3.915~\mathrm{fb}$} \\
\multicolumn{8}{l}{\textbf{Couplings}} \\
\hline
\multicolumn{2}{l}{$C_{h_sb\bar b},C_{h_st\bar t},C_{h_sVV},C_{h_s\gamma\gamma},C_{h_sgg}$} & \multicolumn{2}{l|}{$0.084,0.200,0.138,0.202,0.221$} & \multicolumn{2}{l}{$C_{h_sb\bar b},C_{h_st\bar t},C_{h_sVV},C_{h_s\gamma\gamma},C_{h_sgg}$} & \multicolumn{2}{l}{$0.092,0.199,0.136,0.196,0.221$} \\
\multicolumn{2}{l}{$C_{hb\bar b},C_{ht\bar t},C_{hVV},C_{h\gamma\gamma},C_{hgg}$} & \multicolumn{2}{l|}{$1.031,0.979,0.990,0.994,0.980$} & \multicolumn{2}{l}{$C_{hb\bar b},C_{ht\bar t},C_{hVV},C_{h\gamma\gamma},C_{hgg}$} & \multicolumn{2}{l}{$1.033,0.979,0.991,0.994,0.981$} \\
\multicolumn{2}{l}{$C_{Hb\bar b},C_{Ht\bar t},C_{HVV},C_{H\gamma\gamma},C_{Hgg}$} & \multicolumn{2}{l|}{$1.874,0.529,0.005,3.666,0.459$} & \multicolumn{2}{l}{$C_{Hb\bar b},C_{Ht\bar t},C_{HVV},C_{H\gamma\gamma},C_{Hgg}$} & \multicolumn{2}{l}{$1.883,0.527,0.006,3.558,0.458$} \\
\multicolumn{2}{l}{$C_{A_sb\bar b},C_{A_st\bar t},C_{A_s\gamma\gamma},C_{A_sgg}$} & \multicolumn{2}{l|}{$0.261,0.073,0.150,0.110$} & \multicolumn{2}{l}{$C_{A_sb\bar b},C_{A_st\bar t},C_{A_s\gamma\gamma},C_{A_sgg}$} & \multicolumn{2}{l}{$0.041,0.011,2.974,0.012$} \\
\multicolumn{2}{l}{$C_{A_Hb\bar b},C_{A_Ht\bar t},C_{A_H\gamma\gamma},C_{A_Hgg}$} & \multicolumn{2}{l|}{$1.875,0.523,4.236,0.568$} & \multicolumn{2}{l}{$C_{A_Hb\bar b},C_{A_Ht\bar t},C_{A_H\gamma\gamma},C_{A_Hgg}$} & \multicolumn{2}{l}{$1.903,0.525,3.835,0.574$} \\
\multicolumn{8}{l}{\textbf{Mixing}} \\
\hline
\multicolumn{2}{l}{$N_{11},N_{12},N_{13},N_{14},N_{15}$} & \multicolumn{2}{l|}{$\phantom{-}0.025,\;-0.041,\;\phantom{-}0.157,\;-0.386,\;\phantom{-}0.908$} & \multicolumn{2}{l}{$N_{11},N_{12},N_{13},N_{14},N_{15}$} & \multicolumn{2}{l}{$\phantom{-}0.059,\;-0.101,\;\phantom{-}0.638,\;-0.698,\;\phantom{-}0.305$} \\
\multicolumn{2}{l}{$N_{21},N_{22},N_{23},N_{24},N_{25}$} & \multicolumn{2}{l|}{$-0.056,\;\phantom{-}0.097,\;-0.693,\;\phantom{-}0.602,\;\phantom{-}0.381$} & \multicolumn{2}{l}{$N_{21},N_{22},N_{23},N_{24},N_{25}$} & \multicolumn{2}{l}{$-0.009,\;\phantom{-}0.017,\;\phantom{-}0.703,\;\phantom{-}0.698,\;\phantom{-}0.133$} \\
\multicolumn{2}{l}{$N_{31},N_{32},N_{33},N_{34},N_{35}$} & \multicolumn{2}{l|}{$-0.009,\;\phantom{-}0.016,\;\phantom{-}0.700,\;\phantom{-}0.692,\;\phantom{-}0.174$} & \multicolumn{2}{l}{$N_{31},N_{32},N_{33},N_{34},N_{35}$} & \multicolumn{2}{l}{$\phantom{-}0.018,\;-0.033,\;\phantom{-}0.305,\;-0.127,\;-0.943$} \\
\multicolumn{2}{l}{$N_{41},N_{42},N_{43},N_{44},N_{45}$} & \multicolumn{2}{l|}{$\phantom{-}0.972,\;\phantom{-}0.231,\;-0.021,\;\phantom{-}0.029,\;\phantom{-}0.000$} & \multicolumn{2}{l}{$N_{41},N_{42},N_{43},N_{44},N_{45}$} & \multicolumn{2}{l}{$-0.971,\;-0.235,\;\phantom{-}0.021,\;-0.029,\;\phantom{-}0.000$} \\
\multicolumn{2}{l}{$N_{51},N_{52},N_{53},N_{54},N_{55}$} & \multicolumn{2}{l|}{$\phantom{-}0.226,\;-0.967,\;-0.069,\;\phantom{-}0.095,\;\phantom{-}0.002$} & \multicolumn{2}{l}{$N_{51},N_{52},N_{53},N_{54},N_{55}$} & \multicolumn{2}{l}{$-0.230,\;\phantom{-}0.966,\;\phantom{-}0.070,\;-0.096,\;-0.003$} \\
\multicolumn{8}{l}{\textbf{Decays}} \\
\hline
\multicolumn{2}{l}{$h_s\to b\bar b/gg/c\bar c/\tau^+\tau^-/\gamma\gamma/W^+W^-$} & \multicolumn{2}{l|}{$58.8/17.9/15.7/6.51/0.636/0.372$} & \multicolumn{2}{l}{$h_s\to b\bar b/gg/c\bar c/\tau^+\tau^-/\gamma\gamma/W^+W^-$} & \multicolumn{2}{l}{$62.8/15.5/14.0/6.93/0.519/0.202$} \\
\multicolumn{2}{l}{$h\to b\bar b/W^+W^-/\tau^+\tau^-/gg/c\bar c/\gamma\gamma$} & \multicolumn{2}{l|}{$56.8/26.6/6.62/4.22/2.42/0.238$} & \multicolumn{2}{l}{$h\to b\bar b/W^+W^-/\tau^+\tau^-/gg/c\bar c/\gamma\gamma$} & \multicolumn{2}{l}{$58.9/24.4/6.85/4.29/2.50/0.239$} \\
\multicolumn{2}{l}{$H\to t\bar t/hh_s/h_sh_s/\tilde\chi^0_1\tilde\chi^0_1/\tilde\chi^0_1\tilde\chi^0_2/\tilde\chi^0_1\tilde\chi^0_3$} & \multicolumn{2}{l|}{$75.3/9.99/4.18/3.67/1.68/1.40$} & \multicolumn{2}{l}{$H\to t\bar t/hh_s/h_sh_s/b\bar b/hh/gg$} & \multicolumn{2}{l}{$83.7/11.8/3.65/0.437/0.231/0.095$} \\
\multicolumn{2}{l}{$A_s\to t\bar t/Zh_s/b\bar b/gg/Zh/\tau^+\tau^-$} & \multicolumn{2}{l|}{$94.1/4.96/0.481/0.218/0.138/0.067$} & \multicolumn{2}{l}{$A_s\to \tilde\chi^+_1\tilde\chi^-_1/\tilde\chi^0_1\tilde\chi^0_1/t\bar t/\tilde\chi^0_1\tilde\chi^0_2/Zh_s/\gamma\gamma$} & \multicolumn{2}{l}{$59.5/40.2/0.168/0.031/0.027/0.003$} \\
\multicolumn{2}{l}{$A_H\to t\bar t/Zh_s/\tilde\chi^0_1\tilde\chi^0_1/\tilde\chi^0_1\tilde\chi^0_3/\tilde\chi^0_1\tilde\chi^0_2/Zh$} & \multicolumn{2}{l|}{$78.4/9.79/5.78/1.23/0.547/0.305$} & \multicolumn{2}{l}{$A_H\to t\bar t/Zh_s/Zh/b\bar b/gg/\tau^+\tau^-$} & \multicolumn{2}{l}{$88.0/11.2/0.354/0.347/0.116/0.051$} \\
\multicolumn{2}{l}{$H^+\to t\bar b/h_sW^+/\tilde\chi^0_1\tilde\chi^\pm_1/A_sW^+/hW^+/\tau\nu_\tau$} & \multicolumn{2}{l|}{$80.5/11.4/5.18/2.60/0.354/0.054$} & \multicolumn{2}{l}{$H^+\to t\bar b/h_sW^+/hW^+/\tau\nu_\tau$} & \multicolumn{2}{l}{$86.9/12.7/0.402/0.059$} \\
\multicolumn{2}{l}{$\tilde\chi^0_2\to \tilde\chi^0_1Z^*$} & \multicolumn{2}{l|}{$100$} & \multicolumn{2}{l}{$\tilde\chi^0_2\to \tilde\chi^0_1Z^*/\tilde\chi^\pm_1W^{\mp *}$} & \multicolumn{2}{l}{$81.3/18.7$} \\
\multicolumn{2}{l}{$\tilde\chi^0_3\to \tilde\chi^0_1Z/\tilde\chi^0_1h_s$} & \multicolumn{2}{l|}{$97.9/2.00$} & \multicolumn{2}{l}{$\tilde\chi^0_3\to \tilde\chi^\pm_1W^\mp/\tilde\chi^0_1Z/\tilde\chi^0_1h_s$} & \multicolumn{2}{l}{$66.9/20.9/12.1$} \\
\multicolumn{2}{l}{$\tilde\chi^0_4\to \tilde\chi^\pm_1W^\mp/\tilde\chi^0_3Z/\tilde\chi^0_2h/\tilde\chi^\pm_1H^\mp$} & \multicolumn{2}{l|}{$80.4/6.40/5.80/3.90$} & \multicolumn{2}{l}{$\tilde\chi^0_4\to \tilde\chi^\pm_1W^\mp/\tilde\chi^0_2Z/\tilde\chi^0_1h/\tilde\chi^\pm_1H^\mp$} & \multicolumn{2}{l}{$81.0/6.40/6.10/3.90$} \\
\multicolumn{2}{l}{$\tilde\chi^0_5\to \tilde\chi^\pm_1W^\mp/\tilde\chi^0_3Z/\tilde\chi^0_2h/\tilde\chi^0_1h$} & \multicolumn{2}{l|}{$35.0/27.3/24.1/4.60$} & \multicolumn{2}{l}{$\tilde\chi^0_5\to \tilde\chi^\pm_1W^\mp/\tilde\chi^0_2Z/\tilde\chi^0_1h/\tilde\chi^0_3h$} & \multicolumn{2}{l}{$34.9/28.1/26.2/2.60$} \\
\multicolumn{2}{l}{$\tilde\chi^\pm_1\to \tilde\chi^0_1W^*$} & \multicolumn{2}{l|}{$100$} & \multicolumn{2}{l}{$\tilde\chi^\pm_1\to \tilde\chi^0_1W^*$} & \multicolumn{2}{l}{$100$} \\
\multicolumn{2}{l}{$\tilde\chi^\pm_2\to \tilde\chi^\pm_1Z/\tilde\chi^\pm_1h/\tilde\chi^0_3W^\mp/\tilde\chi^0_2W^\mp$} & \multicolumn{2}{l|}{$23.9/23.5/21.7/19.1$} & \multicolumn{2}{l}{$\tilde\chi^\pm_2\to \tilde\chi^\pm_1Z/\tilde\chi^\pm_1h/\tilde\chi^0_2W^\mp/\tilde\chi^0_1W^\mp$} & \multicolumn{2}{l}{$24.0/23.6/22.3/21.5$} \\
\multicolumn{8}{l}{\textbf{Tests}} \\
\hline
\multicolumn{2}{l}{$S,T,U$} & \multicolumn{2}{l|}{$0.048,-0.013,0.000$} & \multicolumn{2}{l}{$S,T,U$} & \multicolumn{2}{l}{$0.046,-0.015,0.000$} \\
\multicolumn{2}{l}{$\chi^2_{\rm HiggsSignals}$} & \multicolumn{2}{l|}{$158.119$} & \multicolumn{2}{l}{$\chi^2_{\rm HiggsSignals}$} & \multicolumn{2}{l}{$158.878$} \\
\multicolumn{2}{l}{$V_{\rm tree-level}$} & \multicolumn{2}{l|}{long-lived} & \multicolumn{2}{l}{$V_{\rm tree-level}$} & \multicolumn{2}{l}{long-lived} \\
\multicolumn{2}{l}{$V_{\rm 1-loop}$} & \multicolumn{2}{l|}{long-lived} & \multicolumn{2}{l}{$V_{\rm 1-loop}$} & \multicolumn{2}{l}{long-lived} \\
\multicolumn{2}{l}{$R_{\rm SModelS}$} & \multicolumn{2}{l|}{$0.01840$} & \multicolumn{2}{l}{$R_{\rm SModelS}$} & \multicolumn{2}{l}{$0.00871$} \\
\multicolumn{2}{l}{Signal Region} & \multicolumn{2}{l|}{(combined) in CMS-SUS-19-006-agg~\cite{CMS:SUS19006}} & \multicolumn{2}{l}{Signal Region} & \multicolumn{2}{l}{\texttt{4j\_Meff\_3000} in ATLAS-SUSY-2016-07~\cite{ATLAS:SUSY201607}} \\
\multicolumn{8}{l}{\textbf{Gravitino}} \\
\hline
\multicolumn{2}{l}{$m_{3/2}\,[\mathrm{MeV}]$} & \multicolumn{2}{l|}{$0.02723$} & \multicolumn{2}{l}{$m_{3/2}\,[\mathrm{MeV}]$} & \multicolumn{2}{l}{$\text{\ttfamily --}$} \\
\multicolumn{2}{l}{$\Gamma(\tilde\chi^0_1\to\gamma\tilde G)\,[\mathrm{GeV}]$} & \multicolumn{2}{l|}{$4.234\times 10^{-24}$} & \multicolumn{2}{l}{$\Gamma(\tilde\chi^0_1\to\gamma\tilde G)\,[\mathrm{GeV}]$} & \multicolumn{2}{l}{$\text{\ttfamily --}$} \\
\multicolumn{2}{l}{$\Gamma(\tilde\chi^0_1\to Z\tilde G)\,[\mathrm{GeV}]$} & \multicolumn{2}{l|}{$1.192\times 10^{-19}$} & \multicolumn{2}{l}{$\Gamma(\tilde\chi^0_1\to Z\tilde G)\,[\mathrm{GeV}]$} & \multicolumn{2}{l}{$\text{\ttfamily --}$} \\
\multicolumn{2}{l}{$\Gamma(\tilde\chi^0_1\to h\tilde G)\,[\mathrm{GeV}]$} & \multicolumn{2}{l|}{$1.136\times 10^{-20}$} & \multicolumn{2}{l}{$\Gamma(\tilde\chi^0_1\to h\tilde G)\,[\mathrm{GeV}]$} & \multicolumn{2}{l}{$\text{\ttfamily --}$} \\
\hline\hline
\end{tabular}}
\end{table}

\clearpage

\begin{table}[!p]
\centering
\caption{\label{tab:benchmark-p3-p4} Details of benchmark points P3 and P4 in \textbf{Scenario II} with $\mu<0$. Input dimensional parameters and cross sections are given in GeV and fb, respectively; branching ratios are in percent. \BPCascadeNote}
\BPTableSkip
\resizebox{1\textwidth}{!}{%
\begin{tabular}{lrlr|lrlr}
\hline\hline
\multicolumn{4}{c|}{\bf Benchmark Point P3 (purple)} & \multicolumn{4}{c}{\bf Benchmark Point P4 (black)} \\
\hline
\multicolumn{8}{l}{\textbf{Parameters}} \\
\hline
$\lambda$ & $0.381$ & $\kappa$ & $-0.210$ & $\lambda$ & $0.509$ & $\kappa$ & $0.261$ \\
$\delta$ & $0.029$ & $\tan\beta$ & $1.953$ & $\delta$ & $0.043$ & $\tan\beta$ & $1.847$ \\
$\mu_{\rm tot}$ & $417.0~\mathrm{GeV}$ & $\mu$ & $-428.6~\mathrm{GeV}$ & $\mu_{\rm tot}$ & $371.0~\mathrm{GeV}$ & $\mu$ & $-782.6~\mathrm{GeV}$ \\
$\mu_{\rm eff}$ & $845.6~\mathrm{GeV}$ & $B_\mu$ & $-1125778.0~\mathrm{GeV}$ & $\mu_{\rm eff}$ & $1153.6~\mathrm{GeV}$ & $B_\mu$ & $-157180.6~\mathrm{GeV}$ \\
$A_\lambda$ & $1928.2~\mathrm{GeV}$ & $A_\kappa$ & $1848.8~\mathrm{GeV}$ & $A_\lambda$ & $-333.6~\mathrm{GeV}$ & $A_\kappa$ & $-2349.7~\mathrm{GeV}$ \\
$A_t$ & $2916.3~\mathrm{GeV}$ & $m_N$ & $-929.8~\mathrm{GeV}$ & $A_t$ & $2608.3~\mathrm{GeV}$ & $m_N$ & $1181.3~\mathrm{GeV}$ \\
$m_A$ & $524.6~\mathrm{GeV}$ & $m_B$ & $106.4~\mathrm{GeV}$ & $m_A$ & $576.9~\mathrm{GeV}$ & $m_B$ & $109.6~\mathrm{GeV}$ \\
\multicolumn{8}{l}{\textbf{Spectrum}} \\
\hline
$m_{\tilde\chi^0_1}$ & $413.8~\mathrm{GeV}$ & $m_{h_s}$ & $97.8~\mathrm{GeV}$ & $m_{\tilde\chi^0_1}$ & $366.3~\mathrm{GeV}$ & $m_{h_s}$ & $98.0~\mathrm{GeV}$ \\
$m_{\tilde\chi^0_2}$ & $-417.6~\mathrm{GeV}$ & $m_h$ & $124.0~\mathrm{GeV}$ & $m_{\tilde\chi^0_2}$ & $-382.9~\mathrm{GeV}$ & $m_h$ & $127.9~\mathrm{GeV}$ \\
$m_{\tilde\chi^0_3}$ & $-931.5~\mathrm{GeV}$ & $m_H$ & $629.0~\mathrm{GeV}$ & $m_{\tilde\chi^0_3}$ & $1006.7~\mathrm{GeV}$ & $m_H$ & $640.2~\mathrm{GeV}$ \\
$m_{\tilde\chi^0_4}$ & $1006.6~\mathrm{GeV}$ & $m_{A_s}$ & $1619.6~\mathrm{GeV}$ & $m_{\tilde\chi^0_4}$ & $1040.5~\mathrm{GeV}$ & $m_{A_s}$ & $2043.6~\mathrm{GeV}$ \\
$m_{\tilde\chi^0_5}$ & $1040.8~\mathrm{GeV}$ & $m_{A_H}$ & $574.9~\mathrm{GeV}$ & $m_{\tilde\chi^0_5}$ & $1174.6~\mathrm{GeV}$ & $m_{A_H}$ & $620.3~\mathrm{GeV}$ \\
$m_{\tilde\chi^\pm_1}$ & $416.9~\mathrm{GeV}$ & $m_{H^\pm}$ & $591.6~\mathrm{GeV}$ & $m_{\tilde\chi^\pm_1}$ & $370.4~\mathrm{GeV}$ & $m_{H^\pm}$ & $621.2~\mathrm{GeV}$ \\
$m_{\tilde\chi^\pm_2}$ & $1040.1~\mathrm{GeV}$ & \multicolumn{2}{l|}{} & $m_{\tilde\chi^\pm_2}$ & $1039.9~\mathrm{GeV}$ & \multicolumn{2}{l}{} \\
\multicolumn{8}{l}{\textbf{Signals}} \\
\hline
\multicolumn{2}{l}{$\mu_{\gamma\gamma}$} & \multicolumn{2}{l|}{$0.080$} & \multicolumn{2}{l}{$\mu_{\gamma\gamma}$} & \multicolumn{2}{l}{$0.055$} \\
\multicolumn{2}{l}{$\mu_{b\bar b}$} & \multicolumn{2}{l|}{$0.197$} & \multicolumn{2}{l}{$\mu_{b\bar b}$} & \multicolumn{2}{l}{$0.166$} \\
\multicolumn{2}{l}{$\sigma(gg\to H\to h h_s\to \gamma\gamma b\bar b)$} & \multicolumn{2}{l|}{$0.060~\mathrm{fb}$} & \multicolumn{2}{l}{$\sigma(gg\to H\to h h_s\to \gamma\gamma b\bar b)$} & \multicolumn{2}{l}{$0.048~\mathrm{fb}$} \\
\multicolumn{2}{l}{$\sigma(gg\to H\to h h_s\to b\bar b\tau\bar{\tau})$} & \multicolumn{2}{l|}{$1.429~\mathrm{fb}$} & \multicolumn{2}{l}{$\sigma(gg\to H\to h h_s\to b\bar b\tau\bar{\tau})$} & \multicolumn{2}{l}{$1.049~\mathrm{fb}$} \\
\multicolumn{2}{l}{$\sigma(gg\to H\to h h_s\to b\bar b b\bar b)$} & \multicolumn{2}{l|}{$12.334~\mathrm{fb}$} & \multicolumn{2}{l}{$\sigma(gg\to H\to h h_s\to b\bar b b\bar b)$} & \multicolumn{2}{l}{$9.010~\mathrm{fb}$} \\
\multicolumn{2}{l}{$\sigma(gg\to H\to h_s h_s\to b\bar b\tau\bar{\tau})$} & \multicolumn{2}{l|}{$0.300~\mathrm{fb}$} & \multicolumn{2}{l}{$\sigma(gg\to H\to h_s h_s\to b\bar b\tau\bar{\tau})$} & \multicolumn{2}{l}{$0.377~\mathrm{fb}$} \\
\multicolumn{2}{l}{$\sigma(gg\to H\to h_s h_s\to b\bar b b\bar b)$} & \multicolumn{2}{l|}{$2.694~\mathrm{fb}$} & \multicolumn{2}{l}{$\sigma(gg\to H\to h_s h_s\to b\bar b b\bar b)$} & \multicolumn{2}{l}{$3.385~\mathrm{fb}$} \\
\multicolumn{2}{l}{$\sigma(gg\to A_2\to h A_1\to b\bar b b\bar b)$} & \multicolumn{2}{l|}{$\text{\ttfamily -----}$} & \multicolumn{2}{l}{$\sigma(gg\to A_2\to h A_1\to b\bar b b\bar b)$} & \multicolumn{2}{l}{$\text{\ttfamily -----}$} \\
\multicolumn{2}{l}{$\sigma(gg\to A_H\to Zh\to Zb\bar b)$} & \multicolumn{2}{l|}{$3.031~\mathrm{fb}$} & \multicolumn{2}{l}{$\sigma(gg\to A_H\to Zh\to Zb\bar b)$} & \multicolumn{2}{l}{$1.602~\mathrm{fb}$} \\
\multicolumn{2}{l}{$\sigma(gg\to A_H\to Zh_s\to \ell\ell b\bar b)$} & \multicolumn{2}{l|}{$3.412~\mathrm{fb}$} & \multicolumn{2}{l}{$\sigma(gg\to A_H\to Zh_s\to \ell\ell b\bar b)$} & \multicolumn{2}{l}{$2.966~\mathrm{fb}$} \\
\multicolumn{8}{l}{\textbf{Couplings}} \\
\hline
\multicolumn{2}{l}{$C_{h_sb\bar b},C_{h_st\bar t},C_{h_sVV},C_{h_s\gamma\gamma},C_{h_sgg}$} & \multicolumn{2}{l|}{$0.586,0.384,0.426,0.402,0.387$} & \multicolumn{2}{l}{$C_{h_sb\bar b},C_{h_st\bar t},C_{h_sVV},C_{h_s\gamma\gamma},C_{h_sgg}$} & \multicolumn{2}{l}{$0.550,0.344,0.390,0.351,0.345$} \\
\multicolumn{2}{l}{$C_{hb\bar b},C_{ht\bar t},C_{hVV},C_{h\gamma\gamma},C_{hgg}$} & \multicolumn{2}{l|}{$0.845,0.920,0.905,0.924,0.925$} & \multicolumn{2}{l}{$C_{hb\bar b},C_{ht\bar t},C_{hVV},C_{h\gamma\gamma},C_{hgg}$} & \multicolumn{2}{l}{$0.871,0.935,0.921,0.947,0.937$} \\
\multicolumn{2}{l}{$C_{Hb\bar b},C_{Ht\bar t},C_{HVV},C_{H\gamma\gamma},C_{Hgg}$} & \multicolumn{2}{l|}{$1.938,0.517,0.007,3.380,0.448$} & \multicolumn{2}{l}{$C_{Hb\bar b},C_{Ht\bar t},C_{HVV},C_{H\gamma\gamma},C_{Hgg}$} & \multicolumn{2}{l}{$1.830,0.549,0.009,3.658,0.475$} \\
\multicolumn{2}{l}{$C_{A_sb\bar b},C_{A_st\bar t},C_{A_s\gamma\gamma},C_{A_sgg}$} & \multicolumn{2}{l|}{$0.162,0.042,0.185,0.039$} & \multicolumn{2}{l}{$C_{A_sb\bar b},C_{A_st\bar t},C_{A_s\gamma\gamma},C_{A_sgg}$} & \multicolumn{2}{l}{$0.062,0.018,0.146,0.017$} \\
\multicolumn{2}{l}{$C_{A_Hb\bar b},C_{A_Ht\bar t},C_{A_H\gamma\gamma},C_{A_Hgg}$} & \multicolumn{2}{l|}{$1.946,0.510,2.269,0.577$} & \multicolumn{2}{l}{$C_{A_Hb\bar b},C_{A_Ht\bar t},C_{A_H\gamma\gamma},C_{A_Hgg}$} & \multicolumn{2}{l}{$1.846,0.541,3.933,0.591$} \\
\multicolumn{8}{l}{\textbf{Mixing}} \\
\hline
\multicolumn{2}{l}{$N_{11},N_{12},N_{13},N_{14},N_{15}$} & \multicolumn{2}{l|}{$-0.070,\;\phantom{-}0.120,\;-0.704,\;\phantom{-}0.696,\;\phantom{-}0.015$} & \multicolumn{2}{l}{$N_{11},N_{12},N_{13},N_{14},N_{15}$} & \multicolumn{2}{l}{$-0.066,\;\phantom{-}0.113,\;-0.701,\;\phantom{-}0.700,\;-0.031$} \\
\multicolumn{2}{l}{$N_{21},N_{22},N_{23},N_{24},N_{25}$} & \multicolumn{2}{l|}{$-0.010,\;\phantom{-}0.017,\;\phantom{-}0.699,\;\phantom{-}0.705,\;-0.122$} & \multicolumn{2}{l}{$N_{21},N_{22},N_{23},N_{24},N_{25}$} & \multicolumn{2}{l}{$\phantom{-}0.009,\;-0.016,\;-0.706,\;-0.706,\;-0.054$} \\
\multicolumn{2}{l}{$N_{31},N_{32},N_{33},N_{34},N_{35}$} & \multicolumn{2}{l|}{$\phantom{-}0.000,\;-0.001,\;-0.096,\;-0.076,\;-0.992$} & \multicolumn{2}{l}{$N_{31},N_{32},N_{33},N_{34},N_{35}$} & \multicolumn{2}{l}{$\phantom{-}0.970,\;\phantom{-}0.241,\;-0.023,\;\phantom{-}0.029,\;\phantom{-}0.000$} \\
\multicolumn{2}{l}{$N_{41},N_{42},N_{43},N_{44},N_{45}$} & \multicolumn{2}{l|}{$\phantom{-}0.969,\;\phantom{-}0.245,\;-0.024,\;\phantom{-}0.031,\;\phantom{-}0.000$} & \multicolumn{2}{l}{$N_{41},N_{42},N_{43},N_{44},N_{45}$} & \multicolumn{2}{l}{$\phantom{-}0.235,\;-0.964,\;-0.076,\;\phantom{-}0.101,\;-0.011$} \\
\multicolumn{2}{l}{$N_{51},N_{52},N_{53},N_{54},N_{55}$} & \multicolumn{2}{l|}{$\phantom{-}0.238,\;-0.962,\;-0.081,\;\phantom{-}0.107,\;\phantom{-}0.001$} & \multicolumn{2}{l}{$N_{51},N_{52},N_{53},N_{54},N_{55}$} & \multicolumn{2}{l}{$\phantom{-}0.000,\;-0.008,\;-0.060,\;-0.015,\;\phantom{-}0.998$} \\
\multicolumn{8}{l}{\textbf{Decays}} \\
\hline
\multicolumn{2}{l}{$h_s\to b\bar b/\tau^+\tau^-/c\bar c/gg/W^+W^-/\gamma\gamma$} & \multicolumn{2}{l|}{$86.5/9.62/1.76/1.76/0.234/0.082$} & \multicolumn{2}{l}{$h_s\to b\bar b/\tau^+\tau^-/c\bar c/gg/W^+W^-/\gamma\gamma$} & \multicolumn{2}{l}{$86.8/9.66/1.60/1.59/0.236/0.072$} \\
\multicolumn{2}{l}{$h\to b\bar b/W^+W^-/\tau^+\tau^-/gg/c\bar c/\gamma\gamma$} & \multicolumn{2}{l|}{$58.9/22.8/6.82/5.43/3.29/0.286$} & \multicolumn{2}{l}{$h\to b\bar b/W^+W^-/\tau^+\tau^-/gg/c\bar c/\gamma\gamma$} & \multicolumn{2}{l}{$52.4/29.8/6.11/4.99/2.85/0.280$} \\
\multicolumn{2}{l}{$H\to t\bar t/hh_s/hh/h_sh_s/b\bar b/gg$} & \multicolumn{2}{l|}{$87.8/7.66/2.61/1.14/0.503/0.099$} & \multicolumn{2}{l}{$H\to t\bar t/hh_s/hh/h_sh_s/b\bar b/gg$} & \multicolumn{2}{l}{$89.4/6.07/2.49/1.38/0.398/0.099$} \\
\multicolumn{2}{l}{$A_s\to \tilde\chi^+_1\tilde\chi^-_1/\tilde\chi^0_1\tilde\chi^0_1/\tilde\chi^0_2\tilde\chi^0_2/t\bar t/Zh_s/\tilde\chi^0_2\tilde\chi^0_3$} & \multicolumn{2}{l|}{$12.1/6.07/5.73/0.409/0.198/0.028$} & \multicolumn{2}{l}{$A_s\to \tilde\chi^+_1\tilde\chi^-_1/\tilde\chi^0_2\tilde\chi^0_2/\tilde\chi^0_1\tilde\chi^0_1/\tilde\chi^\pm_1\tilde\chi^\mp_2/\tilde\chi^0_1\tilde\chi^0_4/\tilde\chi^0_2\tilde\chi^0_5$} & \multicolumn{2}{l}{$31.9/16.2/15.8/0.899/0.531/0.176$} \\
\multicolumn{2}{l}{$A_H\to t\bar t/Zh_s/Zh/b\bar b/gg/\tau^+\tau^-$} & \multicolumn{2}{l|}{$93.8/4.93/0.643/0.420/0.134/0.061$} & \multicolumn{2}{l}{$A_H\to t\bar t/Zh_s/Zh/b\bar b/gg/\tau^+\tau^-$} & \multicolumn{2}{l}{$93.2/5.82/0.520/0.326/0.123/0.048$} \\
\multicolumn{2}{l}{$H^+\to t\bar b/h_sW^+/hW^+/\tau\nu_\tau$} & \multicolumn{2}{l|}{$93.1/6.08/0.796/0.071$} & \multicolumn{2}{l}{$H^+\to t\bar b/h_sW^+/hW^+/\tau\nu_\tau$} & \multicolumn{2}{l}{$92.6/6.74/0.603/0.056$} \\
\multicolumn{2}{l}{$\tilde\chi^0_2\to \tilde\chi^0_1Z^*$} & \multicolumn{2}{l|}{$99.9$} & \multicolumn{2}{l}{$\tilde\chi^0_2\to \tilde\chi^0_1Z^*/\tilde\chi^\pm_1W^{\mp *}$} & \multicolumn{2}{l}{$61.4/38.6$} \\
\multicolumn{2}{l}{$\tilde\chi^0_3\to \tilde\chi^\pm_1W^\mp/\tilde\chi^0_1Z/\tilde\chi^0_2h/\tilde\chi^0_2h_s$} & \multicolumn{2}{l|}{$49.3/24.8/19.8/4.80$} & \multicolumn{2}{l}{$\tilde\chi^0_3\to \tilde\chi^\pm_1W^\mp/\tilde\chi^0_2Z/\tilde\chi^0_1h/\tilde\chi^0_1h_s$} & \multicolumn{2}{l}{$84.0/6.60/4.70/1.90$} \\
\multicolumn{2}{l}{$\tilde\chi^0_4\to \tilde\chi^\pm_1W^\mp/\tilde\chi^0_2Z/\tilde\chi^0_1h/\tilde\chi^0_1h_s$} & \multicolumn{2}{l|}{$84.8/6.60/4.50/2.00$} & \multicolumn{2}{l}{$\tilde\chi^0_4\to \tilde\chi^\pm_1W^\mp/\tilde\chi^0_2Z/\tilde\chi^0_1h/\tilde\chi^0_1h_s$} & \multicolumn{2}{l}{$35.7/29.8/21.0/8.50$} \\
\multicolumn{2}{l}{$\tilde\chi^0_5\to \tilde\chi^\pm_1W^\mp/\tilde\chi^0_2Z/\tilde\chi^0_1h/\tilde\chi^0_1h_s$} & \multicolumn{2}{l|}{$35.2/30.7/20.8/8.90$} & \multicolumn{2}{l}{$\tilde\chi^0_5\to \tilde\chi^\pm_1H^\mp/\tilde\chi^\pm_1W^\mp/\tilde\chi^0_2A_H/\tilde\chi^0_1H$} & \multicolumn{2}{l}{$30.2/19.9/14.5/13.9$} \\
\multicolumn{2}{l}{$\tilde\chi^\pm_1\to \tilde\chi^0_1W^*$} & \multicolumn{2}{l|}{$100$} & \multicolumn{2}{l}{$\tilde\chi^\pm_1\to \tilde\chi^0_1W^*$} & \multicolumn{2}{l}{$100$} \\
\multicolumn{2}{l}{$\tilde\chi^\pm_2\to \tilde\chi^\pm_1Z/\tilde\chi^0_2W^\mp/\tilde\chi^0_1W^\mp/\tilde\chi^\pm_1h$} & \multicolumn{2}{l|}{$24.6/24.1/23.7/17.1$} & \multicolumn{2}{l}{$\tilde\chi^\pm_2\to \tilde\chi^\pm_1Z/\tilde\chi^0_1W^\mp/\tilde\chi^0_2W^\mp/\tilde\chi^\pm_1h$} & \multicolumn{2}{l}{$24.6/23.7/23.6/17.5$} \\
\multicolumn{8}{l}{\textbf{Tests}} \\
\hline
\multicolumn{2}{l}{$S,T,U$} & \multicolumn{2}{l|}{$0.014,-0.003,0.000$} & \multicolumn{2}{l}{$S,T,U$} & \multicolumn{2}{l}{$0.031,-0.011,0.000$} \\
\multicolumn{2}{l}{$\chi^2_{\rm HiggsSignals}$} & \multicolumn{2}{l|}{$154.799$} & \multicolumn{2}{l}{$\chi^2_{\rm HiggsSignals}$} & \multicolumn{2}{l}{$151.932$} \\
\multicolumn{2}{l}{$V_{\rm tree-level}$} & \multicolumn{2}{l|}{short-lived} & \multicolumn{2}{l}{$V_{\rm tree-level}$} & \multicolumn{2}{l}{short-lived} \\
\multicolumn{2}{l}{$V_{\rm 1-loop}$} & \multicolumn{2}{l|}{long-lived} & \multicolumn{2}{l}{$V_{\rm 1-loop}$} & \multicolumn{2}{l}{short-lived} \\
\multicolumn{2}{l}{$R_{\rm SModelS}$} & \multicolumn{2}{l|}{$0.43444$} & \multicolumn{2}{l}{$R_{\rm SModelS}$} & \multicolumn{2}{l}{$0.42046$} \\
\multicolumn{2}{l}{Signal Region} & \multicolumn{2}{l|}{\texttt{4j\_Meff\_3000} in ATLAS-SUSY-2016-07~\cite{ATLAS:SUSY201607}} & \multicolumn{2}{l}{Signal Region} & \multicolumn{2}{l}{\texttt{4j\_Meff\_3000} in ATLAS-SUSY-2016-07~\cite{ATLAS:SUSY201607}} \\
\multicolumn{8}{l}{\textbf{Gravitino}} \\
\hline
\multicolumn{2}{l}{$m_{3/2}\,[\mathrm{MeV}]$} & \multicolumn{2}{l|}{$\text{\ttfamily --}$} & \multicolumn{2}{l}{$m_{3/2}\,[\mathrm{MeV}]$} & \multicolumn{2}{l}{$\text{\ttfamily --}$} \\
\multicolumn{2}{l}{$\Gamma(\tilde\chi^0_1\to\gamma\tilde G)\,[\mathrm{GeV}]$} & \multicolumn{2}{l|}{$\text{\ttfamily --}$} & \multicolumn{2}{l}{$\Gamma(\tilde\chi^0_1\to\gamma\tilde G)\,[\mathrm{GeV}]$} & \multicolumn{2}{l}{$\text{\ttfamily --}$} \\
\multicolumn{2}{l}{$\Gamma(\tilde\chi^0_1\to Z\tilde G)\,[\mathrm{GeV}]$} & \multicolumn{2}{l|}{$\text{\ttfamily --}$} & \multicolumn{2}{l}{$\Gamma(\tilde\chi^0_1\to Z\tilde G)\,[\mathrm{GeV}]$} & \multicolumn{2}{l}{$\text{\ttfamily --}$} \\
\multicolumn{2}{l}{$\Gamma(\tilde\chi^0_1\to h\tilde G)\,[\mathrm{GeV}]$} & \multicolumn{2}{l|}{$\text{\ttfamily --}$} & \multicolumn{2}{l}{$\Gamma(\tilde\chi^0_1\to h\tilde G)\,[\mathrm{GeV}]$} & \multicolumn{2}{l}{$\text{\ttfamily --}$} \\
\hline\hline
\end{tabular}}
\end{table}

\clearpage

\begin{table}[!p]
\centering
\caption{\label{tab:benchmark-p5-p6} Details of benchmark points P5 and P6 in \textbf{Scenario II} with $\mu>0$. Input dimensional parameters and cross sections are given in GeV and fb, respectively; branching ratios are in percent. \BPCascadeNote}
\BPTableSkip
\resizebox{1\textwidth}{!}{%
\begin{tabular}{lrlr|lrlr}
\hline\hline
\multicolumn{4}{c|}{\bf Benchmark Point P5 (cyan)} & \multicolumn{4}{c}{\bf Benchmark Point P6 (green)} \\
\hline
\multicolumn{8}{l}{\textbf{Parameters}} \\
\hline
$\lambda$ & $0.566$ & $\kappa$ & $0.363$ & $\lambda$ & $0.415$ & $\kappa$ & $0.389$ \\
$\delta$ & $0.018$ & $\tan\beta$ & $1.543$ & $\delta$ & $0.011$ & $\tan\beta$ & $2.247$ \\
$\mu_{\rm tot}$ & $458.1~\mathrm{GeV}$ & $\mu$ & $100.6~\mathrm{GeV}$ & $\mu_{\rm tot}$ & $177.5~\mathrm{GeV}$ & $\mu$ & $2.7~\mathrm{GeV}$ \\
$\mu_{\rm eff}$ & $357.5~\mathrm{GeV}$ & $B_\mu$ & $-87877.5~\mathrm{GeV}$ & $\mu_{\rm eff}$ & $174.8~\mathrm{GeV}$ & $B_\mu$ & $85250.5~\mathrm{GeV}$ \\
$A_\lambda$ & $526.9~\mathrm{GeV}$ & $A_\kappa$ & $-894.7~\mathrm{GeV}$ & $A_\lambda$ & $145.3~\mathrm{GeV}$ & $A_\kappa$ & $-608.1~\mathrm{GeV}$ \\
$A_t$ & $1614.2~\mathrm{GeV}$ & $m_N$ & $458.5~\mathrm{GeV}$ & $A_t$ & $2294.7~\mathrm{GeV}$ & $m_N$ & $327.5~\mathrm{GeV}$ \\
$m_A$ & $632.2~\mathrm{GeV}$ & $m_B$ & $94.2~\mathrm{GeV}$ & $m_A$ & $612.3~\mathrm{GeV}$ & $m_B$ & $95.9~\mathrm{GeV}$ \\
\multicolumn{8}{l}{\textbf{Spectrum}} \\
\hline
$m_{\tilde\chi^0_1}$ & $436.6~\mathrm{GeV}$ & $m_{h_s}$ & $97.2~\mathrm{GeV}$ & $m_{\tilde\chi^0_1}$ & $169.8~\mathrm{GeV}$ & $m_{h_s}$ & $94.9~\mathrm{GeV}$ \\
$m_{\tilde\chi^0_2}$ & $-475.4~\mathrm{GeV}$ & $m_h$ & $124.9~\mathrm{GeV}$ & $m_{\tilde\chi^0_2}$ & $-191.0~\mathrm{GeV}$ & $m_h$ & $124.7~\mathrm{GeV}$ \\
$m_{\tilde\chi^0_3}$ & $478.7~\mathrm{GeV}$ & $m_H$ & $665.4~\mathrm{GeV}$ & $m_{\tilde\chi^0_3}$ & $335.6~\mathrm{GeV}$ & $m_H$ & $640.5~\mathrm{GeV}$ \\
$m_{\tilde\chi^0_4}$ & $1006.9~\mathrm{GeV}$ & $m_{A_s}$ & $787.8~\mathrm{GeV}$ & $m_{\tilde\chi^0_4}$ & $1006.4~\mathrm{GeV}$ & $m_{A_s}$ & $546.7~\mathrm{GeV}$ \\
$m_{\tilde\chi^0_5}$ & $1042.1~\mathrm{GeV}$ & $m_{A_H}$ & $660.9~\mathrm{GeV}$ & $m_{\tilde\chi^0_5}$ & $1038.9~\mathrm{GeV}$ & $m_{A_H}$ & $635.1~\mathrm{GeV}$ \\
$m_{\tilde\chi^\pm_1}$ & $455.9~\mathrm{GeV}$ & $m_{H^\pm}$ & $657.9~\mathrm{GeV}$ & $m_{\tilde\chi^\pm_1}$ & $176.5~\mathrm{GeV}$ & $m_{H^\pm}$ & $635.4~\mathrm{GeV}$ \\
$m_{\tilde\chi^\pm_2}$ & $1041.3~\mathrm{GeV}$ & \multicolumn{2}{l|}{} & $m_{\tilde\chi^\pm_2}$ & $1038.6~\mathrm{GeV}$ & \multicolumn{2}{l}{} \\
\multicolumn{8}{l}{\textbf{Signals}} \\
\hline
\multicolumn{2}{l}{$\mu_{\gamma\gamma}$} & \multicolumn{2}{l|}{$0.085$} & \multicolumn{2}{l}{$\mu_{\gamma\gamma}$} & \multicolumn{2}{l}{$0.069$} \\
\multicolumn{2}{l}{$\mu_{b\bar b}$} & \multicolumn{2}{l|}{$0.205$} & \multicolumn{2}{l}{$\mu_{b\bar b}$} & \multicolumn{2}{l}{$0.188$} \\
\multicolumn{2}{l}{$\sigma(gg\to H\to h h_s\to \gamma\gamma b\bar b)$} & \multicolumn{2}{l|}{$0.038~\mathrm{fb}$} & \multicolumn{2}{l}{$\sigma(gg\to H\to h h_s\to \gamma\gamma b\bar b)$} & \multicolumn{2}{l}{$0.011~\mathrm{fb}$} \\
\multicolumn{2}{l}{$\sigma(gg\to H\to h h_s\to b\bar b\tau\bar{\tau})$} & \multicolumn{2}{l|}{$0.896~\mathrm{fb}$} & \multicolumn{2}{l}{$\sigma(gg\to H\to h h_s\to b\bar b\tau\bar{\tau})$} & \multicolumn{2}{l}{$0.254~\mathrm{fb}$} \\
\multicolumn{2}{l}{$\sigma(gg\to H\to h h_s\to b\bar b b\bar b)$} & \multicolumn{2}{l|}{$7.727~\mathrm{fb}$} & \multicolumn{2}{l}{$\sigma(gg\to H\to h h_s\to b\bar b b\bar b)$} & \multicolumn{2}{l}{$2.190~\mathrm{fb}$} \\
\multicolumn{2}{l}{$\sigma(gg\to H\to h_s h_s\to b\bar b\tau\bar{\tau})$} & \multicolumn{2}{l|}{$0.090~\mathrm{fb}$} & \multicolumn{2}{l}{$\sigma(gg\to H\to h_s h_s\to b\bar b\tau\bar{\tau})$} & \multicolumn{2}{l}{$0.325~\mathrm{fb}$} \\
\multicolumn{2}{l}{$\sigma(gg\to H\to h_s h_s\to b\bar b b\bar b)$} & \multicolumn{2}{l|}{$0.812~\mathrm{fb}$} & \multicolumn{2}{l}{$\sigma(gg\to H\to h_s h_s\to b\bar b b\bar b)$} & \multicolumn{2}{l}{$2.934~\mathrm{fb}$} \\
\multicolumn{2}{l}{$\sigma(gg\to A_2\to h A_1\to b\bar b b\bar b)$} & \multicolumn{2}{l|}{$0.000~\mathrm{fb}$} & \multicolumn{2}{l}{$\sigma(gg\to A_2\to h A_1\to b\bar b b\bar b)$} & \multicolumn{2}{l}{$\text{\ttfamily -----}$} \\
\multicolumn{2}{l}{$\sigma(gg\to A_H\to Zh\to Zb\bar b)$} & \multicolumn{2}{l|}{$1.846~\mathrm{fb}$} & \multicolumn{2}{l}{$\sigma(gg\to A_H\to Zh\to Zb\bar b)$} & \multicolumn{2}{l}{$0.546~\mathrm{fb}$} \\
\multicolumn{2}{l}{$\sigma(gg\to A_H\to Zh_s\to \ell\ell b\bar b)$} & \multicolumn{2}{l|}{$2.307~\mathrm{fb}$} & \multicolumn{2}{l}{$\sigma(gg\to A_H\to Zh_s\to \ell\ell b\bar b)$} & \multicolumn{2}{l}{$1.222~\mathrm{fb}$} \\
\multicolumn{8}{l}{\textbf{Couplings}} \\
\hline
\multicolumn{2}{l}{$C_{h_sb\bar b},C_{h_st\bar t},C_{h_sVV},C_{h_s\gamma\gamma},C_{h_sgg}$} & \multicolumn{2}{l|}{$0.566,0.380,0.435,0.402,0.385$} & \multicolumn{2}{l}{$C_{h_sb\bar b},C_{h_st\bar t},C_{h_sVV},C_{h_s\gamma\gamma},C_{h_sgg}$} & \multicolumn{2}{l}{$0.553,0.391,0.418,0.345,0.398$} \\
\multicolumn{2}{l}{$C_{hb\bar b},C_{ht\bar t},C_{hVV},C_{h\gamma\gamma},C_{hgg}$} & \multicolumn{2}{l|}{$0.854,0.920,0.900,0.926,0.926$} & \multicolumn{2}{l}{$C_{hb\bar b},C_{ht\bar t},C_{hVV},C_{h\gamma\gamma},C_{hgg}$} & \multicolumn{2}{l}{$0.873,0.915,0.908,0.954,0.920$} \\
\multicolumn{2}{l}{$C_{Hb\bar b},C_{Ht\bar t},C_{HVV},C_{H\gamma\gamma},C_{Hgg}$} & \multicolumn{2}{l|}{$1.526,0.656,0.010,3.715,0.566$} & \multicolumn{2}{l}{$C_{Hb\bar b},C_{Ht\bar t},C_{HVV},C_{H\gamma\gamma},C_{Hgg}$} & \multicolumn{2}{l}{$2.232,0.455,0.011,3.314,0.394$} \\
\multicolumn{2}{l}{$C_{A_sb\bar b},C_{A_st\bar t},C_{A_s\gamma\gamma},C_{A_sgg}$} & \multicolumn{2}{l|}{$0.088,0.037,0.678,0.037$} & \multicolumn{2}{l}{$C_{A_sb\bar b},C_{A_st\bar t},C_{A_s\gamma\gamma},C_{A_sgg}$} & \multicolumn{2}{l}{$0.216,0.043,1.300,0.050$} \\
\multicolumn{2}{l}{$C_{A_Hb\bar b},C_{A_Ht\bar t},C_{A_H\gamma\gamma},C_{A_Hgg}$} & \multicolumn{2}{l|}{$1.540,0.647,4.316,0.689$} & \multicolumn{2}{l}{$C_{A_Hb\bar b},C_{A_Ht\bar t},C_{A_H\gamma\gamma},C_{A_Hgg}$} & \multicolumn{2}{l}{$2.237,0.443,3.595,0.480$} \\
\multicolumn{8}{l}{\textbf{Mixing}} \\
\hline
\multicolumn{2}{l}{$N_{11},N_{12},N_{13},N_{14},N_{15}$} & \multicolumn{2}{l|}{$\phantom{-}0.061,\;-0.104,\;\phantom{-}0.513,\;-0.599,\;\phantom{-}0.603$} & \multicolumn{2}{l}{$N_{11},N_{12},N_{13},N_{14},N_{15}$} & \multicolumn{2}{l}{$-0.049,\;\phantom{-}0.085,\;-0.681,\;\phantom{-}0.710,\;-0.153$} \\
\multicolumn{2}{l}{$N_{21},N_{22},N_{23},N_{24},N_{25}$} & \multicolumn{2}{l|}{$-0.006,\;\phantom{-}0.011,\;\phantom{-}0.704,\;\phantom{-}0.703,\;\phantom{-}0.101$} & \multicolumn{2}{l}{$N_{21},N_{22},N_{23},N_{24},N_{25}$} & \multicolumn{2}{l}{$\phantom{-}0.013,\;-0.023,\;-0.703,\;-0.699,\;-0.128$} \\
\multicolumn{2}{l}{$N_{31},N_{32},N_{33},N_{34},N_{35}$} & \multicolumn{2}{l|}{$-0.047,\;\phantom{-}0.082,\;-0.481,\;\phantom{-}0.366,\;\phantom{-}0.791$} & \multicolumn{2}{l}{$N_{31},N_{32},N_{33},N_{34},N_{35}$} & \multicolumn{2}{l}{$\phantom{-}0.006,\;-0.011,\;\phantom{-}0.198,\;-0.020,\;-0.980$} \\
\multicolumn{2}{l}{$N_{41},N_{42},N_{43},N_{44},N_{45}$} & \multicolumn{2}{l|}{$\phantom{-}0.965,\;\phantom{-}0.258,\;-0.027,\;\phantom{-}0.031,\;\phantom{-}0.000$} & \multicolumn{2}{l}{$N_{41},N_{42},N_{43},N_{44},N_{45}$} & \multicolumn{2}{l}{$-0.976,\;-0.214,\;\phantom{-}0.015,\;-0.027,\;\phantom{-}0.000$} \\
\multicolumn{2}{l}{$N_{51},N_{52},N_{53},N_{54},N_{55}$} & \multicolumn{2}{l|}{$\phantom{-}0.250,\;-0.957,\;-0.096,\;\phantom{-}0.113,\;\phantom{-}0.003$} & \multicolumn{2}{l}{$N_{51},N_{52},N_{53},N_{54},N_{55}$} & \multicolumn{2}{l}{$-0.211,\;\phantom{-}0.973,\;\phantom{-}0.049,\;-0.084,\;-0.001$} \\
\multicolumn{8}{l}{\textbf{Decays}} \\
\hline
\multicolumn{2}{l}{$h_s\to b\bar b/\tau^+\tau^-/c\bar c/gg/W^+W^-/\gamma\gamma$} & \multicolumn{2}{l|}{$86.3/9.60/1.84/1.83/0.226/0.087$} & \multicolumn{2}{l}{$h_s\to b\bar b/\tau^+\tau^-/c\bar c/gg/W^+W^-/\gamma\gamma$} & \multicolumn{2}{l}{$86.2/9.54/2.04/1.94/0.109/0.062$} \\
\multicolumn{2}{l}{$h\to b\bar b/W^+W^-/\tau^+\tau^-/gg/c\bar c/\gamma\gamma$} & \multicolumn{2}{l|}{$57.9/24.0/6.71/5.33/3.17/0.284$} & \multicolumn{2}{l}{$h\to b\bar b/W^+W^-/\tau^+\tau^-/gg/c\bar c/\gamma\gamma$} & \multicolumn{2}{l}{$59.0/23.2/6.84/5.10/3.06/0.292$} \\
\multicolumn{2}{l}{$H\to t\bar t/hh_s/hh/h_sh_s/b\bar b/gg$} & \multicolumn{2}{l|}{$92.9/4.03/2.35/0.284/0.194/0.099$} & \multicolumn{2}{l}{$H\to t\bar t/\tilde\chi^0_1\tilde\chi^0_3/\tilde\chi^0_2\tilde\chi^0_3/hh_s/h_sh_s/b\bar b$} & \multicolumn{2}{l}{$83.4/7.30/2.67/1.93/1.76/0.804$} \\
\multicolumn{2}{l}{$A_s\to t\bar t/Zh_s/Zh/b\bar b/gg/\gamma\gamma$} & \multicolumn{2}{l|}{$63.2/4.46/0.541/0.103/0.066/0.025$} & \multicolumn{2}{l}{$A_s\to \tilde\chi^+_1\tilde\chi^-_1/\tilde\chi^0_1\tilde\chi^0_1/\tilde\chi^0_2\tilde\chi^0_2/t\bar t/\tilde\chi^0_1\tilde\chi^0_3/\tilde\chi^0_2\tilde\chi^0_3$} & \multicolumn{2}{l}{$49.2/26.9/21.5/1.56/0.530/0.087$} \\
\multicolumn{2}{l}{$A_H\to t\bar t/Zh_s/Zh/b\bar b/gg/\tau^+\tau^-$} & \multicolumn{2}{l|}{$94.6/4.56/0.544/0.159/0.117/0.024$} & \multicolumn{2}{l}{$A_H\to t\bar t/\tilde\chi^0_2\tilde\chi^0_3/Zh_s/\tilde\chi^0_1\tilde\chi^0_3/\tilde\chi^0_1\tilde\chi^0_1/b\bar b$} & \multicolumn{2}{l}{$86.7/4.69/4.10/1.80/0.821/0.661$} \\
\multicolumn{2}{l}{$H^+\to t\bar b/h_sW^+/hW^+/\tau\nu_\tau$} & \multicolumn{2}{l|}{$94.2/5.19/0.619/0.027$} & \multicolumn{2}{l}{$H^+\to t\bar b/\tilde\chi^0_3\tilde\chi^\pm_1/h_sW^+/\tilde\chi^0_1\tilde\chi^\pm_1/hW^+/\tilde\chi^0_2\tilde\chi^\pm_1$} & \multicolumn{2}{l}{$85.6/8.49/4.67/0.696/0.306/0.134$} \\
\multicolumn{2}{l}{$\tilde\chi^0_2\to \tilde\chi^0_1Z^*/\tilde\chi^\pm_1W^{\mp *}$} & \multicolumn{2}{l|}{$89.4/10.6$} & \multicolumn{2}{l}{$\tilde\chi^0_2\to \tilde\chi^0_1Z^*/\tilde\chi^\pm_1W^{\mp *}$} & \multicolumn{2}{l}{$72.3/27.7$} \\
\multicolumn{2}{l}{$\tilde\chi^0_3\to \tilde\chi^\pm_1W^{\mp *}/\tilde\chi^0_1Z^*$} & \multicolumn{2}{l|}{$94.6/5.30$} & \multicolumn{2}{l}{$\tilde\chi^0_3\to \tilde\chi^\pm_1W^\mp/\tilde\chi^0_1Z/\tilde\chi^0_1h/\tilde\chi^0_2Z$} & \multicolumn{2}{l}{$61.1/17.6/11.0/8.20$} \\
\multicolumn{2}{l}{$\tilde\chi^0_4\to \tilde\chi^\pm_1W^\mp/\tilde\chi^0_2Z/\tilde\chi^0_1h/\tilde\chi^0_1h_s$} & \multicolumn{2}{l|}{$87.2/5.70/2.30/1.40$} & \multicolumn{2}{l}{$\tilde\chi^0_4\to \tilde\chi^\pm_1W^\mp/\tilde\chi^\pm_1H^\mp/\tilde\chi^0_2Z/\tilde\chi^0_1h$} & \multicolumn{2}{l}{$70.5/12.0/6.50/4.80$} \\
\multicolumn{2}{l}{$\tilde\chi^0_5\to \tilde\chi^\pm_1W^\mp/\tilde\chi^0_2Z/\tilde\chi^0_1h/\tilde\chi^0_3h$} & \multicolumn{2}{l|}{$36.0/31.0/12.7/7.40$} & \multicolumn{2}{l}{$\tilde\chi^0_5\to \tilde\chi^\pm_1W^\mp/\tilde\chi^0_2Z/\tilde\chi^0_1h/\tilde\chi^0_1h_s$} & \multicolumn{2}{l}{$32.1/24.1/17.4/6.30$} \\
\multicolumn{2}{l}{$\tilde\chi^\pm_1\to \tilde\chi^0_1W^*$} & \multicolumn{2}{l|}{$100$} & \multicolumn{2}{l}{$\tilde\chi^\pm_1\to \tilde\chi^0_1W^*$} & \multicolumn{2}{l}{$100$} \\
\multicolumn{2}{l}{$\tilde\chi^\pm_2\to \tilde\chi^\pm_1Z/\tilde\chi^0_2W^\mp/\tilde\chi^\pm_1h/\tilde\chi^0_1W^\mp$} & \multicolumn{2}{l|}{$25.6/23.7/16.1/15.8$} & \multicolumn{2}{l}{$\tilde\chi^\pm_2\to \tilde\chi^\pm_1Z/\tilde\chi^0_1W^\mp/\tilde\chi^0_2W^\mp/\tilde\chi^\pm_1h$} & \multicolumn{2}{l}{$21.4/20.7/20.4/15.8$} \\
\multicolumn{8}{l}{\textbf{Tests}} \\
\hline
\multicolumn{2}{l}{$S,T,U$} & \multicolumn{2}{l|}{$0.045,-0.016,-0.001$} & \multicolumn{2}{l}{$S,T,U$} & \multicolumn{2}{l}{$0.015,0.002,0.003$} \\
\multicolumn{2}{l}{$\chi^2_{\rm HiggsSignals}$} & \multicolumn{2}{l|}{$156.058$} & \multicolumn{2}{l}{$\chi^2_{\rm HiggsSignals}$} & \multicolumn{2}{l}{$156.545$} \\
\multicolumn{2}{l}{$V_{\rm tree-level}$} & \multicolumn{2}{l|}{long-lived} & \multicolumn{2}{l}{$V_{\rm tree-level}$} & \multicolumn{2}{l}{long-lived} \\
\multicolumn{2}{l}{$V_{\rm 1-loop}$} & \multicolumn{2}{l|}{long-lived} & \multicolumn{2}{l}{$V_{\rm 1-loop}$} & \multicolumn{2}{l}{long-lived} \\
\multicolumn{2}{l}{$R_{\rm SModelS}$} & \multicolumn{2}{l|}{$0.00298$} & \multicolumn{2}{l}{$R_{\rm SModelS}$} & \multicolumn{2}{l}{$0.03214$} \\
\multicolumn{2}{l}{Signal Region} & \multicolumn{2}{l|}{\texttt{4j\_Meff\_2200} in ATLAS-SUSY-2016-07~\cite{ATLAS:SUSY201607}} & \multicolumn{2}{l}{Signal Region} & \multicolumn{2}{l}{\texttt{SR\_ewk\_2l\_high} in ATLAS-SUSY-2018-16-hino~\cite{ATLAS:2019lng}} \\
\multicolumn{8}{l}{\textbf{Gravitino}} \\
\hline
\multicolumn{2}{l}{$m_{3/2}\,[\mathrm{MeV}]$} & \multicolumn{2}{l|}{$1.186$} & \multicolumn{2}{l}{$m_{3/2}\,[\mathrm{MeV}]$} & \multicolumn{2}{l}{$0.04398$} \\
\multicolumn{2}{l}{$\Gamma(\tilde\chi^0_1\to\gamma\tilde G)\,[\mathrm{GeV}]$} & \multicolumn{2}{l|}{$2.152\times 10^{-25}$} & \multicolumn{2}{l}{$\Gamma(\tilde\chi^0_1\to\gamma\tilde G)\,[\mathrm{GeV}]$} & \multicolumn{2}{l}{$3.679\times 10^{-25}$} \\
\multicolumn{2}{l}{$\Gamma(\tilde\chi^0_1\to Z\tilde G)\,[\mathrm{GeV}]$} & \multicolumn{2}{l|}{$6.494\times 10^{-21}$} & \multicolumn{2}{l}{$\Gamma(\tilde\chi^0_1\to Z\tilde G)\,[\mathrm{GeV}]$} & \multicolumn{2}{l}{$2.018\times 10^{-20}$} \\
\multicolumn{2}{l}{$\Gamma(\tilde\chi^0_1\to h\tilde G)\,[\mathrm{GeV}]$} & \multicolumn{2}{l|}{$2.313\times 10^{-22}$} & \multicolumn{2}{l}{$\Gamma(\tilde\chi^0_1\to h\tilde G)\,[\mathrm{GeV}]$} & \multicolumn{2}{l}{$2.338\times 10^{-22}$} \\
\hline\hline
\end{tabular}}
\end{table}

\clearpage

\begin{table}[!p]
\centering
\caption{\label{tab:benchmark-p7-p8} Details of benchmark points P7 and P8 in \textbf{Scenario II} with $\mu>0$. Input dimensional parameters and cross sections are given in GeV and fb, respectively; branching ratios are in percent. \BPCascadeNote}
\BPTableSkip
\resizebox{1\textwidth}{!}{%
\begin{tabular}{lrlr|lrlr}
\hline\hline
\multicolumn{4}{c|}{\bf Benchmark Point P7 (grey)} & \multicolumn{4}{c}{\bf Benchmark Point P8 (yellow)} \\
\hline
\multicolumn{8}{l}{\textbf{Parameters}} \\
\hline
$\lambda$ & $0.574$ & $\kappa$ & $-0.352$ & $\lambda$ & $0.561$ & $\kappa$ & $-0.386$ \\
$\delta$ & $-0.016$ & $\tan\beta$ & $1.565$ & $\delta$ & $0.013$ & $\tan\beta$ & $1.525$ \\
$\mu_{\rm tot}$ & $294.5~\mathrm{GeV}$ & $\mu$ & $1427.1~\mathrm{GeV}$ & $\mu_{\rm tot}$ & $454.7~\mathrm{GeV}$ & $\mu$ & $1386.4~\mathrm{GeV}$ \\
$\mu_{\rm eff}$ & $-1132.6~\mathrm{GeV}$ & $B_\mu$ & $144901.1~\mathrm{GeV}$ & $\mu_{\rm eff}$ & $-931.7~\mathrm{GeV}$ & $B_\mu$ & $483895.8~\mathrm{GeV}$ \\
$A_\lambda$ & $-730.0~\mathrm{GeV}$ & $A_\kappa$ & $-2780.8~\mathrm{GeV}$ & $A_\lambda$ & $-304.0~\mathrm{GeV}$ & $A_\kappa$ & $-2568.5~\mathrm{GeV}$ \\
$A_t$ & $1073.5~\mathrm{GeV}$ & $m_N$ & $1389.4~\mathrm{GeV}$ & $A_t$ & $1542.5~\mathrm{GeV}$ & $m_N$ & $1283.1~\mathrm{GeV}$ \\
$m_A$ & $638.3~\mathrm{GeV}$ & $m_B$ & $118.8~\mathrm{GeV}$ & $m_A$ & $607.9~\mathrm{GeV}$ & $m_B$ & $118.6~\mathrm{GeV}$ \\
\multicolumn{8}{l}{\textbf{Spectrum}} \\
\hline
$m_{\tilde\chi^0_1}$ & $289.7~\mathrm{GeV}$ & $m_{h_s}$ & $95.1~\mathrm{GeV}$ & $m_{\tilde\chi^0_1}$ & $448.8~\mathrm{GeV}$ & $m_{h_s}$ & $94.3~\mathrm{GeV}$ \\
$m_{\tilde\chi^0_2}$ & $-306.1~\mathrm{GeV}$ & $m_h$ & $123.7~\mathrm{GeV}$ & $m_{\tilde\chi^0_2}$ & $-467.8~\mathrm{GeV}$ & $m_h$ & $122.8~\mathrm{GeV}$ \\
$m_{\tilde\chi^0_3}$ & $1006.7~\mathrm{GeV}$ & $m_H$ & $654.6~\mathrm{GeV}$ & $m_{\tilde\chi^0_3}$ & $1006.9~\mathrm{GeV}$ & $m_H$ & $643.5~\mathrm{GeV}$ \\
$m_{\tilde\chi^0_4}$ & $1040.2~\mathrm{GeV}$ & $m_{A_s}$ & $2408.4~\mathrm{GeV}$ & $m_{\tilde\chi^0_4}$ & $1042.1~\mathrm{GeV}$ & $m_{A_s}$ & $2220.6~\mathrm{GeV}$ \\
$m_{\tilde\chi^0_5}$ & $1379.2~\mathrm{GeV}$ & $m_{A_H}$ & $646.2~\mathrm{GeV}$ & $m_{\tilde\chi^0_5}$ & $1270.3~\mathrm{GeV}$ & $m_{A_H}$ & $635.5~\mathrm{GeV}$ \\
$m_{\tilde\chi^\pm_1}$ & $293.1~\mathrm{GeV}$ & $m_{H^\pm}$ & $649.1~\mathrm{GeV}$ & $m_{\tilde\chi^\pm_1}$ & $453.0~\mathrm{GeV}$ & $m_{H^\pm}$ & $634.9~\mathrm{GeV}$ \\
$m_{\tilde\chi^\pm_2}$ & $1039.7~\mathrm{GeV}$ & \multicolumn{2}{l|}{} & $m_{\tilde\chi^\pm_2}$ & $1041.3~\mathrm{GeV}$ & \multicolumn{2}{l}{} \\
\multicolumn{8}{l}{\textbf{Signals}} \\
\hline
\multicolumn{2}{l}{$\mu_{\gamma\gamma}$} & \multicolumn{2}{l|}{$0.074$} & \multicolumn{2}{l}{$\mu_{\gamma\gamma}$} & \multicolumn{2}{l}{$0.068$} \\
\multicolumn{2}{l}{$\mu_{b\bar b}$} & \multicolumn{2}{l|}{$0.179$} & \multicolumn{2}{l}{$\mu_{b\bar b}$} & \multicolumn{2}{l}{$0.186$} \\
\multicolumn{2}{l}{$\sigma(gg\to H\to h h_s\to \gamma\gamma b\bar b)$} & \multicolumn{2}{l|}{$0.039~\mathrm{fb}$} & \multicolumn{2}{l}{$\sigma(gg\to H\to h h_s\to \gamma\gamma b\bar b)$} & \multicolumn{2}{l}{$0.097~\mathrm{fb}$} \\
\multicolumn{2}{l}{$\sigma(gg\to H\to h h_s\to b\bar b\tau\bar{\tau})$} & \multicolumn{2}{l|}{$0.971~\mathrm{fb}$} & \multicolumn{2}{l}{$\sigma(gg\to H\to h h_s\to b\bar b\tau\bar{\tau})$} & \multicolumn{2}{l}{$2.449~\mathrm{fb}$} \\
\multicolumn{2}{l}{$\sigma(gg\to H\to h h_s\to b\bar b b\bar b)$} & \multicolumn{2}{l|}{$8.385~\mathrm{fb}$} & \multicolumn{2}{l}{$\sigma(gg\to H\to h h_s\to b\bar b b\bar b)$} & \multicolumn{2}{l}{$21.171~\mathrm{fb}$} \\
\multicolumn{2}{l}{$\sigma(gg\to H\to h_s h_s\to b\bar b\tau\bar{\tau})$} & \multicolumn{2}{l|}{$0.022~\mathrm{fb}$} & \multicolumn{2}{l}{$\sigma(gg\to H\to h_s h_s\to b\bar b\tau\bar{\tau})$} & \multicolumn{2}{l}{$0.241~\mathrm{fb}$} \\
\multicolumn{2}{l}{$\sigma(gg\to H\to h_s h_s\to b\bar b b\bar b)$} & \multicolumn{2}{l|}{$0.203~\mathrm{fb}$} & \multicolumn{2}{l}{$\sigma(gg\to H\to h_s h_s\to b\bar b b\bar b)$} & \multicolumn{2}{l}{$2.179~\mathrm{fb}$} \\
\multicolumn{2}{l}{$\sigma(gg\to A_2\to h A_1\to b\bar b b\bar b)$} & \multicolumn{2}{l|}{$\text{\ttfamily -----}$} & \multicolumn{2}{l}{$\sigma(gg\to A_2\to h A_1\to b\bar b b\bar b)$} & \multicolumn{2}{l}{$\text{\ttfamily -----}$} \\
\multicolumn{2}{l}{$\sigma(gg\to A_H\to Zh\to Zb\bar b)$} & \multicolumn{2}{l|}{$0.754~\mathrm{fb}$} & \multicolumn{2}{l}{$\sigma(gg\to A_H\to Zh\to Zb\bar b)$} & \multicolumn{2}{l}{$2.239~\mathrm{fb}$} \\
\multicolumn{2}{l}{$\sigma(gg\to A_H\to Zh_s\to \ell\ell b\bar b)$} & \multicolumn{2}{l|}{$1.405~\mathrm{fb}$} & \multicolumn{2}{l}{$\sigma(gg\to A_H\to Zh_s\to \ell\ell b\bar b)$} & \multicolumn{2}{l}{$3.161~\mathrm{fb}$} \\
\multicolumn{8}{l}{\textbf{Couplings}} \\
\hline
\multicolumn{2}{l}{$C_{h_sb\bar b},C_{h_st\bar t},C_{h_sVV},C_{h_s\gamma\gamma},C_{h_sgg}$} & \multicolumn{2}{l|}{$0.507,0.367,0.408,0.347,0.374$} & \multicolumn{2}{l}{$C_{h_sb\bar b},C_{h_st\bar t},C_{h_sVV},C_{h_s\gamma\gamma},C_{h_sgg}$} & \multicolumn{2}{l}{$0.555,0.353,0.414,0.380,0.358$} \\
\multicolumn{2}{l}{$C_{hb\bar b},C_{ht\bar t},C_{hVV},C_{h\gamma\gamma},C_{hgg}$} & \multicolumn{2}{l|}{$0.885,0.924,0.913,0.949,0.931$} & \multicolumn{2}{l}{$C_{hb\bar b},C_{ht\bar t},C_{hVV},C_{h\gamma\gamma},C_{hgg}$} & \multicolumn{2}{l}{$0.865,0.930,0.910,0.933,0.938$} \\
\multicolumn{2}{l}{$C_{Hb\bar b},C_{Ht\bar t},C_{HVV},C_{H\gamma\gamma},C_{Hgg}$} & \multicolumn{2}{l|}{$1.552,0.647,0.010,4.125,0.560$} & \multicolumn{2}{l}{$C_{Hb\bar b},C_{Ht\bar t},C_{HVV},C_{H\gamma\gamma},C_{Hgg}$} & \multicolumn{2}{l}{$1.507,0.663,0.011,4.341,0.575$} \\
\multicolumn{2}{l}{$C_{A_sb\bar b},C_{A_st\bar t},C_{A_s\gamma\gamma},C_{A_sgg}$} & \multicolumn{2}{l|}{$0.061,0.025,0.120,0.022$} & \multicolumn{2}{l}{$C_{A_sb\bar b},C_{A_st\bar t},C_{A_s\gamma\gamma},C_{A_sgg}$} & \multicolumn{2}{l}{$0.051,0.022,0.154,0.020$} \\
\multicolumn{2}{l}{$C_{A_Hb\bar b},C_{A_Ht\bar t},C_{A_H\gamma\gamma},C_{A_Hgg}$} & \multicolumn{2}{l|}{$1.564,0.638,5.296,0.685$} & \multicolumn{2}{l}{$C_{A_Hb\bar b},C_{A_Ht\bar t},C_{A_H\gamma\gamma},C_{A_Hgg}$} & \multicolumn{2}{l}{$1.525,0.655,4.991,0.707$} \\
\multicolumn{8}{l}{\textbf{Mixing}} \\
\hline
\multicolumn{2}{l}{$N_{11},N_{12},N_{13},N_{14},N_{15}$} & \multicolumn{2}{l|}{$\phantom{-}0.060,\;-0.104,\;\phantom{-}0.702,\;-0.701,\;\phantom{-}0.019$} & \multicolumn{2}{l}{$N_{11},N_{12},N_{13},N_{14},N_{15}$} & \multicolumn{2}{l}{$-0.077,\;\phantom{-}0.131,\;-0.699,\;\phantom{-}0.698,\;-0.023$} \\
\multicolumn{2}{l}{$N_{21},N_{22},N_{23},N_{24},N_{25}$} & \multicolumn{2}{l|}{$\phantom{-}0.007,\;-0.013,\;-0.706,\;-0.706,\;-0.057$} & \multicolumn{2}{l}{$N_{21},N_{22},N_{23},N_{24},N_{25}$} & \multicolumn{2}{l}{$\phantom{-}0.006,\;-0.011,\;-0.706,\;-0.706,\;-0.054$} \\
\multicolumn{2}{l}{$N_{31},N_{32},N_{33},N_{34},N_{35}$} & \multicolumn{2}{l|}{$\phantom{-}0.971,\;\phantom{-}0.235,\;-0.021,\;\phantom{-}0.027,\;\phantom{-}0.000$} & \multicolumn{2}{l}{$N_{31},N_{32},N_{33},N_{34},N_{35}$} & \multicolumn{2}{l}{$\phantom{-}0.965,\;\phantom{-}0.259,\;-0.027,\;\phantom{-}0.031,\;\phantom{-}0.000$} \\
\multicolumn{2}{l}{$N_{41},N_{42},N_{43},N_{44},N_{45}$} & \multicolumn{2}{l|}{$\phantom{-}0.229,\;-0.966,\;-0.071,\;\phantom{-}0.091,\;-0.003$} & \multicolumn{2}{l}{$N_{41},N_{42},N_{43},N_{44},N_{45}$} & \multicolumn{2}{l}{$\phantom{-}0.250,\;-0.957,\;-0.095,\;\phantom{-}0.112,\;-0.007$} \\
\multicolumn{2}{l}{$N_{51},N_{52},N_{53},N_{54},N_{55}$} & \multicolumn{2}{l|}{$-0.001,\;-0.002,\;-0.054,\;-0.027,\;\phantom{-}0.998$} & \multicolumn{2}{l}{$N_{51},N_{52},N_{53},N_{54},N_{55}$} & \multicolumn{2}{l}{$\phantom{-}0.000,\;\phantom{-}0.004,\;\phantom{-}0.055,\;\phantom{-}0.021,\;-0.998$} \\
\multicolumn{8}{l}{\textbf{Decays}} \\
\hline
\multicolumn{2}{l}{$h_s\to b\bar b/\tau^+\tau^-/c\bar c/gg/W^+W^-/\gamma\gamma$} & \multicolumn{2}{l|}{$86.0/9.52/2.14/2.06/0.133/0.076$} & \multicolumn{2}{l}{$h_s\to b\bar b/\tau^+\tau^-/c\bar c/gg/W^+W^-/\gamma\gamma$} & \multicolumn{2}{l}{$86.9/9.61/1.67/1.56/0.089/0.075$} \\
\multicolumn{2}{l}{$h\to b\bar b/W^+W^-/\tau^+\tau^-/gg/c\bar c/\gamma\gamma$} & \multicolumn{2}{l|}{$60.9/21.2/7.05/5.17/3.13/0.282$} & \multicolumn{2}{l}{$h\to b\bar b/W^+W^-/\tau^+\tau^-/gg/c\bar c/\gamma\gamma$} & \multicolumn{2}{l}{$61.5/20.2/7.12/5.46/3.36/0.282$} \\
\multicolumn{2}{l}{$H\to t\bar t/hh_s/hh/b\bar b/gg/W^+W^-$} & \multicolumn{2}{l|}{$93.8/3.94/1.61/0.212/0.101/0.076$} & \multicolumn{2}{l}{$H\to t\bar t/hh_s/hh/h_sh_s/b\bar b/gg$} & \multicolumn{2}{l}{$87.4/8.50/3.03/0.619/0.180/0.096$} \\
\multicolumn{2}{l}{$A_s\to \tilde\chi^+_1\tilde\chi^-_1/\tilde\chi^0_2\tilde\chi^0_2/\tilde\chi^0_1\tilde\chi^0_1/\tilde\chi^\pm_1\tilde\chi^\mp_2/\tilde\chi^0_1\tilde\chi^0_4/t\bar t$} & \multicolumn{2}{l|}{$27.4/14.1/13.6/0.773/0.456/0.128$} & \multicolumn{2}{l}{$A_s\to \tilde\chi^+_1\tilde\chi^-_1/\tilde\chi^0_2\tilde\chi^0_2/\tilde\chi^0_1\tilde\chi^0_1/\tilde\chi^\pm_1\tilde\chi^\mp_2/\tilde\chi^0_1\tilde\chi^0_4/t\bar t$} & \multicolumn{2}{l}{$31.5/16.4/15.5/1.22/0.729/0.129$} \\
\multicolumn{2}{l}{$A_H\to t\bar t/Zh_s/\tilde\chi^+_1\tilde\chi^-_1/\tilde\chi^0_1\tilde\chi^0_1/Zh/b\bar b$} & \multicolumn{2}{l|}{$95.7/2.52/0.602/0.500/0.191/0.171$} & \multicolumn{2}{l}{$A_H\to t\bar t/Zh_s/Zh/b\bar b/gg/\tau^+\tau^-$} & \multicolumn{2}{l}{$94.4/4.85/0.485/0.153/0.121/0.023$} \\
\multicolumn{2}{l}{$H^+\to t\bar b/h_sW^+/hW^+/\tau\nu_\tau/\tilde\chi^0_1\tilde\chi^\pm_1/\tilde\chi^0_2\tilde\chi^\pm_1$} & \multicolumn{2}{l|}{$96.7/2.97/0.226/0.030/0.015/0.012$} & \multicolumn{2}{l}{$H^+\to t\bar b/h_sW^+/hW^+/\tau\nu_\tau$} & \multicolumn{2}{l}{$93.8/5.58/0.559/0.026$} \\
\multicolumn{2}{l}{$\tilde\chi^0_2\to \tilde\chi^0_1Z^*/\tilde\chi^\pm_1W^{\mp *}$} & \multicolumn{2}{l|}{$56.0/44.0$} & \multicolumn{2}{l}{$\tilde\chi^0_2\to \tilde\chi^0_1Z^*/\tilde\chi^\pm_1W^{\mp *}$} & \multicolumn{2}{l}{$58.2/41.8$} \\
\multicolumn{2}{l}{$\tilde\chi^0_3\to \tilde\chi^\pm_1W^\mp/\tilde\chi^0_2Z/\tilde\chi^0_1h/\tilde\chi^\pm_1H^\mp$} & \multicolumn{2}{l|}{$81.8/6.80/4.80/2.80$} & \multicolumn{2}{l}{$\tilde\chi^0_3\to \tilde\chi^\pm_1W^\mp/\tilde\chi^0_2Z/\tilde\chi^0_1h/\tilde\chi^0_1h_s$} & \multicolumn{2}{l}{$87.2/5.80/3.80/2.10$} \\
\multicolumn{2}{l}{$\tilde\chi^0_4\to \tilde\chi^\pm_1W^\mp/\tilde\chi^0_2Z/\tilde\chi^0_1h/\tilde\chi^0_1h_s$} & \multicolumn{2}{l|}{$35.2/29.6/20.5/8.60$} & \multicolumn{2}{l}{$\tilde\chi^0_4\to \tilde\chi^\pm_1W^\mp/\tilde\chi^0_2Z/\tilde\chi^0_1h/\tilde\chi^0_1h_s$} & \multicolumn{2}{l}{$36.0/31.5/20.7/10.6$} \\
\multicolumn{2}{l}{$\tilde\chi^0_5\to \tilde\chi^\pm_1H^\mp/\tilde\chi^\pm_1W^\mp/\tilde\chi^0_2A_H/\tilde\chi^0_1H$} & \multicolumn{2}{l|}{$29.5/19.9/14.6/14.5$} & \multicolumn{2}{l}{$\tilde\chi^0_5\to \tilde\chi^\pm_1H^\mp/\tilde\chi^0_1H/\tilde\chi^0_2A_H/\tilde\chi^\pm_1W^\mp$} & \multicolumn{2}{l}{$33.9/16.5/16.5/16.1$} \\
\multicolumn{2}{l}{$\tilde\chi^\pm_1\to \tilde\chi^0_1W^*$} & \multicolumn{2}{l|}{$100$} & \multicolumn{2}{l}{$\tilde\chi^\pm_1\to \tilde\chi^0_1W^*$} & \multicolumn{2}{l}{$100$} \\
\multicolumn{2}{l}{$\tilde\chi^\pm_2\to \tilde\chi^\pm_1Z/\tilde\chi^0_1W^\mp/\tilde\chi^0_2W^\mp/\tilde\chi^\pm_1h$} & \multicolumn{2}{l|}{$24.2/23.4/23.3/16.9$} & \multicolumn{2}{l}{$\tilde\chi^\pm_2\to \tilde\chi^\pm_1Z/\tilde\chi^0_1W^\mp/\tilde\chi^0_2W^\mp/\tilde\chi^\pm_1h$} & \multicolumn{2}{l}{$25.5/24.6/24.2/16.6$} \\
\multicolumn{8}{l}{\textbf{Tests}} \\
\hline
\multicolumn{2}{l}{$S,T,U$} & \multicolumn{2}{l|}{$0.008,0.001,0.001$} & \multicolumn{2}{l}{$S,T,U$} & \multicolumn{2}{l}{$0.009,0.000,0.000$} \\
\multicolumn{2}{l}{$\chi^2_{\rm HiggsSignals}$} & \multicolumn{2}{l|}{$156.823$} & \multicolumn{2}{l}{$\chi^2_{\rm HiggsSignals}$} & \multicolumn{2}{l}{$155.652$} \\
\multicolumn{2}{l}{$V_{\rm tree-level}$} & \multicolumn{2}{l|}{short-lived} & \multicolumn{2}{l}{$V_{\rm tree-level}$} & \multicolumn{2}{l}{short-lived} \\
\multicolumn{2}{l}{$V_{\rm 1-loop}$} & \multicolumn{2}{l|}{short-lived} & \multicolumn{2}{l}{$V_{\rm 1-loop}$} & \multicolumn{2}{l}{short-lived} \\
\multicolumn{2}{l}{$R_{\rm SModelS}$} & \multicolumn{2}{l|}{$0.42560$} & \multicolumn{2}{l}{$R_{\rm SModelS}$} & \multicolumn{2}{l}{$0.44679$} \\
\multicolumn{2}{l}{Signal Region} & \multicolumn{2}{l|}{\texttt{4j\_Meff\_3000} in ATLAS-SUSY-2016-07~\cite{ATLAS:SUSY201607}} & \multicolumn{2}{l}{Signal Region} & \multicolumn{2}{l}{\texttt{4j\_Meff\_3000} in ATLAS-SUSY-2016-07~\cite{ATLAS:SUSY201607}} \\
\multicolumn{8}{l}{\textbf{Gravitino}} \\
\hline
\multicolumn{2}{l}{$m_{3/2}\,[\mathrm{MeV}]$} & \multicolumn{2}{l|}{$16.58$} & \multicolumn{2}{l}{$m_{3/2}\,[\mathrm{MeV}]$} & \multicolumn{2}{l}{$16.48$} \\
\multicolumn{2}{l}{$\Gamma(\tilde\chi^0_1\to\gamma\tilde G)\,[\mathrm{GeV}]$} & \multicolumn{2}{l|}{$7.107\times 10^{-29}$} & \multicolumn{2}{l}{$\Gamma(\tilde\chi^0_1\to\gamma\tilde G)\,[\mathrm{GeV}]$} & \multicolumn{2}{l}{$1.586\times 10^{-27}$} \\
\multicolumn{2}{l}{$\Gamma(\tilde\chi^0_1\to Z\tilde G)\,[\mathrm{GeV}]$} & \multicolumn{2}{l|}{$5.163\times 10^{-24}$} & \multicolumn{2}{l}{$\Gamma(\tilde\chi^0_1\to Z\tilde G)\,[\mathrm{GeV}]$} & \multicolumn{2}{l}{$5.884\times 10^{-23}$} \\
\multicolumn{2}{l}{$\Gamma(\tilde\chi^0_1\to h\tilde G)\,[\mathrm{GeV}]$} & \multicolumn{2}{l|}{$8.566\times 10^{-26}$} & \multicolumn{2}{l}{$\Gamma(\tilde\chi^0_1\to h\tilde G)\,[\mathrm{GeV}]$} & \multicolumn{2}{l}{$1.269\times 10^{-24}$} \\
\hline\hline
\end{tabular}}
\end{table}

\clearpage
\noindent In the pseudoscalar cascade row, $A_2$ and $A_1$ follow the mass-ordered convention used in Fig.~\ref{fig:pseudoscalar-plane}; a dash means that the parent pseudoscalar exceeds $1~{\rm TeV}$, outside the range used for this calculation, or that $A_2\to h A_1$ is kinematically forbidden.
The vacuum-stability entries should be read as an additional check rather than a scan cut.
P1, P2, P5 and P6 are long-lived for both $V_{\rm tree-level}$ and $V_{\rm 1-loop}$, P3 becomes long-lived after including the one-loop effective potential, while P4, P7 and P8 are short-lived in the one-loop test and would be disfavored if vacuum stability were imposed strictly. 
The block labelled Mixing gives the neutralino diagonalization matrix $N_{ij}$. 
Its basis follows Eq.~(\ref{eq:neutralino-mixing}), namely $(\tilde B,\tilde W^0,\tilde H_d^0,\tilde H_u^0,\tilde S)$. 
For the positive-$\mu$ points P1 and P5--P8, the tables also list the inferred $m_{3/2}$ and the three partial widths in Fig.~\ref{fig:gravitino-widths}; these entries are left blank for the negative-$\mu$ benchmarks because the supergravity-motivated relation is not imposed there.

\section{Conclusion}
\label{conclusion}
We have examined the scalar resonance interpretation of the $\mu$-term extended NMSSM in light of the persistent $95~{\rm GeV}$ hints and the present status of heavy resonance searches. 
The scan was designed to keep a singlet-like CP-even scalar near $95~{\rm GeV}$ and a heavier doublet-like scalar in the range relevant for the CMS $X\to H_{\rm 125}Y$ searches. 
After imposing the Higgs measurements, extra Higgs limits, flavor data, electroweak precision observables and direct SUSY search bounds, we find viable regions that reproduce both the diphoton and $b\bar b$ excesses near $95~{\rm GeV}$ within the $2\sigma$ ranges. 
This shows that the restricted $\mu$NMSSM still has enough freedom to describe the light scalar data, even though it is much less general than the full GNMSSM.

The viable samples are not organized around a single solution. 
They separate into two patterns controlled mainly by the reduced couplings of the light scalar. 
\textbf{Scenario I} has $\mu_{\gamma\gamma}>0.11$ and a suppressed $h_s b\bar b$ coupling. 
This reduces the total width of $h_s$ and gives a larger diphoton rate, while the LEP $b\bar b$ signal remains small. 
\textbf{Scenario II} has $\mu_{\gamma\gamma}<0.11$ and a larger doublet component of $h_s$. 
It gives a better description of the LEP excess, but the larger bottom width lowers the diphoton branching fraction. 
The benchmark points illustrate that lighter Higgsino-like charginos can enhance the loop-induced $h_s\gamma\gamma$ coupling, but this effect cannot compensate for a large $h_s b\bar b$ width in \textbf{Scenario II}.

The CP-even state $H$ typically lies around $625$--$675~{\rm GeV}$ and can decay through $H\to h h_s$ and $H\to h_s h_s$. 
The predicted $H\to h h_s\to \gamma\gamma b\bar b$ rate is generally below the original CMS best-fit value at $(650,95)~{\rm GeV}$, reflecting the tension with later searches and with the $b\bar b\tau\bar{\tau}$ and $4b$ limits. 
Nevertheless, rates of order $10^{-2}$ to $10^{-1}~{\rm fb}$ remain possible in the $\gamma\gamma b\bar b$ channel, while the associated $4b$ and $b\bar b\tau\bar{\tau}$ modes are sufficiently close to current limits that Run~3 data may test the upper part of the allowed region. 
The latest CMS $4b$ analysis~\cite{CMS:2026mwf} also changes the emphasis of the heavy sector. 
It does not support a strong signal at the earlier $(650,95)~{\rm GeV}$ point, but it shows its largest local deviation near $(600,400)~{\rm GeV}$. 
In our scan this mass region can be approached by the CP-odd cascade $A_2\to h A_1\to b\bar b b\bar b$, where the ordering of $A_H$ and $A_s$ determines the parent and daughter states. 
The predicted rates remain below the current CMS limits, but a fit tailored to this topology would be a useful test of the model.

We also studied the positive-$\mu$ subset motivated by the inflation-inspired superconformal embedding of the Higgs sector into supergravity. 
The special feature is not the gravitino LSP itself, which is already familiar in supersymmetric models. 
Rather, in this construction the same nonminimal Higgs-gravity coupling that can support Higgs-sector inflation also induces the low energy $\mu$ term and relates it to the gravitino mass. 
The viable positive-$\mu$ samples then correspond to a gravitino LSP from the keV scale to several tens of MeV, while the lightest neutralino becomes the NLSP. 
Its decay widths into $\gamma\tilde G$, $Z\tilde G$ and $h\tilde G$ are usually small enough that the neutralino is long-lived on collider scales, so present prompt electroweakino searches have little impact. 
Future displaced, delayed and missing energy searches can probe part of this region, while the relic abundance points to reheating temperatures that depend sensitively on the gravitino mass. 
Warm dark matter and Big Bang nucleosynthesis constraints have not been imposed as full cuts here and require a separate cosmological analysis.

The benchmark tables show that the representative points satisfy the collider and precision constraints used in the scan, but they also indicate where further theory checks matter. 
The vacuum stability test was applied only after the scan, and some benchmark points become short-lived when this requirement is imposed strictly. 
Taken together, these results show that the $\mu$NMSSM gives a viable and testable framework for the $95~{\rm GeV}$ anomalies and for correlated heavy scalar searches, while a sharper statement about the preferred region will require the next round of dedicated $4b$, $\gamma\gamma b\bar b$, $b\bar b\tau\bar{\tau}$ and long-lived particle analyses.

\section*{Acknowledgements}
This work is supported by the National Natural Science Foundation of China (NSFC) under Grant No.~12447140 and the Natural Science Foundation of Henan Province, China under Grant No.~252300421771. 
This work is also supported by the Henan Postdoctoral Research Project under Grant No.~HN2026054.

\bibliographystyle{CitationStyle}
\bibliography{LianRef}

@article{Cao:2024axg,
    author = "Cao, Junjie and Jia, Xinglong and Lian, Jingwei",
    title = "{Unified interpretation of the muon g-2 anomaly, the 95~GeV diphoton, and bb{\textasciimacron} excesses in the general next-to-minimal supersymmetric standard model}",
    eprint = "2402.15847",
    archivePrefix = "arXiv",
    primaryClass = "hep-ph",
    doi = "10.1103/PhysRevD.110.115039",
    journal = "Phys. Rev. D",
    volume = "110",
    number = "11",
    pages = "115039",
    year = "2024"
}

@article{Cao:2023gkc,
    author = "Cao, Junjie and Jia, Xinglong and Lian, Jingwei and Meng, Lei",
    title = "{95~GeV diphoton and bb{\textasciimacron} excesses in the general next-to-minimal supersymmetric standard model}",
    eprint = "2310.08436",
    archivePrefix = "arXiv",
    primaryClass = "hep-ph",
    doi = "10.1103/PhysRevD.109.075001",
    journal = "Phys. Rev. D",
    volume = "109",
    number = "7",
    pages = "075001",
    year = "2024"
}

@article{Zhou:2021pit,
    author = "Zhou, Haijing and Cao, Junjie and Lian, Jingwei and Zhang, Di",
    title = "{Singlino-dominated dark matter in Z3-symmetric NMSSM}",
    eprint = "2102.05309",
    archivePrefix = "arXiv",
    primaryClass = "hep-ph",
    doi = "10.1103/PhysRevD.104.015017",
    journal = "Phys. Rev. D",
    volume = "104",
    number = "1",
    pages = "015017",
    year = "2021"
}

@article{Cao:2019ofo,
    author = "Cao, Junjie and Jia, Xinglong and Yue, Yuanfang and Zhou, Haijing and Zhu, Pengxuan",
    title = "{96 GeV diphoton excess in seesaw extensions of the natural NMSSM}",
    eprint = "1908.07206",
    archivePrefix = "arXiv",
    primaryClass = "hep-ph",
    doi = "10.1103/PhysRevD.101.055008",
    journal = "Phys. Rev. D",
    volume = "101",
    number = "5",
    pages = "055008",
    year = "2020"
}

@article{Cao:2018rix,
    author = "Cao, Junjie and He, Yangle and Shang, Liangliang and Zhang, Yang and Zhu, Pengxuan",
    title = "{Current status of a natural NMSSM in light of LHC 13 TeV data and XENON-1T results}",
    eprint = "1810.09143",
    archivePrefix = "arXiv",
    primaryClass = "hep-ph",
    reportNumber = "CoEPP-MN-18-8",
    doi = "10.1103/PhysRevD.99.075020",
    journal = "Phys. Rev. D",
    volume = "99",
    number = "7",
    pages = "075020",
    year = "2019"
}

@article{Cao:2016uwt,
    author = "Cao, Junjie and Guo, Xiaofei and He, Yangle and Wu, Peiwen and Zhang, Yang",
    title = "{Diphoton signal of the light Higgs boson in natural NMSSM}",
    eprint = "1612.08522",
    archivePrefix = "arXiv",
    primaryClass = "hep-ph",
    doi = "10.1103/PhysRevD.95.116001",
    journal = "Phys. Rev. D",
    volume = "95",
    number = "11",
    pages = "116001",
    year = "2017"
}

@article{Cao:2012fz,
    author = "Cao, Jun-Jie and Heng, Zhao-Xia and Yang, Jin Min and Zhang, Yan-Ming and Zhu, Jing-Ya",
    title = "{A SM-like Higgs near 125 GeV in low energy SUSY: a comparative study for MSSM and NMSSM}",
    eprint = "1202.5821",
    archivePrefix = "arXiv",
    primaryClass = "hep-ph",
    doi = "10.1007/JHEP03(2012)086",
    journal = "JHEP",
    volume = "03",
    pages = "086",
    year = "2012"
}

@article{Lian:2024smg,
    author = "Lian, Jingwei",
    title = "{95~GeV excesses in the Z3-symmetric next-to-minimal supersymmetric standard model}",
    eprint = "2406.10969",
    archivePrefix = "arXiv",
    primaryClass = "hep-ph",
    doi = "10.1103/PhysRevD.110.115018",
    journal = "Phys. Rev. D",
    volume = "110",
    number = "11",
    pages = "115018",
    year = "2024"
}

@article{Lian:2025zoi,
    author = "Lian, Jingwei and Liu, Yao-Bei",
    title = "{Scalar resonances near 650 and 95 GeV in the GNMSSM with correct dark matter relic abundance}",
    eprint = "2511.04968",
    archivePrefix = "arXiv",
    primaryClass = "hep-ph",
    doi = "10.1007/JHEP03(2026)103",
    journal = "JHEP",
    volume = "03",
    pages = "103",
    year = "2026"
}

@article{Bahl:2022igd,
    author = {Bahl, Henning and Biek\"otter, Thomas and Heinemeyer, Sven and Li, Cheng and Paasch, Steven and Weiglein, Georg and Wittbrodt, Jonas},
    title = "{HiggsTools: BSM scalar phenomenology with new versions of HiggsBounds and HiggsSignals}",
    eprint = "2210.09332",
    archivePrefix = "arXiv",
    primaryClass = "hep-ph",
    doi = "10.1016/j.cpc.2023.108803",
    journal = "Comput. Phys. Commun.",
    volume = "291",
    pages = "108803",
    year = "2023"
}

@article{Biekotter:2023jld,
    author = {Biek{\"o}tter, Thomas and Heinemeyer, Sven and Weiglein, Georg},
    title = "{The CMS di-photon excess at 95 GeV in view of the LHC Run 2 results}",
    eprint = "2303.12018",
    archivePrefix = "arXiv",
    primaryClass = "hep-ph",
    reportNumber = "KA-TP-03-2023, DESY-23-033, IFT-UAM/CSIC-23-028",
    doi = "10.1016/j.physletb.2023.138217",
    journal = "Phys. Lett. B",
    volume = "846",
    pages = "138217",
    year = "2023"
}

@article{Aguilar-Saavedra:2023vpd,
    author = "Aguilar-Saavedra, J. A. and C\^amara, H. B. and Joaquim, F. R. and Seabra, J. F.",
    title = "{Confronting the 95 GeV excesses within the UN2HDM}",
    eprint = "2307.03768",
    archivePrefix = "arXiv",
    primaryClass = "hep-ph",
    reportNumber = "IFT-UAM-CSIC-23-86",
    month = "7",
    year = "2023"
}

@article{CMS:2023boe,
    author = "Tumasyan, Armen and others",
    collaboration = "CMS",
    title = "{Search for a new resonance decaying into two spin-0 bosons in a final state with two photons and two bottom quarks in proton-proton collisions at $ \sqrt{s} $ = 13 TeV}",
    eprint = "2310.01643",
    archivePrefix = "arXiv",
    primaryClass = "hep-ex",
    reportNumber = "CMS-HIG-21-011, CERN-EP-2023-132",
    doi = "10.1007/JHEP05(2024)316",
    journal = "JHEP",
    volume = "05",
    pages = "316",
    year = "2024"
}

@article{CMS:2025qit,
    author = "Hayrapetyan, Aram and others",
    collaboration = "CMS",
    title = "{Search for a new scalar resonance decaying to a Higgs boson and another new scalar particle in the final state with two bottom quarks and two photons in proton-proton collisions at $\sqrt{s}=13$ TeV}",
    eprint = "2508.11494",
    archivePrefix = "arXiv",
    primaryClass = "hep-ex",
    reportNumber = "CMS-B2G-24-001, CERN-EP-2025-160",
    doi = "10.1007/JHEP12(2025)178",
    journal = "JHEP",
    volume = "12",
    pages = "178",
    year = "2025"
}

@article{ATLAS:2024auw,
    author = "Aad, Georges and others",
    collaboration = "ATLAS",
    title = "{Search for a resonance decaying into a scalar particle and a Higgs boson in the final state with two bottom quarks and two photons in proton{\textendash}proton collisions at $ \sqrt{s} $ = 13 TeV with the ATLAS detector}",
    eprint = "2404.12915",
    archivePrefix = "arXiv",
    primaryClass = "hep-ex",
    reportNumber = "CERN-EP-2024-072",
    doi = "10.1007/JHEP11(2024)047",
    journal = "JHEP",
    volume = "11",
    pages = "047",
    year = "2024"
}

@article{ATLAS:2025nda,
    author = "Aad, Georges and others",
    collaboration = "ATLAS",
    title = "{Search for a resonance decaying into a scalar particle and a Higgs boson in the final state with two bottom quarks and two photons with 199 fb$^{-1}$ of data collected at $\sqrt{s}$=13 TeV and $\sqrt{s}$=13.6 TeV with the ATLAS detector}",
    eprint = "2510.02857",
    archivePrefix = "arXiv",
    primaryClass = "hep-ex",
    reportNumber = "CERN-EP-2025-204",
    doi = "10.1016/j.physletb.2026.140425",
    journal = "Phys. Lett. B",
    volume = "877",
    pages = "140425",
    year = "2026"
}

@article{CMS:2026mwf,
    author = "Hayrapetyan, Aram and others",
    collaboration = "CMS",
    title = "{Search for a new heavy scalar resonance decaying into the Higgs boson and a new scalar particle in the $\mathrm{b}\bar{\mathrm{b}}\mathrm{b}\bar{\mathrm{b}}$ final state using proton-proton collisions at $\sqrt{s}$ = 13 TeV}",
    eprint = "2605.02848",
    archivePrefix = "arXiv",
    primaryClass = "hep-ex",
    reportNumber = "CMS-HIG-20-012, CERN-EP-2026-010",
    month = "5",
    year = "2026"
}

@article{CMS:2025tqi,
    author = "Hayrapetyan, Aram and others",
    collaboration = "CMS",
    title = "{Search for the nonresonant and resonant production of a Higgs boson in association with an additional scalar boson in the $\gamma\gamma\tau\tau$ final state in proton-proton collisions at $\sqrt{s}$ = 13 TeV}",
    eprint = "2506.23012",
    archivePrefix = "arXiv",
    primaryClass = "hep-ex",
    reportNumber = "CMS-HIG-22-012, CERN-EP-2025-116",
    doi = "10.1007/JHEP05(2026)186",
    journal = "JHEP",
    volume = "05",
    pages = "186",
    year = "2026"
}

@article{Ahriche:2023hho,
    author = "Ahriche, Amine and Bellilet, Mohamed Lamine and Khojali, Mohammed Omer and Kumar, Mukesh and Mulaudzi, Anza-Tshildzi",
    title = "{Scale invariant scotogenic model: CDF-II W-boson mass and the 95~GeV excesses}",
    eprint = "2311.08297",
    archivePrefix = "arXiv",
    primaryClass = "hep-ph",
    doi = "10.1103/PhysRevD.110.015025",
    journal = "Phys. Rev. D",
    volume = "110",
    number = "1",
    pages = "015025",
    year = "2024"
}

@article{Banik:2023ecr,
    author = "Banik, Sumit and Crivellin, Andreas and Iguro, Syuhei and Kitahara, Teppei",
    title = "{Asymmetric di-Higgs signals of the next-to-minimal 2HDM with a U(1) symmetry}",
    eprint = "2303.11351",
    archivePrefix = "arXiv",
    primaryClass = "hep-ph",
    reportNumber = "PSI-PR-23-7, ZU-TH 15/23, P3H-23-015, TTP23-010, KEK-TH-2506",
    doi = "10.1103/PhysRevD.108.075011",
    journal = "Phys. Rev. D",
    volume = "108",
    number = "7",
    pages = "075011",
    year = "2023"
}

@article{Iguro:2022fel,
    author = "Iguro, Syuhei and Kitahara, Teppei and Omura, Yuji and Zhang, Hantian",
    title = "{Chasing the two-Higgs doublet model in the di-Higgs boson production}",
    eprint = "2211.00011",
    archivePrefix = "arXiv",
    primaryClass = "hep-ph",
    reportNumber = "P3H-22-102, TTP22-061, KEK-TH-2468",
    doi = "10.1103/PhysRevD.107.075017",
    journal = "Phys. Rev. D",
    volume = "107",
    number = "7",
    pages = "075017",
    year = "2023"
}

@article{Dutta:2023cig,
    author = "Dutta, Juhi and Lahiri, Jayita and Li, Cheng and Moortgat-Pick, Gudrid and Tabira, Sheikh Farah and Ziegler, Julia Anabell",
    title = "{Dark matter phenomenology in 2HDMS in light of the 95 GeV excess}",
    eprint = "2308.05653",
    archivePrefix = "arXiv",
    primaryClass = "hep-ph",
    reportNumber = "DESY-23-114",
    doi = "10.1140/epjc/s10052-024-13176-9",
    journal = "Eur. Phys. J. C",
    volume = "84",
    number = "9",
    pages = "926",
    year = "2024"
}

@article{Ellwanger:2024txc,
    author = "Ellwanger, Ulrich and Hugonie, Cyril",
    title = "{Nmssm with correct relic density and an additional 95~GeV Higgs boson}",
    eprint = "2403.16884",
    archivePrefix = "arXiv",
    primaryClass = "hep-ph",
    doi = "10.1140/epjc/s10052-024-12886-4",
    journal = "Eur. Phys. J. C",
    volume = "84",
    number = "5",
    pages = "526",
    year = "2024"
}

@article{Ellwanger:2023zjc,
    author = "Ellwanger, Ulrich and Hugonie, Cyril",
    title = "{Additional Higgs Bosons near 95 and 650 GeV in the NMSSM}",
    eprint = "2309.07838",
    archivePrefix = "arXiv",
    primaryClass = "hep-ph",
    doi = "10.1140/epjc/s10052-023-12315-y",
    journal = "Eur. Phys. J. C",
    volume = "83",
    number = "12",
    pages = "1138",
    year = "2023"
}

@article{Ellwanger:2024vvs,
    author = "Ellwanger, Ulrich and Hugonie, Cyril and King, Stephen F. and Moretti, Stefano",
    title = "{NMSSM explanation for excesses in the search for neutralinos and charginos and a 95 GeV Higgs boson}",
    eprint = "2404.19338",
    archivePrefix = "arXiv",
    primaryClass = "hep-ph",
    doi = "10.1140/epjc/s10052-024-13129-2",
    journal = "Eur. Phys. J. C",
    volume = "84",
    number = "8",
    pages = "788",
    year = "2024"
}

@article{King:1995vk,
    author = "King, S. F. and White, P. L.",
    title = "{Resolving the constrained minimal and next-to-minimal supersymmetric standard models}",
    eprint = "hep-ph/9505326",
    archivePrefix = "arXiv",
    reportNumber = "SHEP-95-17, OUTP-95-19-P, OUTP-9519P",
    doi = "10.1103/PhysRevD.52.4183",
    journal = "Phys. Rev. D",
    volume = "52",
    pages = "4183--4216",
    year = "1995"
}

@article{Masip:1998jc,
    author = "Masip, M. and Munoz-Tapia, R. and Pomarol, A.",
    title = "{Limits on the mass of the lightest Higgs in supersymmetric models}",
    eprint = "hep-ph/9801437",
    archivePrefix = "arXiv",
    reportNumber = "UG-FT-84-97, UAB-FT-436",
    doi = "10.1103/PhysRevD.57.R5340",
    journal = "Phys. Rev. D",
    volume = "57",
    pages = "R5340",
    year = "1998"
}

@article{CMS:2021yci,
    author = "Tumasyan, Armen and others",
    collaboration = "CMS",
    title = "{Search for a heavy Higgs boson decaying into two lighter Higgs bosons in the $\tau\tau$bb final state at 13 TeV}",
    eprint = "2106.10361",
    archivePrefix = "arXiv",
    primaryClass = "hep-ex",
    reportNumber = "CMS-HIG-20-014, CERN-EP-2021-094",
    doi = "10.1007/JHEP11(2021)057",
    journal = "JHEP",
    volume = "11",
    pages = "057",
    year = "2021"
}

@article{CMS:2024yhz,
    author = "Hayrapetyan, Aram and others",
    collaboration = "CMS",
    title = "{Search for a standard model-like Higgs boson in the mass range between 70 and 110~GeV in the diphoton final state in proton-proton collisions at $\sqrt{s}=13~\mathrm{TeV}$}",
    eprint = "2405.18149",
    archivePrefix = "arXiv",
    primaryClass = "hep-ex",
    reportNumber = "CMS-HIG-20-002, CERN-EP-2024-088",
    doi = "10.1016/j.physletb.2024.139067",
    journal = "Phys. Lett. B",
    volume = "860",
    pages = "139067",
    year = "2025"
}

@article{CMS:2022dwd,
    author = "Tumasyan, Armen and others",
    collaboration = "CMS",
    title = "{A portrait of the Higgs boson by the CMS experiment ten years after the discovery}",
    eprint = "2207.00043",
    archivePrefix = "arXiv",
    primaryClass = "hep-ex",
    reportNumber = "CMS-HIG-22-001, CERN-EP-2022-039",
    doi = "10.1038/s41586-022-04892-x",
    journal = "Nature",
    volume = "607",
    number = "7917",
    pages = "60--68",
    year = "2022"
}

@article{ATLAS:2022vkf,
    collaboration = "ATLAS",
    title = "{A detailed map of Higgs boson interactions by the ATLAS experiment ten years after the discovery}",
    eprint = "2207.00092",
    archivePrefix = "arXiv",
    primaryClass = "hep-ex",
    reportNumber = "CERN-EP-2022-057",
    doi = "10.1038/s41586-022-04893-w",
    journal = "Nature",
    volume = "607",
    number = "7917",
    pages = "52--59",
    year = "2022",
    note = "[Erratum: Nature 612, E24 (2022)]"
}

@article{CMS:2022goy,
    author = "Tumasyan, Armen and others",
    collaboration = "CMS",
    title = "{Searches for additional Higgs bosons and for vector leptoquarks in $\tau\tau$ final states in proton-proton collisions at $\sqrt{s}$ = 13 TeV}",
    eprint = "2208.02717",
    archivePrefix = "arXiv",
    primaryClass = "hep-ex",
    reportNumber = "CMS-HIG-21-001, CERN-EP-2022-137",
    doi = "10.1007/JHEP07(2023)073",
    journal = "JHEP",
    volume = "07",
    pages = "073",
    year = "2023"
}

@misc{VPP2014,
author = "O'Leary, B. and Camargo-Molina, J. E.",
title = "$\textsf{VevaciousPlusPlus}$",
howpublished = {\url{https://github.com/JoseEliel/VevaciousPlusPlus}},
year = "2014"
}

@article{Camargo-Molina:2013qva,
    author = "Camargo-Molina, J. E. and O'Leary, B. and Porod, W. and Staub, F.",
    title = "{$\mathbf{Vevacious}$: A Tool For Finding The Global Minima Of One-Loop Effective Potentials With Many Scalars}",
    eprint = "1307.1477",
    archivePrefix = "arXiv",
    primaryClass = "hep-ph",
    doi = "10.1140/epjc/s10052-013-2588-2",
    journal = "Eur. Phys. J. C",
    volume = "73",
    number = "10",
    pages = "2588",
    year = "2013"
}

@article{Khosa:2020zar,
	author = "Khosa, Charanjit K. and Kraml, Sabine and Lessa, Andre and Neuhuber, Philipp and Waltenberger, Wolfgang",
	title = "{SModelS database update v1.2.3}",
	eprint = "2005.00555",
	archivePrefix = "arXiv",
	primaryClass = "hep-ph",
	doi = "10.31526/lhep.2020.158",
	month = "5",
	year = "2020"
}

@article{Altakach:2024jwk,
    author = "Altakach, Mohammad Mahdi and Kraml, Sabine and Lessa, Andre and Narasimha, Sahana and Pascal, Timoth{\'e}e and Ramos, Camila and Villamizar, Yoxara and Waltenberger, Wolfgang",
    title = "{SModelS v3: going beyond $ \mathcal{Z} _{2}$ topologies}",
    eprint = "2409.12942",
    archivePrefix = "arXiv",
    primaryClass = "hep-ph",
    doi = "10.1007/JHEP11(2024)074",
    journal = "JHEP",
    volume = "11",
    pages = "074",
    year = "2024"
}

@article{ATLAS:2018nud,
	author = "Aaboud, Morad and others",
	collaboration = "ATLAS",
	title = "{Search for photonic signatures of gauge-mediated supersymmetry in 13 TeV $pp$ collisions with the ATLAS detector}",
	eprint = "1802.03158",
	archivePrefix = "arXiv",
	primaryClass = "hep-ex",
	reportNumber = "CERN-EP-2017-323",
	doi = "10.1103/PhysRevD.97.092006",
	journal = "Phys. Rev. D",
	volume = "97",
	number = "9",
	pages = "092006",
	year = "2018"
}

@article{Biekotter:2023oen,
    author = {Biek{\"o}tter, Thomas and Heinemeyer, Sven and Weiglein, Georg},
    title = "{95.4~GeV diphoton excess at ATLAS and CMS}",
    eprint = "2306.03889",
    archivePrefix = "arXiv",
    primaryClass = "hep-ph",
    reportNumber = "KA-TP-11-2023, DESY-23-071, IFT--UAM/CSIC-23-062",
    doi = "10.1103/PhysRevD.109.035005",
    journal = "Phys. Rev. D",
    volume = "109",
    number = "3",
    pages = "035005",
    year = "2024"
}

@article{Iguro:2022dok,
    author = "Iguro, Syuhei and Kitahara, Teppei and Omura, Yuji",
    title = "{Scrutinizing the 95\textendash{}100 GeV di-tau excess in the top associated process}",
    eprint = "2205.03187",
    archivePrefix = "arXiv",
    primaryClass = "hep-ph",
    reportNumber = "P3H-22-047, TTP22-027, KEK-TH-2424",
    doi = "10.1140/epjc/s10052-022-11028-y",
    journal = "Eur. Phys. J. C",
    volume = "82",
    number = "11",
    pages = "1053",
    year = "2022"
}

@misc{CMS:2015ocq,
    collaboration = "CMS",
    title = "{Search for new resonances in the diphoton final state in the mass range between 80 and 115 GeV in pp collisions at $\sqrt{s}=8$ TeV}",
    howpublished = {\href{https://cds.cern.ch/record/2063739}{\rm CMS-PAS-HIG-14-037}},
    year = "2015"
}

@article{Azatov:2012bz,
    author = "Azatov, Aleksandr and Contino, Roberto and Galloway, Jamison",
    title = "{Model-Independent Bounds on a Light Higgs}",
    eprint = "1202.3415",
    archivePrefix = "arXiv",
    primaryClass = "hep-ph",
    doi = "10.1007/JHEP04(2012)127",
    journal = "JHEP",
    volume = "04",
    pages = "127",
    year = "2012",
    note = "[Erratum: JHEP 04, 140 (2013)]"
}

@article{LEPWorkingGroupforHiggsbosonsearches:2003ing,
    author = "Barate, R. and others",
    collaboration = "LEP Working Group for Higgs boson searches, ALEPH, DELPHI, L3, OPAL",
    title = "{Search for the standard model Higgs boson at LEP}",
    eprint = "hep-ex/0306033",
    archivePrefix = "arXiv",
    reportNumber = "CERN-EP-2003-011",
    doi = "10.1016/S0370-2693(03)00614-2",
    journal = "Phys. Lett. B",
    volume = "565",
    pages = "61--75",
    year = "2003"
}

@misc{Arcangeletti,
    author = {Arcangeletti, Chiara},
    howpublished = "{on behalf of ATLAS collaboration, LHC Seminar} \url{https://indico.cern.ch/event/1281604/attachments/2660420/4608571/LHCSeminarArcangeletti_final.pdf}",
    month = "7$^{th}$ of June",
    year = "2023"
}

@article{Wang:2018vxp,
    author = "Wang, Kun and Wang, Fei and Zhu, Jingya and Jie, Quanlin",
    title = "{The semi-constrained NMSSM in light of muon g-2, LHC, and dark matter constraints}",
    eprint = "1811.04435",
    archivePrefix = "arXiv",
    primaryClass = "hep-ph",
    doi = "10.1088/1674-1137/42/10/103109",
    journal = "Chin. Phys. C",
    volume = "42",
    number = "10",
    pages = "103109--103109",
    year = "2018"
}

@article{Heinemeyer:2018wzl,
    author = "Heinemeyer, Sven and Stefaniak, T.",
    title = "{A Higgs Boson at 96 GeV?!}",
    eprint = "1812.05864",
    archivePrefix = "arXiv",
    primaryClass = "hep-ph",
    reportNumber = "IFT-UAM/CSIC-18-125, DESY-18-221",
    doi = "10.22323/1.339.0016",
    journal = "PoS",
    volume = "CHARGED2018",
    pages = "016",
    year = "2019"
}

@article{Heinemeyer:2018jcd,
    author = "Heinemeyer, S.",
    title = "{A Higgs boson below 125 GeV?!}",
    doi = "10.1142/S0217751X18440062",
    journal = "Int. J. Mod. Phys. A",
    volume = "33",
    number = "31",
    pages = "1844006",
    year = "2018"
}

@article{Sachdeva:2019hvk,
    author = "Sachdeva, Divya and Sadhukhan, Soumya",
    title = "{Discussing 125 GeV and 95 GeV excess in light radion model}",
    eprint = "1908.01668",
    archivePrefix = "arXiv",
    primaryClass = "hep-ph",
    doi = "10.1103/PhysRevD.101.055045",
    journal = "Phys. Rev. D",
    volume = "101",
    number = "5",
    pages = "055045",
    year = "2020"
}

@article{Vega:2018ddp,
    author = "Vega, Roberto and Vega-Morales, Roberto and Xie, Keping",
    title = "{Light (and darkness) from a light hidden Higgs}",
    eprint = "1805.01970",
    archivePrefix = "arXiv",
    primaryClass = "hep-ph",
    reportNumber = "UG-FT-327-18, CAFPE-197-18, UG-FT 327/18, CAFPE 197/18, SMU-HEP-18-08, FERMILAB-PUB-18-159-T",
    doi = "10.1007/JHEP06(2018)137",
    journal = "JHEP",
    volume = "06",
    pages = "137",
    year = "2018"
}

@article{Fox:2017uwr,
    author = "Fox, Patrick J. and Weiner, Neal",
    title = "{Light Signals from a Lighter Higgs}",
    eprint = "1710.07649",
    archivePrefix = "arXiv",
    primaryClass = "hep-ph",
    reportNumber = "FERMILAB-PUB-17-468-T",
    doi = "10.1007/JHEP08(2018)025",
    journal = "JHEP",
    volume = "08",
    pages = "025",
    year = "2018"
}

@article{Kundu:2019nqo,
    author = "Kundu, Anirban and Maharana, Suvam and Mondal, Poulami",
    title = "{A 96 GeV scalar tagged to dark matter models}",
    eprint = "1907.12808",
    archivePrefix = "arXiv",
    primaryClass = "hep-ph",
    doi = "10.1016/j.nuclphysb.2020.115057",
    journal = "Nucl. Phys. B",
    volume = "955",
    pages = "115057",
    year = "2020"
}

@article{Fan:2013gjf,
    author = "Fan, Jia-Wei and Tao, Jun-Quan and Shen, Yu-Qiao and Chen, Guo-Ming and Chen, He-Sheng and Gascon-Shotkin, S. and Lethuillier, M. and Sgandurra, L. and Soulet, P.",
    title = "{Study of diphoton decays of the lightest scalar Higgs boson in the Next-to-Minimal Supersymmetric Standard Model}",
    eprint = "1309.6394",
    archivePrefix = "arXiv",
    primaryClass = "hep-ph",
    doi = "10.1088/1674-1137/38/7/073101",
    journal = "Chin. Phys. C",
    volume = "38",
    pages = "073101",
    year = "2014"
}

@article{Belyaev:2023xnv,
    author = "Belyaev, Alexander and Benbrik, Rachid and Boukidi, Mohammed and Chakraborti, Manimala and Moretti, Stefano and Semlali, Souad",
    title = "{Explanation of the hints for a 95 GeV Higgs boson within a 2-Higgs Doublet Model}",
    eprint = "2306.09029",
    archivePrefix = "arXiv",
    primaryClass = "hep-ph",
    doi = "10.1007/JHEP05(2024)209",
    journal = "JHEP",
    volume = "05",
    pages = "209",
    year = "2024"
}

@article{Azevedo:2023zkg,
    author = {Azevedo, Duarte and Biek{\"o}tter, Thomas and Ferreira, P. M.},
    title = "{2HDM interpretations of the CMS diphoton excess at 95 GeV}",
    eprint = "2305.19716",
    archivePrefix = "arXiv",
    primaryClass = "hep-ph",
    reportNumber = "KA-TP-10-2023",
    doi = "10.1007/JHEP11(2023)017",
    journal = "JHEP",
    volume = "11",
    pages = "017",
    year = "2023"
}

@article{Benbrik:2022dja,
    author = "Benbrik, Rachid and Boukidi, Mohammed and Manaut, Bouzid",
    title = "{Interpreting the W-mass and muon $(g_{\mu} - 2)$ anomalies within a 2-Higgs doublet model}",
    eprint = "2204.11755",
    archivePrefix = "arXiv",
    primaryClass = "hep-ph",
    doi = "10.1016/j.nuclphysb.2024.116593",
    journal = "Nucl. Phys. B",
    volume = "1005",
    pages = "116593",
    year = "2024"
}

@article{Benbrik:2022azi,
    author = "Benbrik, Rachid and Boukidi, Mohammed and Moretti, Stefano and Semlali, Souad",
    title = "{Explaining the 96 GeV Di-photon anomaly in a generic 2HDM Type-III}",
    eprint = "2204.07470",
    archivePrefix = "arXiv",
    primaryClass = "hep-ph",
    doi = "10.1016/j.physletb.2022.137245",
    journal = "Phys. Lett. B",
    volume = "832",
    pages = "137245",
    year = "2022"
}

@article{Haisch:2017gql,
    author = "Haisch, Ulrich and Malinauskas, Augustinas",
    title = "{Let there be light from a second light Higgs doublet}",
    eprint = "1712.06599",
    archivePrefix = "arXiv",
    primaryClass = "hep-ph",
    reportNumber = "CERN-TH-2017-276",
    doi = "10.1007/JHEP03(2018)135",
    journal = "JHEP",
    volume = "03",
    pages = "135",
    year = "2018"
}

@article{Ashanujjaman:2023etj,
    author = "Ashanujjaman, Saiyad and Banik, Sumit and Coloretti, Guglielmo and Crivellin, Andreas and Mellado, Bruce and Mulaudzi, Anza-Tshilidzi",
    title = "{SU(2)L triplet scalar as the origin of the 95~GeV excess?}",
    eprint = "2306.15722",
    archivePrefix = "arXiv",
    primaryClass = "hep-ph",
    reportNumber = "PSI-PR-23-20, ZU-TH 28/23, ICPP-71",
    doi = "10.1103/PhysRevD.108.L091704",
    journal = "Phys. Rev. D",
    volume = "108",
    number = "9",
    pages = "L091704",
    year = "2023"
}

@article{Aguilar-Saavedra:2020wrj,
    author = "Aguilar-Saavedra, Juan Antonio and Joaquim, Filipe Rafael",
    title = "{Multiphoton signals of a (96 GeV?) stealth boson}",
    eprint = "2002.07697",
    archivePrefix = "arXiv",
    primaryClass = "hep-ph",
    reportNumber = "IFT-UAM/CSIC-19-153, CFTP/20-002",
    doi = "10.1140/epjc/s10052-020-7952-4",
    journal = "Eur. Phys. J. C",
    volume = "80",
    number = "5",
    pages = "403",
    year = "2020"
}

@article{Biekotter:2022jyr,
    author = {Biek\"otter, Thomas and Heinemeyer, Sven and Weiglein, Georg},
    title = "{Mounting evidence for a 95 GeV Higgs boson}",
    eprint = "2203.13180",
    archivePrefix = "arXiv",
    primaryClass = "hep-ph",
    reportNumber = "DESY 22-057, IFT-UAM/CSIC-22-033, IFT-UAM/CSIC--22--033",
    doi = "10.1007/JHEP08(2022)201",
    journal = "JHEP",
    volume = "08",
    pages = "201",
    year = "2022"
}

@phdthesis{Li:2023hsr,
    author = "Li, Cheng",
    title = "{Phenomenology of extended Two-Higgs-Doublets models}",
    doi = "10.3204/PUBDB-2023-03151",
    school = "Hamburg U.",
    year = "2023"
}

@article{Biekotter:2021ovi,
    author = {Biek\"otter, Thomas and Olea-Romacho, Mar\'\i{}a Olalla},
    title = "{Reconciling Higgs physics and pseudo-Nambu-Goldstone dark matter in the S2HDM using a genetic algorithm}",
    eprint = "2108.10864",
    archivePrefix = "arXiv",
    primaryClass = "hep-ph",
    reportNumber = "DESY-21-125",
    doi = "10.1007/JHEP10(2021)215",
    journal = "JHEP",
    volume = "10",
    pages = "215",
    year = "2021"
}

@article{Heinemeyer:2021msz,
    author = "Heinemeyer, S. and Li, C. and Lika, F. and Moortgat-Pick, G. and Paasch, S.",
    title = "{Phenomenology of a 96~GeV Higgs boson in the 2HDM with an additional singlet}",
    eprint = "2112.11958",
    archivePrefix = "arXiv",
    primaryClass = "hep-ph",
    reportNumber = "DESY 21-230, IFT-UAM/CSIC-21-158",
    doi = "10.1103/PhysRevD.106.075003",
    journal = "Phys. Rev. D",
    volume = "106",
    number = "7",
    pages = "075003",
    year = "2022"
}

@article{Biekotter:2019mib,
    author = {Biek\"otter, Thomas and Chakraborti, M. and Heinemeyer, Sven},
    editor = "Anagnostopoulos, Konstantinos and others",
    title = "{An N2HDM Solution for the possible 96 GeV Excess}",
    eprint = "1905.03280",
    archivePrefix = "arXiv",
    primaryClass = "hep-ph",
    reportNumber = "IFT-UAM/CSIC-19-064",
    doi = "10.22323/1.347.0015",
    journal = "PoS",
    volume = "CORFU2018",
    pages = "015",
    year = "2019"
}

@article{Biekotter:2019kde,
    author = {Biek\"otter, T. and Chakraborti, M. and Heinemeyer, S.},
    title = "{A 96 GeV Higgs boson in the N2HDM}",
    eprint = "1903.11661",
    archivePrefix = "arXiv",
    primaryClass = "hep-ph",
    reportNumber = "IFT-UAM/CSIC-19-034",
    doi = "10.1140/epjc/s10052-019-7561-2",
    journal = "Eur. Phys. J. C",
    volume = "80",
    number = "1",
    pages = "2",
    year = "2020"
}

@article{Biekotter:2019gtq,
    author = {Biek\"otter, T. and Heinemeyer, S. and Mu\~noz, C.},
    title = "{Precise prediction for the Higgs-Boson masses in the $\mu \nu $SSM with three right-handed neutrino superfields}",
    eprint = "1906.06173",
    archivePrefix = "arXiv",
    primaryClass = "hep-ph",
    reportNumber = "IFT--UAM/CSIC--19-030",
    doi = "10.1140/epjc/s10052-019-7175-8",
    journal = "Eur. Phys. J. C",
    volume = "79",
    number = "8",
    pages = "667",
    year = "2019"
}

@article{Biekotter:2017xmf,
    author = {Biek\"otter, T. and Heinemeyer, S. and Mu\~noz, C.},
    title = "{Precise prediction for the Higgs-boson masses in the $\mu \nu $ SSM}",
    eprint = "1712.07475",
    archivePrefix = "arXiv",
    primaryClass = "hep-ph",
    reportNumber = "IFT--UAM-CSIC--17-118",
    doi = "10.1140/epjc/s10052-018-5978-7",
    journal = "Eur. Phys. J. C",
    volume = "78",
    number = "6",
    pages = "504",
    year = "2018"
}

@article{Biekotter:2020cjs,
    author = {Biek\"otter, T. and Chakraborti, M. and Heinemeyer, S.},
    title = "{The \textquotedblleft{}96 GeV excess\textquotedblright{} at the LHC}",
    eprint = "2003.05422",
    archivePrefix = "arXiv",
    primaryClass = "hep-ph",
    reportNumber = "IFT-UAM/CSIC-20-041, DESY 20-047, DESY-20-047",
    doi = "10.1142/S0217751X21420185",
    journal = "Int. J. Mod. Phys. A",
    volume = "36",
    number = "22",
    pages = "2142018",
    year = "2021"
}

@article{Biekotter:2021qbc,
    author = {Biek\"otter, Thomas and Grohsjean, Alexander and Heinemeyer, Sven and Schwanenberger, Christian and Weiglein, Georg},
    title = "{Possible indications for new Higgs bosons in the reach of the LHC: N2HDM and NMSSM interpretations}",
    eprint = "2109.01128",
    archivePrefix = "arXiv",
    primaryClass = "hep-ph",
    reportNumber = "IFT--UAM/CSIC--21-041, IFT-UAM/CSIC-21-041, DESY 21-132",
    doi = "10.1140/epjc/s10052-022-10099-1",
    journal = "Eur. Phys. J. C",
    volume = "82",
    number = "2",
    pages = "178",
    year = "2022"
}

@article{Carena:2015moc,
    author = "Carena, Marcela and Haber, Howard E. and Low, Ian and Shah, Nausheen R. and Wagner, Carlos E. M.",
    title = "{Alignment limit of the NMSSM Higgs sector}",
    eprint = "1510.09137",
    archivePrefix = "arXiv",
    primaryClass = "hep-ph",
    reportNumber = "FERMILAB-PUB-15-407-T, EFI-15-32, MCTP-15-15, SCIPP-15-12, WSU-HEP-1505",
    doi = "10.1103/PhysRevD.93.035013",
    journal = "Phys. Rev. D",
    volume = "93",
    number = "3",
    pages = "035013",
    year = "2016"
}

@article{Hollik:2018yek,
    author = "Hollik, Wolfgang Gregor and Liebler, Stefan and Moortgat-Pick, Gudrid and Pa\ss{}ehr, Sebastian and Weiglein, Georg",
    title = "{Phenomenology of the inflation-inspired NMSSM at the electroweak scale}",
    eprint = "1809.07371",
    archivePrefix = "arXiv",
    primaryClass = "hep-ph",
    reportNumber = "DESY-17-075, KA-TP-26-2018, TTP18--035",
    doi = "10.1140/epjc/s10052-019-6561-6",
    journal = "Eur. Phys. J. C",
    volume = "79",
    number = "1",
    pages = "75",
    year = "2019"
}

@article{Hollik:2020plc,
    author = "Hollik, Wolfgang Gregor and Li, Cheng and Moortgat-Pick, Gudrid and Paasch, Steven",
    title = "{Phenomenology of a Supersymmetric Model Inspired by Inflation}",
    eprint = "2004.14852",
    archivePrefix = "arXiv",
    primaryClass = "hep-ph",
    reportNumber = "DESY-20-059, TTP-2020-017, P3H-20-013",
    doi = "10.1140/epjc/s10052-021-08869-4",
    journal = "Eur. Phys. J. C",
    volume = "81",
    number = "2",
    pages = "141",
    year = "2021"
}

@article{Lee:2010hj,
    author = "Lee, Hyun Min",
    title = "{Chaotic inflation in Jordan frame supergravity}",
    eprint = "1005.2735",
    archivePrefix = "arXiv",
    primaryClass = "hep-ph",
    reportNumber = "CERN-PH-TH-2010-102",
    doi = "10.1088/1475-7516/2010/08/003",
    journal = "JCAP",
    volume = "08",
    pages = "003",
    year = "2010"
}

@article{Domingo:2018uim,
    author = "Domingo, Florian and Heinemeyer, Sven and Pa\ss{}ehr, Sebastian and Weiglein, Georg",
    title = "{Decays of the neutral Higgs bosons into SM fermions and gauge bosons in the $\mathcal{CP}$-violating NMSSM}",
    eprint = "1807.06322",
    archivePrefix = "arXiv",
    primaryClass = "hep-ph",
    reportNumber = "DESY-18-084, IFT-UAM/CSIC-17-125, DESY--18--084, IFT--UAM/CSIC--17--125",
    doi = "10.1140/epjc/s10052-018-6400-1",
    journal = "Eur. Phys. J. C",
    volume = "78",
    number = "11",
    pages = "942",
    year = "2018"
}

@article{Li:2022etb,
    author = "Li, Weichao and Qiao, Haoxue and Zhu, Jingya",
    title = "{Light Higgs boson in the NMSSM confronted with the CMS di-photon and di-tau excesses*}",
    eprint = "2212.11739",
    archivePrefix = "arXiv",
    primaryClass = "hep-ph",
    doi = "10.1088/1674-1137/acfaf1",
    journal = "Chin. Phys. C",
    volume = "47",
    number = "12",
    pages = "123102",
    year = "2023"
}

@article{Abdelalim:2020xfk,
    author = "Abdelalim, Ahmed Ali and Das, Biswaranjan and Khalil, Shaaban and Moretti, Stefano",
    title = "{Di-photon decay of a light Higgs state in the BLSSM}",
    eprint = "2012.04952",
    archivePrefix = "arXiv",
    primaryClass = "hep-ph",
    doi = "10.1016/j.nuclphysb.2022.116013",
    journal = "Nucl. Phys. B",
    volume = "985",
    pages = "116013",
    year = "2022"
}

@article{Beskidt:2017dil,
    author = "Beskidt, C. and de Boer, W. and Kazakov, D. I.",
    title = "{Can we discover a light singlet-like NMSSM Higgs boson at the LHC?}",
    eprint = "1712.02531",
    archivePrefix = "arXiv",
    primaryClass = "hep-ph",
    doi = "10.1016/j.physletb.2018.04.067",
    journal = "Phys. Lett. B",
    volume = "782",
    pages = "69--76",
    year = "2018"
}

@article{Choi:2019yrv,
    author = "Choi, Kiwoon and Im, Sang Hui and Jeong, Kwang Sik and Park, Chan Beom",
    title = "{Light Higgs bosons in the general NMSSM}",
    eprint = "1906.03389",
    archivePrefix = "arXiv",
    primaryClass = "hep-ph",
    reportNumber = "CTPU-PTC-19-17, PNUTP-19-A11",
    doi = "10.1140/epjc/s10052-019-7473-1",
    journal = "Eur. Phys. J. C",
    volume = "79",
    number = "11",
    pages = "956",
    year = "2019"
}

@article{Martin:1997ns,
  author        = {Martin, Stephen P.},
  title         = {{A Supersymmetry primer}},
  year          = {1997},
  pages         = {1-98},
  doi           = {10.1142/9789812839657_0001, 10.1142/9789814307505_0001},
  note          = {[Adv. Ser. Direct. High Energy Phys.18,1(1998)]},
  eprint        = {hep-ph/9709356},
  archiveprefix = {arXiv},
  primaryclass  = {hep-ph},
  reportnumber  = {FERMILAB-PUB-97-425-T},
  slaccitation  = {%%CITATION = HEP-PH/9709356;%%}
}

@article{Maniatis:2009re,
    author = "Maniatis, M.",
    title = "{The Next-to-Minimal Supersymmetric extension of the Standard Model reviewed}",
    eprint = "0906.0777",
    archivePrefix = "arXiv",
    primaryClass = "hep-ph",
    reportNumber = "HD-THEP-09-9",
    doi = "10.1142/S0217751X10049827",
    journal = "Int. J. Mod. Phys. A",
    volume = "25",
    pages = "3505--3602",
    year = "2010"
}

@article{Abel:1996cr,
    author = "Abel, S. A.",
    title = "{Destabilizing divergences in the NMSSM}",
    eprint = "hep-ph/9609323",
    archivePrefix = "arXiv",
    reportNumber = "ULB-TH-96-16",
    doi = "10.1016/S0550-3213(96)00470-1",
    journal = "Nucl. Phys. B",
    volume = "480",
    pages = "55--72",
    year = "1996"
}

@article{Panagiotakopoulos:1998yw,
    author = "Panagiotakopoulos, C. and Tamvakis, K.",
    title = "{Stabilized NMSSM without domain walls}",
    eprint = "hep-ph/9809475",
    archivePrefix = "arXiv",
    doi = "10.1016/S0370-2693(98)01493-2",
    journal = "Phys. Lett. B",
    volume = "446",
    pages = "224--227",
    year = "1999"
}

@article{Kolda:1998rm,
    author = "Kolda, Christopher F. and Pokorski, Stefan and Polonsky, Nir",
    title = "{Stabilized singlets in supergravity as a source of the mu - parameter}",
    eprint = "hep-ph/9803310",
    archivePrefix = "arXiv",
    reportNumber = "IASSNS-HEP-97-137, CERN-TH-98-75, RU-97-97",
    doi = "10.1103/PhysRevLett.80.5263",
    journal = "Phys. Rev. Lett.",
    volume = "80",
    pages = "5263--5266",
    year = "1998"
}

@article{Ellwanger:1983mg,
    author = "Ellwanger, U.",
    title = "{Nonrenormalizable interactions from supergravity, quantum corrections and effecive low-energy theories}",
    doi = "10.1016/0370-2693(83)90557-9",
    journal = "Phys. Lett. B",
    volume = "133",
    pages = "187--191",
    year = "1983"
}

@article{Ellwanger:2009dp,
    author = "Ellwanger, Ulrich and Hugonie, Cyril and Teixeira, Ana M.",
    title = "{The Next-to-Minimal Supersymmetric Standard Model}",
    eprint = "0910.1785",
    archivePrefix = "arXiv",
    primaryClass = "hep-ph",
    reportNumber = "LPT-ORSAY-09-76, CFTP-09-032, LPTA-09-066",
    doi = "10.1016/j.physrep.2010.07.001",
    journal = "Phys. Rept.",
    volume = "496",
    pages = "1--77",
    year = "2010"
}

@article{Miller:2003ay,
    author = "Miller, D. J. and Nevzorov, R. and Zerwas, P. M.",
    title = "{The Higgs sector of the next-to-minimal supersymmetric standard model}",
    eprint = "hep-ph/0304049",
    archivePrefix = "arXiv",
    reportNumber = "CERN-TH-2003-077, DESY-03-066, ITEP-5-03",
    doi = "10.1016/j.nuclphysb.2003.12.021",
    journal = "Nucl. Phys. B",
    volume = "681",
    pages = "3--30",
    year = "2004"
}

@article{Fowlie:2016hew,
    author = "Fowlie, Andrew and Bardsley, Michael Hugh",
    title = "{Superplot: a graphical interface for plotting and analysing MultiNest output}",
    eprint = "1603.00555",
    archivePrefix = "arXiv",
    primaryClass = "physics.data-an",
    reportNumber = "COEPP-MN-16-5",
    doi = "10.1140/epjp/i2016-16391-0",
    journal = "Eur. Phys. J. Plus",
    volume = "131",
    number = "11",
    pages = "391",
    year = "2016"
}

@article{SARAH_Staub2015,
    author = "Staub, Florian",
    title = "{Exploring new models in all detail with SARAH}",
    eprint = "1503.04200",
    archivePrefix = "arXiv",
    primaryClass = "hep-ph",
    reportNumber = "CERN-PH-TH-2015-051",
    doi = "10.1155/2015/840780",
    journal = "Adv. High Energy Phys.",
    volume = "2015",
    pages = "840780",
    year = "2015"
}

@article{SARAH4_Staub2013,
    author = "Staub, Florian",
    title = "{SARAH 4 : A tool for (not only SUSY) model builders}",
    eprint = "1309.7223",
    archivePrefix = "arXiv",
    primaryClass = "hep-ph",
    reportNumber = "BONN-TH-2013-17",
    doi = "10.1016/j.cpc.2014.02.018",
    journal = "Comput. Phys. Commun.",
    volume = "185",
    pages = "1773--1790",
    year = "2014"
}

@article{SARAH3_Staub2012,
    author = "Staub, Florian",
    title = "{SARAH 3.2: Dirac Gauginos, UFO output, and more}",
    eprint = "1207.0906",
    archivePrefix = "arXiv",
    primaryClass = "hep-ph",
    reportNumber = "BONN-TH-2012-17",
    doi = "10.1016/j.cpc.2013.02.019",
    journal = "Comput. Phys. Commun.",
    volume = "184",
    pages = "1792--1809",
    year = "2013"
}

@article{SARAH_Staub2008,
    author = "Staub, F.",
    title = "{SARAH}",
    eprint = "0806.0538",
    archivePrefix = "arXiv",
    primaryClass = "hep-ph",
    month = "6",
    year = "2008"
}

@article{Porod2003SPheno,
    author = "Porod, Werner",
    title = "{SPheno, a program for calculating supersymmetric spectra, SUSY particle decays and SUSY particle production at e+ e- colliders}",
    eprint = "hep-ph/0301101",
    archivePrefix = "arXiv",
    reportNumber = "ZU-TH-01-03",
    doi = "10.1016/S0010-4655(03)00222-4",
    journal = "Comput. Phys. Commun.",
    volume = "153",
    pages = "275--315",
    year = "2003"
}

@article{Porod2011SPheno3,
    author = "Porod, W. and Staub, F.",
    title = "{SPheno 3.1: Extensions including flavour, CP-phases and models beyond the MSSM}",
    eprint = "1104.1573",
    archivePrefix = "arXiv",
    primaryClass = "hep-ph",
    doi = "10.1016/j.cpc.2012.05.021",
    journal = "Comput. Phys. Commun.",
    volume = "183",
    pages = "2458--2469",
    year = "2012"
}

@article{Porod:2014xia,
    author = "Porod, Werner and Staub, Florian and Vicente, Avelino",
    title = "{A Flavor Kit for BSM models}",
    eprint = "1405.1434",
    archivePrefix = "arXiv",
    primaryClass = "hep-ph",
    reportNumber = "BONN-TH-2014-07",
    doi = "10.1140/epjc/s10052-014-2992-2",
    journal = "Eur. Phys. J. C",
    volume = "74",
    number = "8",
    pages = "2992",
    year = "2014"
}

@article{MultiNest2009,
    author = "Feroz, F. and Hobson, M. P. and Bridges, M.",
    title = "{MultiNest: an efficient and robust Bayesian inference tool for cosmology and particle physics}",
    eprint = "0809.3437",
    archivePrefix = "arXiv",
    primaryClass = "astro-ph",
    doi = "10.1111/j.1365-2966.2009.14548.x",
    journal = "Mon. Not. Roy. Astron. Soc.",
    volume = "398",
    pages = "1601--1614",
    year = "2009"
}

@article{Importance2019,
    author = "Feroz, F. and Hobson, M. P. and Cameron, E. and Pettitt, A. N.",
    title = "{Importance Nested Sampling and the MultiNest Algorithm}",
    eprint = "1306.2144",
    archivePrefix = "arXiv",
    primaryClass = "astro-ph.IM",
    doi = "10.21105/astro.1306.2144",
    journal = "Open J. Astrophys.",
    volume = "2",
    number = "1",
    pages = "10",
    year = "2019"
}

@article{HB2008jh,
    author = "Bechtle, Philip and Brein, Oliver and Heinemeyer, Sven and Weiglein, Georg and Williams, Karina E.",
    title = "{HiggsBounds: Confronting Arbitrary Higgs Sectors with Exclusion Bounds from LEP and the Tevatron}",
    eprint = "0811.4169",
    archivePrefix = "arXiv",
    primaryClass = "hep-ph",
    reportNumber = "DCPT-08-172, IPPP-08-86, BONN-TH-2008-17",
    doi = "10.1016/j.cpc.2009.09.003",
    journal = "Comput. Phys. Commun.",
    volume = "181",
    pages = "138--167",
    year = "2010"
}

@article{HB2011sb,
    author = "Bechtle, Philip and Brein, Oliver and Heinemeyer, Sven and Weiglein, Georg and Williams, Karina E.",
    title = "{HiggsBounds 2.0.0: Confronting Neutral and Charged Higgs Sector Predictions with Exclusion Bounds from LEP and the Tevatron}",
    eprint = "1102.1898",
    archivePrefix = "arXiv",
    primaryClass = "hep-ph",
    reportNumber = "FR-PHENO-2011-002, BONN-TH-2011-02, DESY-11-016",
    doi = "10.1016/j.cpc.2011.07.015",
    journal = "Comput. Phys. Commun.",
    volume = "182",
    pages = "2605--2631",
    year = "2011"
}

@article{HBHS2012lvg,
    author = "Bechtle, Philip and Brein, Oliver and Heinemeyer, Sven and Stal, Oscar and Stefaniak, Tim and Weiglein, Georg and Williams, Karina",
    editor = "Enberg, Rikard and Ferrari, Arnaud",
    title = "{Recent Developments in HiggsBounds and a Preview of HiggsSignals}",
    eprint = "1301.2345",
    archivePrefix = "arXiv",
    primaryClass = "hep-ph",
    reportNumber = "BONN-TH-2013-01, DESY-13-004",
    doi = "10.22323/1.156.0024",
    journal = "PoS",
    volume = "CHARGED2012",
    pages = "024",
    year = "2012"
}

@article{HB2013wla,
    author = "Bechtle, Philip and Brein, Oliver and Heinemeyer, Sven and St\r{a}l, Oscar and Stefaniak, Tim and Weiglein, Georg and Williams, Karina E.",
    title = "{$\mathsf{HiggsBounds}-4$: Improved Tests of Extended Higgs Sectors against Exclusion Bounds from LEP, the Tevatron and the LHC}",
    eprint = "1311.0055",
    archivePrefix = "arXiv",
    primaryClass = "hep-ph",
    reportNumber = "BONN-TH-2013-21, DESY-13-110",
    doi = "10.1140/epjc/s10052-013-2693-2",
    journal = "Eur. Phys. J. C",
    volume = "74",
    number = "3",
    pages = "2693",
    year = "2014"
}

@article{HB2020pkv,
    author = "Bechtle, Philip and Dercks, Daniel and Heinemeyer, Sven and Klingl, Tobias and Stefaniak, Tim and Weiglein, Georg and Wittbrodt, Jonas",
    title = "{HiggsBounds-5: Testing Higgs Sectors in the LHC 13 TeV Era}",
    eprint = "2006.06007",
    archivePrefix = "arXiv",
    primaryClass = "hep-ph",
    reportNumber = "BONN-TH-2020-03, DESY 20-093, DESY-20-093, IFT-UAM/CSIC-20-072, LU 20-27",
    doi = "10.1140/epjc/s10052-020-08557-9",
    journal = "Eur. Phys. J. C",
    volume = "80",
    number = "12",
    pages = "1211",
    year = "2020"
}

@article{HS2013xfa,
    author = "Bechtle, Philip and Heinemeyer, Sven and St\r{a}l, Oscar and Stefaniak, Tim and Weiglein, Georg",
    title = "{$HiggsSignals$: Confronting arbitrary Higgs sectors with measurements at the Tevatron and the LHC}",
    eprint = "1305.1933",
    archivePrefix = "arXiv",
    primaryClass = "hep-ph",
    reportNumber = "BONN-TH-2013-07, DESY-13-078",
    doi = "10.1140/epjc/s10052-013-2711-4",
    journal = "Eur. Phys. J. C",
    volume = "74",
    number = "2",
    pages = "2711",
    year = "2014"
}

@article{HSConstraining2013hwa,
    author = "St\r{a}l, Oscar and Stefaniak, Tim",
    title = "{Constraining extended Higgs sectors with HiggsSignals}",
    eprint = "1310.4039",
    archivePrefix = "arXiv",
    primaryClass = "hep-ph",
    reportNumber = "BONN-TH-2013-20",
    doi = "10.22323/1.180.0314",
    journal = "PoS",
    volume = "EPS-HEP2013",
    pages = "314",
    year = "2013"
}

@article{HS2014ewa,
    author = "Bechtle, Philip and Heinemeyer, Sven and St\r{a}l, Oscar and Stefaniak, Tim and Weiglein, Georg",
    title = "{Probing the Standard Model with Higgs signal rates from the Tevatron, the LHC and a future ILC}",
    eprint = "1403.1582",
    archivePrefix = "arXiv",
    primaryClass = "hep-ph",
    reportNumber = "DESY-14-026, BONN-TH-2014-05",
    doi = "10.1007/JHEP11(2014)039",
    journal = "JHEP",
    volume = "11",
    pages = "039",
    year = "2014"
}

@article{HS2020uwn,
    author = "Bechtle, Philip and Heinemeyer, Sven and Klingl, Tobias and Stefaniak, Tim and Weiglein, Georg and Wittbrodt, Jonas",
    title = "{HiggsSignals-2: Probing new physics with precision Higgs measurements in the LHC 13 TeV era}",
    eprint = "2012.09197",
    archivePrefix = "arXiv",
    primaryClass = "hep-ph",
    reportNumber = "BONN-TH-2020-09, DESY-20-228, DESY 20-228, IFT-UAM/CSIC-20-081, LU TP 20-53",
    doi = "10.1140/epjc/s10052-021-08942-y",
    journal = "Eur. Phys. J. C",
    volume = "81",
    number = "2",
    pages = "145",
    year = "2021"
}

@article{ParticleDataGroup:2024cfk,
    author = "Navas, S. and others",
    collaboration = "Particle Data Group",
    title = "{Review of particle physics}",
    doi = "10.1103/PhysRevD.110.030001",
    journal = "Phys. Rev. D",
    volume = "110",
    number = "3",
    pages = "030001",
    year = "2024"
}

@article{ATLAS:2019lng,
    author = "Aad, Georges and others",
    collaboration = "ATLAS",
    title = "{Searches for electroweak production of supersymmetric particles with compressed mass spectra in $\sqrt{s}=$ 13 TeV $pp$ collisions with the ATLAS detector}",
    eprint = "1911.12606",
    archivePrefix = "arXiv",
    primaryClass = "hep-ex",
    reportNumber = "CERN-EP-2019-242",
    doi = "10.1103/PhysRevD.101.052005",
    journal = "Phys. Rev. D",
    volume = "101",
    number = "5",
    pages = "052005",
    year = "2020"
}

@article{Lee:2010gv,
    author = "Lee, Hyun Min and Raby, Stuart and Ratz, Michael and Ross, Graham G. and Schieren, Roland and Schmidt-Hoberg, Kai and Vaudrevange, Patrick K. S.",
    title = "{A unique $\mathbb{Z}_4^R$ symmetry for the MSSM}",
    eprint = "1009.0905",
    archivePrefix = "arXiv",
    primaryClass = "hep-ph",
    reportNumber = "TUM-HEP-770-10, LMU-ASC-64-10, OHSTPY-HEP-T-10-003, CERN-PH-TH-2010-193, OUTP-10-24P",
    doi = "10.1016/j.physletb.2010.10.038",
    journal = "Phys. Lett. B",
    volume = "694",
    pages = "491--495",
    year = "2011"
}

@article{Lee:2011dya,
    author = "Lee, Hyun Min and Raby, Stuart and Ratz, Michael and Ross, Graham G. and Schieren, Roland and Schmidt-Hoberg, Kai and Vaudrevange, Patrick K. S.",
    title = "{Discrete R symmetries for the MSSM and its singlet extensions}",
    eprint = "1102.3595",
    archivePrefix = "arXiv",
    primaryClass = "hep-ph",
    reportNumber = "TUM-HEP-793-11, LMU-ASC-06-11, OHSTPY-HEP-T-11-001, CERN-PH-TH-2011-022, OUTP-11-33P",
    doi = "10.1016/j.nuclphysb.2011.04.009",
    journal = "Nucl. Phys. B",
    volume = "850",
    pages = "1--30",
    year = "2011"
}

@article{Ross:2011xv,
    author = "Ross, Graham G. and Schmidt-Hoberg, Kai",
    title = "{The Fine-Tuning of the Generalised NMSSM}",
    eprint = "1108.1284",
    archivePrefix = "arXiv",
    primaryClass = "hep-ph",
    reportNumber = "OUTP-11-48P",
    doi = "10.1016/j.nuclphysb.2012.05.007",
    journal = "Nucl. Phys. B",
    volume = "862",
    pages = "710--719",
    year = "2012"
}

@article{Ross:2012nr,
    author = "Ross, Graham G. and Schmidt-Hoberg, Kai and Staub, Florian",
    title = "{The Generalised NMSSM at One Loop: Fine Tuning and Phenomenology}",
    eprint = "1205.1509",
    archivePrefix = "arXiv",
    primaryClass = "hep-ph",
    reportNumber = "OUTP-12-06P, BONN-TH-2012-04",
    doi = "10.1007/JHEP08(2012)074",
    journal = "JHEP",
    volume = "08",
    pages = "074",
    year = "2012"
}

@article{Ferrara:2010in,
    author = "Ferrara, Sergio and Kallosh, Renata and Linde, Andrei and Marrani, Alessio and Van Proeyen, Antoine",
    title = "{Superconformal Symmetry, NMSSM, and Inflation}",
    eprint = "1008.2942",
    archivePrefix = "arXiv",
    primaryClass = "hep-th",
    reportNumber = "CERN-PH-TH-2010-182, SU-ITP-2010-15",
    doi = "10.1103/PhysRevD.83.025008",
    journal = "Phys. Rev. D",
    volume = "83",
    pages = "025008",
    year = "2011"
}

@article{Einhorn:2009bh,
    author = "Einhorn, Martin B. and Jones, D. R. Timothy",
    title = "{Inflation with Non-minimal Gravitational Couplings in Supergravity}",
    eprint = "0912.2718",
    archivePrefix = "arXiv",
    primaryClass = "hep-ph",
    reportNumber = "LTH-860, NSF-KITP-09-216",
    doi = "10.1007/JHEP03(2010)026",
    journal = "JHEP",
    volume = "03",
    pages = "026",
    year = "2010"
}

@article{Dev:2023kzu,
    author = "Dev, P. S. Bhupal and Mohapatra, Rabindra N. and Zhang, Yongchao",
    title = "{Explanation of the 95 GeV {\ensuremath{\gamma}}{\ensuremath{\gamma}} and bb{\textasciimacron} excesses in the minimal left-right symmetric model}",
    eprint = "2312.17733",
    archivePrefix = "arXiv",
    primaryClass = "hep-ph",
    reportNumber = "CETUP-2023-021",
    doi = "10.1016/j.physletb.2024.138481",
    journal = "Phys. Lett. B",
    volume = "849",
    pages = "138481",
    year = "2024"
}

@article{Chen:2023bqr,
    author = "Chen, Ting-Kuo and Chiang, Cheng-Wei and Heinemeyer, Sven and Weiglein, Georg",
    title = "{95~GeV Higgs boson in the Georgi-Machacek model}",
    eprint = "2312.13239",
    archivePrefix = "arXiv",
    primaryClass = "hep-ph",
    doi = "10.1103/PhysRevD.109.075043",
    journal = "Phys. Rev. D",
    volume = "109",
    number = "7",
    pages = "075043",
    year = "2024"
}

@article{Borah:2023hqw,
    author = "Borah, Debasish and Mahapatra, Satyabrata and Paul, Partha Kumar and Sahu, Narendra",
    title = "{Scotogenic U(1)L{\ensuremath{\mu}}-L{\ensuremath{\tau}} origin of (g-2){\ensuremath{\mu}}, W-mass anomaly and 95~GeV excess}",
    eprint = "2310.11953",
    archivePrefix = "arXiv",
    primaryClass = "hep-ph",
    doi = "10.1103/PhysRevD.109.055021",
    journal = "Phys. Rev. D",
    volume = "109",
    number = "5",
    pages = "055021",
    year = "2024"
}

@article{Arcadi:2023smv,
    author = "Arcadi, Giorgio and Busoni, Giorgio and Cabo-Almeida, David and Krishnan, Navneet",
    title = "{Is there a scalar or pseudoscalar at 95~GeV?}",
    eprint = "2311.14486",
    archivePrefix = "arXiv",
    primaryClass = "hep-ph",
    doi = "10.1103/PhysRevD.110.115028",
    journal = "Phys. Rev. D",
    volume = "110",
    number = "11",
    pages = "115028",
    year = "2024"
}

@article{Ahriche:2023wkj,
    author = "Ahriche, Amine",
    title = "{95~GeV excess in the Georgi-Machacek model: Single or twin peak resonance}",
    eprint = "2312.10484",
    archivePrefix = "arXiv",
    primaryClass = "hep-ph",
    doi = "10.1103/PhysRevD.110.035010",
    journal = "Phys. Rev. D",
    volume = "110",
    number = "3",
    pages = "035010",
    year = "2024"
}

@article{Coloretti:2023wng,
    author = "Coloretti, Guglielmo and Crivellin, Andreas and Bhattacharya, Srimoy and Mellado, Bruce",
    title = "{Searching for low-mass resonances decaying into W bosons}",
    eprint = "2302.07276",
    archivePrefix = "arXiv",
    primaryClass = "hep-ph",
    reportNumber = "PSI-PR-23-4, ZU-TH 09/23, ICPP-70",
    doi = "10.1103/PhysRevD.108.035026",
    journal = "Phys. Rev. D",
    volume = "108",
    number = "3",
    pages = "035026",
    year = "2023"
}

@article{Hollik:2018wrr,
    author = "Hollik, Wolfgang G. and Weiglein, Georg and Wittbrodt, Jonas",
    title = "{Impact of Vacuum Stability Constraints on the Phenomenology of Supersymmetric Models}",
    eprint = "1812.04644",
    archivePrefix = "arXiv",
    primaryClass = "hep-ph",
    reportNumber = "DESY 18-148, DESY-18-148, TTP 18-036",
    doi = "10.1007/JHEP03(2019)109",
    journal = "JHEP",
    volume = "03",
    pages = "109",
    year = "2019"
}

@article{Andreassen:2016cvx,
    author = "Andreassen, Anders and Farhi, David and Frost, William and Schwartz, Matthew D.",
    title = "{Precision decay rate calculations in quantum field theory}",
    eprint = "1604.06090",
    archivePrefix = "arXiv",
    primaryClass = "hep-th",
    doi = "10.1103/PhysRevD.95.085011",
    journal = "Phys. Rev. D",
    volume = "95",
    number = "8",
    pages = "085011",
    year = "2017"
}

@article{Liu:2024cbr,
    author = "Liu, Chang-Xin and Zhou, Yang and Zheng, Xiao-Yu and Ma, Jiao and Feng, Tai-Fu and Zhang, Hai-Bin",
    title = "{95~GeV excess in a CP-violating {\ensuremath{\mu}}-from-{\ensuremath{\nu}} SSM}",
    eprint = "2402.00727",
    archivePrefix = "arXiv",
    primaryClass = "hep-ph",
    doi = "10.1103/PhysRevD.109.056001",
    journal = "Phys. Rev. D",
    volume = "109",
    number = "5",
    pages = "056001",
    year = "2024"
}

@article{Wang:2024bkg,
    author = "Wang, Kun and Zhu, Jingya",
    title = "{95 GeV light Higgs in the top-pair-associated diphoton channel at the LHC in the minimal dilaton model*}",
    eprint = "2402.11232",
    archivePrefix = "arXiv",
    primaryClass = "hep-ph",
    doi = "10.1088/1674-1137/ad4268",
    journal = "Chin. Phys. C",
    volume = "48",
    number = "7",
    pages = "073105",
    year = "2024"
}

@article{Dong:2025orv,
    author = "Dong, Yabo and Ruan, Manqi and Wang, Kun and Yang, Haijun and Zhu, Jingya",
    title = "{Testing a 95 GeV Scalar at the CEPC with Machine Learning*}",
    eprint = "2506.21454",
    archivePrefix = "arXiv",
    primaryClass = "hep-ph",
    doi = "10.1088/1674-1137/ae2ebc",
    journal = "Chin. Phys. C",
    volume = "50",
    number = "3",
    pages = "031001",
    year = "2026",
    note = "[Erratum: Chin.Phys.C 50, 059001 (2026)]"
}

@article{Sharma:2024vhv,
    author = "Sharma, Pramod and Mulaudzi, Anza-Tshilidzi and Mosala, Karabo and Mathaha, Thuso and Kumar, Mukesh and Mellado, Bruce and Crivellin, Andreas and Titov, Maxim and Ruan, Manqi and Fang, Yaquan",
    title = "{Discovery potential of future electron-positron colliders for a 95 GeV scalar}",
    eprint = "2407.16806",
    archivePrefix = "arXiv",
    primaryClass = "hep-ph",
    doi = "10.1016/j.physletb.2025.139953",
    journal = "Phys. Lett. B",
    volume = "870",
    pages = "139953",
    year = "2025"
}

@article{Kumar:2025xng,
    author = "Kumar, Mukesh and Sharma, Pramod and Mosala, Karabo and Mellado, Bruce and Crivellin, Andreas",
    title = "{Exploring the Discovery Reach for a 95 GeV Scalar in Future e+e{\ensuremath{-}}e+e{\ensuremath{-}} Collisions}",
    eprint = "2511.03479",
    archivePrefix = "arXiv",
    primaryClass = "hep-ph",
    doi = "10.22323/1.512.0148",
    journal = "PoS",
    volume = "DIS2025",
    pages = "148",
    year = "2025"
}

@article{Robens:2025kaa,
    author = "Robens, Tania",
    title = "{Extended Scalar Sectors from all angles - mostly at lepton colliders -}",
    eprint = "2504.11969",
    archivePrefix = "arXiv",
    primaryClass = "hep-ph",
    reportNumber = "RBI-ThPhys-2025-20, CERN-TH-2025-082",
    doi = "10.22323/1.490.0077",
    journal = "PoS",
    volume = "CORFU2024",
    pages = "077",
    year = "2025"
}

@inproceedings{Robens:2026xfc,
    author = "Robens, Tania",
    title = "{Low mass scalars at $e^+e-$ colliders}",
    booktitle = "{International Workshop on Future Linear Colliders}",
    eprint = "2604.27541",
    archivePrefix = "arXiv",
    primaryClass = "hep-ph",
    reportNumber = "RBI-ThPhys-2026-07",
    month = "4",
    year = "2026"
}

@article{Benbrik:2025hol,
    author = "Benbrik, Rachid and Boukidi, Mohammed and Kahime, Khouloud and Moretti, Stefano and Rahili, Larbi and Taki, Bassim",
    title = "{Exploring potential Higgs resonances at 650 GeV and 95 GeV in the 2HDM Type III}",
    eprint = "2505.07811",
    archivePrefix = "arXiv",
    primaryClass = "hep-ph",
    doi = "10.1016/j.physletb.2025.139688",
    journal = "Phys. Lett. B",
    volume = "868",
    pages = "139688",
    year = "2025"
}

@article{Xu:2025vmy,
    author = "Xu, Haotian and Wang, Yufei and Han, Xiao-Fang and Wang, Lei",
    title = "{95 GeV Higgs boson and nano-Hertz gravitational waves from domain walls in the N2HDM}",
    eprint = "2505.03592",
    archivePrefix = "arXiv",
    primaryClass = "hep-ph",
    month = "5",
    year = "2025"
}

@article{Abbas:2025ser,
    author = "Abbas, Gauhar and Singh, Vartika and Singh, Neelam",
    title = "{Dark-technicolour at colliders}",
    eprint = "2504.21593",
    archivePrefix = "arXiv",
    primaryClass = "hep-ph",
    month = "4",
    year = "2025"
}

@article{Li:2025tkm,
    author = "Li, Zetian and Liu, Ning and Zhu, Bin",
    title = "{Interplay of $95$ GeV Diphoton Excess and Dark Matter in Supersymmetric Triplet Model}",
    eprint = "2504.21273",
    archivePrefix = "arXiv",
    primaryClass = "hep-ph",
    reportNumber = "CPTNP-2025-004",
    month = "4",
    year = "2025"
}

@article{Du:2025eop,
    author = "Du, Xiaokang and Liu, Huiling and Chang, Qin",
    title = "{Interpretation of 95~GeV excess within the Georgi-Machacek model in light of positive definiteness constraints}",
    eprint = "2502.06444",
    archivePrefix = "arXiv",
    primaryClass = "hep-ph",
    doi = "10.1103/w2xw-q49n",
    journal = "Phys. Rev. D",
    volume = "112",
    number = "1",
    pages = "015019",
    year = "2025"
}

@article{Gao:2024ljl,
    author = "Gao, Song and Zhao, Shu-Min and Di, Shuang and Dong, Xing-Xing and Feng, Tai-Fu",
    title = "{A 95 GeV Higgs boson in the U(1)XSSM}",
    eprint = "2411.13261",
    archivePrefix = "arXiv",
    primaryClass = "hep-ph",
    doi = "10.1016/j.nuclphysb.2025.117026",
    journal = "Nucl. Phys. B",
    volume = "1018",
    pages = "117026",
    year = "2025"
}

@article{Gao:2024qag,
    author = "Gao, Jing and Han, Xiao-Fang and Ma, Jinghong and Wang, Lei and Xu, Haotian",
    title = "{95~GeV Higgs boson and spontaneous CP-violation at the finite temperature}",
    eprint = "2408.03705",
    archivePrefix = "arXiv",
    primaryClass = "hep-ph",
    doi = "10.1103/PhysRevD.110.115045",
    journal = "Phys. Rev. D",
    volume = "110",
    number = "11",
    pages = "115045",
    year = "2024"
}

@article{Muhlleitner:2020wwk,
    author = {M{\"u}hlleitner, Margarete and Sampaio, Marco O. P. and Santos, Rui and Wittbrodt, Jonas},
    title = "{ScannerS: parameter scans in extended scalar sectors}",
    eprint = "2007.02985",
    archivePrefix = "arXiv",
    primaryClass = "hep-ph",
    reportNumber = "KA-TP-05-2020, LU TP 20-38",
    doi = "10.1140/epjc/s10052-022-10139-w",
    journal = "Eur. Phys. J. C",
    volume = "82",
    number = "3",
    pages = "198",
    year = "2022"
}

@article{CMS:2019pzc,
    author = "Sirunyan, Albert M and others",
    collaboration = "CMS",
    title = "{Search for heavy Higgs bosons decaying to a top quark pair in proton-proton collisions at $\sqrt{s} =$ 13 TeV}",
    eprint = "1908.01115",
    archivePrefix = "arXiv",
    primaryClass = "hep-ex",
    reportNumber = "CMS-HIG-17-027, CERN-EP-2019-147",
    doi = "10.1007/JHEP04(2020)171",
    journal = "JHEP",
    volume = "04",
    pages = "171",
    year = "2020",
    note = "[Erratum: JHEP 03, 187 (2022)]"
}

@article{Benbrik:2025wkz,
    author = "Benbrik, Rachid and Boukidi, Mohammed and Kahime, Khouloud and Moretti, Stefano and Rahili, Larbi and Taki, Bassim",
    title = "{Interpreting the 650 GeV and 95 GeV Higgs Anomalies in the N2HDM}",
    eprint = "2510.19605",
    archivePrefix = "arXiv",
    primaryClass = "hep-ph",
    month = "10",
    year = "2025"
}

@article{Fayet:1974pd,
    author = "Fayet, Pierre",
    title = "{Supergauge Invariant Extension of the Higgs Mechanism and a Model for the electron and Its Neutrino}",
    reportNumber = "PTENS-74-7",
    doi = "10.1016/0550-3213(75)90636-7",
    journal = "Nucl. Phys. B",
    volume = "90",
    pages = "104--124",
    year = "1975"
}

@article{Fayet:1976et,
    author = "Fayet, Pierre",
    title = "{Supersymmetry and Weak, Electromagnetic and Strong Interactions}",
    reportNumber = "PTENS 76/14",
    doi = "10.1016/0370-2693(76)90319-1",
    journal = "Phys. Lett. B",
    volume = "64",
    pages = "159",
    year = "1976"
}

@article{Fayet:1976cr,
    author = "Fayet, Pierre and Ferrara, S.",
    title = "{Supersymmetry}",
    reportNumber = "LPTENS-76-11",
    doi = "10.1016/0370-1573(77)90066-7",
    journal = "Phys. Rept.",
    volume = "32",
    pages = "249--334",
    year = "1977"
}

@article{Fayet:1977yc,
    author = "Fayet, Pierre",
    title = "{Spontaneously Broken Supersymmetric Theories of Weak, Electromagnetic and Strong Interactions}",
    reportNumber = "LPTENS 77/11",
    doi = "10.1016/0370-2693(77)90852-8",
    journal = "Phys. Lett. B",
    volume = "69",
    pages = "489",
    year = "1977"
}

@article{Constantin:2025mex,
    author = "Constantin, L. and Kraml, S. and Mahmoudi, F.",
    title = "{The LHC has ruled out supersymmetry {\textendash} really?}",
    eprint = "2505.11251",
    archivePrefix = "arXiv",
    primaryClass = "hep-ph",
    reportNumber = "CERN-TH-2025-100",
    doi = "10.1016/j.nuclphysb.2025.117012",
    journal = "Nucl. Phys. B",
    volume = "1018",
    pages = "117012",
    year = "2025"
}

@article{Jeanty:2026etw,
    author = "Jeanty, Laura and Lee, Lawrence",
    title = "{Rare and Experimentally Challenging Supersymmetry Signatures}",
    eprint = "2601.06358",
    archivePrefix = "arXiv",
    primaryClass = "hep-ex",
    month = "1",
    year = "2026"
}

@article{ATLAS:2024lda,
    author = "Aad, Georges and others",
    collaboration = "ATLAS",
    title = "{The quest to discover supersymmetry at the ATLAS experiment}",
    eprint = "2403.02455",
    archivePrefix = "arXiv",
    primaryClass = "hep-ex",
    reportNumber = "CERN-EP-2024-056",
    doi = "10.1016/j.physrep.2024.09.010",
    journal = "Phys. Rept.",
    volume = "1116",
    pages = "261--300",
    year = "2025"
}

@article{Farrar:1978xj,
    author = "Farrar, Glennys R. and Fayet, Pierre",
    title = "{Phenomenology of the Production, Decay, and Detection of New Hadronic States Associated with Supersymmetry}",
    reportNumber = "CALT-68-648",
    doi = "10.1016/0370-2693(78)90858-4",
    journal = "Phys. Lett. B",
    volume = "76",
    pages = "575--579",
    year = "1978"
}

@article{YaserAyazi:2024hpj,
    author = "Yaser Ayazi, Seyed and Hosseini, Mojtaba and Paktinat Mehdiabadi, Saeid and Rouzbehi, Rouzbeh",
    title = "{Vector dark matter and LHC constraints, including a 95~GeV light Higgs boson}",
    eprint = "2405.01132",
    archivePrefix = "arXiv",
    primaryClass = "hep-ph",
    doi = "10.1103/PhysRevD.110.055004",
    journal = "Phys. Rev. D",
    volume = "110",
    number = "5",
    pages = "055004",
    year = "2024"
}

@article{Mondal:2024obd,
    author = "Mondal, Tanmoy and Moretti, Stefano and Sanyal, Prasenjit",
    title = "{On the CP Nature of the `95 GeV' Anomalies}",
    eprint = "2412.00474",
    archivePrefix = "arXiv",
    primaryClass = "hep-ph",
    month = "11",
    year = "2024"
}

@article{Khanna:2024bah,
    author = "Khanna, Akshat and Moretti, Stefano and Sarkar, Agnivo",
    title = "{Explaining 95 (or so) GeV Anomalies in the 2-Higgs Doublet Model Type-I}",
    eprint = "2409.02587",
    archivePrefix = "arXiv",
    primaryClass = "hep-ph",
    reportNumber = "HRI-RECAPP-2024-05",
    month = "9",
    year = "2024"
}

@article{Khanna:2025cwq,
    author = "Khanna, Akshat and Moretti, Stefano and Sarkar, Agnivo",
    title = "{Explaining 650 GeV and 95 GeV Anomalies in the 2-Higgs Doublet Model Type-I}",
    eprint = "2509.06017",
    archivePrefix = "arXiv",
    primaryClass = "hep-ph",
    month = "9",
    year = "2025"
}

@article{Bhatnagar:2025jhh,
    author = "Bhatnagar, Ansh and Croon, Djuna and Schicho, Philipp",
    title = "{Interpreting the 95 GeV resonance in the Two Higgs Doublet Model. Implications for the electroweak phase transition}",
    eprint = "2506.20716",
    archivePrefix = "arXiv",
    primaryClass = "hep-ph",
    reportNumber = "IPPP/25/39",
    doi = "10.1007/JHEP03(2026)014",
    journal = "JHEP",
    volume = "03",
    pages = "014",
    year = "2026"
}

@article{Arcadi:2025grl,
    author = "Arcadi, Giorgio and Djouadi, Abdelhak",
    title = "{Interpreting the current Higgs excesses at the LHC in the 2HD+a framework}",
    eprint = "2512.08807",
    archivePrefix = "arXiv",
    primaryClass = "hep-ph",
    month = "12",
    year = "2025"
}

@article{Chen:2025vtg,
    author = "Chen, Ting-Kuo and Chiang, Cheng-Wei and Heinemeyer, Sven and Weiglein, Georg",
    title = "{Interpretation of LHC excesses at 95 GeV and 152 GeV in an extended Georgi-Machacek model}",
    eprint = "2511.04796",
    archivePrefix = "arXiv",
    primaryClass = "hep-ph",
    reportNumber = "DESY-25-159",
    month = "11",
    year = "2025"
}

@article{Dong:2025exu,
    author = "Dong, Yabo and Ruan, Manqi and Wang, Kun and Yang, Haijun and Zhu, Jingya",
    title = "{Prospects for a 95 GeV Higgs Boson at Future Higgs Factories with Transformer Networks}",
    eprint = "2510.24662",
    archivePrefix = "arXiv",
    primaryClass = "hep-ph",
    month = "10",
    year = "2025"
}

@article{Chang:2025bjt,
    author = "Chang, Qin and Du, Xiaokang and Zhu, Pengxuan",
    title = "{Unified interpretation of 95 GeV di-photon and di-tau Excesses in the Georgi-Machacek Model}",
    eprint = "2509.26155",
    archivePrefix = "arXiv",
    primaryClass = "hep-ph",
    month = "9",
    year = "2025"
}

@article{Hmissou:2025riw,
    author = "Hmissou, Ayoub and Moretti, Stefano and Rahili, Larbi",
    title = "{Could the 650~GeV excess be a pseudoscalar of a three-Higgs-doublet model?}",
    eprint = "2509.06232",
    archivePrefix = "arXiv",
    primaryClass = "hep-ph",
    doi = "10.1103/6qs7-869p",
    journal = "Phys. Rev. D",
    volume = "112",
    number = "7",
    pages = "075049",
    year = "2025"
}

@article{Dutta:2026tbc,
    author = "Dutta, J. and Ferreira, P. M. and Heinemeyer, S.",
    title = "{Pseudoscalar contributions to Zh production at the LHC at 95 GeV and above}",
    eprint = "2603.22494",
    archivePrefix = "arXiv",
    primaryClass = "hep-ph",
    reportNumber = "IFT-UAM/CSIC-26-031",
    month = "3",
    year = "2026"
}

@article{Baek:2024cco,
    author = "Baek, Seungwon and Ko, P. and Omura, Yuji and Yu, Chaehyun",
    title = "{96~GeV scalar boson in the 2HDM with U(1)$_{H}$ gauge symmetry}",
    eprint = "2412.02178",
    archivePrefix = "arXiv",
    primaryClass = "hep-ph",
    doi = "10.1140/epjc/s10052-025-14630-y",
    journal = "Eur. Phys. J. C",
    volume = "85",
    number = "8",
    pages = "908",
    year = "2025"
}

@article{Dong:2024ipo,
    author = "Dong, Yabo and Wang, Kun and Zhu, Jingya",
    title = "{Probing a type I 2HDM light Higgs boson in the top-pair-associated diphoton channel}",
    eprint = "2410.13636",
    archivePrefix = "arXiv",
    primaryClass = "hep-ph",
    doi = "10.1103/2ss7-ygnb",
    journal = "Phys. Rev. D",
    volume = "112",
    number = "5",
    pages = "055013",
    year = "2025"
}

@article{Kalinowski:2024uxe,
    author = "Kalinowski, Jan and Kotlarski, Wojciech",
    title = "{Interpreting 95 GeV di-photon/$ b\overline{b} $ excesses as a lightest Higgs boson of the MRSSM}",
    eprint = "2403.08720",
    archivePrefix = "arXiv",
    primaryClass = "hep-ph",
    doi = "10.1007/JHEP07(2024)037",
    journal = "JHEP",
    volume = "07",
    pages = "037",
    year = "2024"
}

@article{Bolz:2000fu,
    author = "Bolz, Markus and Brandenburg, Alexander and Buchmuller, Wilfried",
    title = "{Thermal production of gravitinos}",
    eprint = "hep-ph/0012052",
    archivePrefix = "arXiv",
    reportNumber = "DESY-00-167",
    doi = "10.1016/S0550-3213(01)00132-8",
    journal = "Nucl. Phys. B",
    volume = "606",
    pages = "518--544",
    year = "2001",
    note = "[Erratum: Nucl. Phys. B 790, 336--337 (2008)]"
}

@article{Pradler:2006qh,
    author = "Pradler, Josef and Steffen, Frank Daniel",
    title = "{Thermal gravitino production and collider tests of leptogenesis}",
    eprint = "hep-ph/0608344",
    archivePrefix = "arXiv",
    reportNumber = "MPP-2006-101",
    doi = "10.1103/PhysRevD.75.023509",
    journal = "Phys. Rev. D",
    volume = "75",
    pages = "023509",
    year = "2007"
}

@article{Feng:2003uy,
    author = "Feng, Jonathan L. and Rajaraman, Arvind and Takayama, Fumihiro",
    title = "{Superweakly interacting massive particles}",
    eprint = "hep-ph/0302215",
    archivePrefix = "arXiv",
    reportNumber = "UCI-TR-2003-07, MCTP-03-07",
    doi = "10.1103/PhysRevLett.91.011302",
    journal = "Phys. Rev. Lett.",
    volume = "91",
    pages = "011302",
    year = "2003"
}

@article{Covi:2009bk,
    author = "Covi, Laura and Hasenkamp, Jasper and Pokorski, Stefan and Roberts, Jonathan",
    title = "{Gravitino Dark Matter and general neutralino NLSP}",
    eprint = "0908.3399",
    archivePrefix = "arXiv",
    primaryClass = "hep-ph",
    reportNumber = "DESY-09-062",
    doi = "10.1088/1126-6708/2009/11/003",
    journal = "JHEP",
    volume = "11",
    pages = "003",
    year = "2009"
}

@mastersthesis{Hasenkamp:2009zz,
    author = "Hasenkamp, Jasper",
    title = "{General neutralino NLSP with gravitino dark matter vs. big bang nucleosynthesis}",
    reportNumber = "DESY-THESIS-2009-016",
    doi = "10.3204/DESY-THESIS-2009-016",
    type = "Other thesis",
    month = "3",
    year = "2009"
}

@article{Ambrosanio:1996jn,
    author = "Ambrosanio, S. and Kane, Gordon L. and Kribs, Graham D. and Martin, Stephen P. and Mrenna, S.",
    title = "{Search for supersymmetry with a light gravitino at the Fermilab Tevatron and CERN LEP colliders}",
    eprint = "hep-ph/9605398",
    archivePrefix = "arXiv",
    reportNumber = "ANL-HEP-PR-97-31",
    doi = "10.1103/PhysRevD.54.5395",
    journal = "Phys. Rev. D",
    volume = "54",
    pages = "5395--5411",
    year = "1996"
}

@article{Kawasaki:2004yh,
    author = "Kawasaki, Masahiro and Kohri, Kazunori and Moroi, Takeo",
    title = "{Big-Bang nucleosynthesis and hadronic decay of long-lived massive particles}",
    eprint = "astro-ph/0408426",
    archivePrefix = "arXiv",
    reportNumber = "ICRR-REPORT-513-2004-5, TU-710, YITP-04-39",
    doi = "10.1103/PhysRevD.71.083502",
    journal = "Phys. Rev. D",
    volume = "71",
    pages = "083502",
    year = "2005"
}

@article{ATLAS:SUSY201607,
    author = "Aaboud, Morad and others",
    collaboration = "ATLAS",
    title = "{Search for squarks and gluinos in final states with jets and missing transverse momentum using 36 fb$^{-1}$ of $\sqrt{s}=13$ TeV pp collision data with the ATLAS detector}",
    eprint = "1712.02332",
    archivePrefix = "arXiv",
    primaryClass = "hep-ex",
    reportNumber = "ATLAS-SUSY-2016-07",
    doi = "10.1103/PhysRevD.97.112001",
    journal = "Phys. Rev. D",
    volume = "97",
    number = "11",
    pages = "112001",
    year = "2018"
}

@article{CMS:SUS19006,
    author = "Sirunyan, Albert M. and others",
    collaboration = "CMS",
    title = "{Search for supersymmetry in proton-proton collisions at 13 TeV in final states with jets and missing transverse momentum}",
    eprint = "1908.04722",
    archivePrefix = "arXiv",
    primaryClass = "hep-ex",
    reportNumber = "CMS-SUS-19-006",
    doi = "10.1007/JHEP10(2019)244",
    journal = "JHEP",
    volume = "10",
    pages = "244",
    year = "2019"
}

@article{Wang:2026rde,
    author = "Wang, Zhi-Chuan and Yang, Jin-Lei and Qin, Qi-Zhen and Zhang, Wen-Hui and Feng, Tai-Fu",
    title = "{95 and 125~GeV Higgs boson excesses in the left-right supersymmetric standard model}",
    eprint = "2602.13976",
    archivePrefix = "arXiv",
    primaryClass = "hep-ph",
    doi = "10.1103/t865-lq98",
    journal = "Phys. Rev. D",
    volume = "113",
    number = "5",
    pages = "055037",
    year = "2026"
}

\end{document}